\documentclass[a4paper,oneside,openany,11pt]{memoir}
\usepackage{geometry}
\usepackage[english]{babel}	
\usepackage{xspace}
\usepackage{microtype}	
\usepackage[utf8]{inputenc}	% Allows for writing special charachters in the tex-file 

\usepackage{amsmath,amssymb,amsfonts,dsfont,bm,mathtools} 	% Standard mathematics 
\allowdisplaybreaks 	%Allows pagebreak during multiline math enviroments
\usepackage{mathrsfs}
\usepackage{xcolor}
\usepackage{hyperref}
\usepackage{slashed}
\usepackage{youngtab}

\usepackage{siunitx}
\usepackage[sort&compress,numbers,merge]{natbib}
\addto\captionsenglish{\renewcommand*{\bibname}{References}}
\makeatletter
\renewcommand{\@memb@bchap}{ \section*{\bibname} \bibmark \prebibhook}
\makeatother

\usepackage{graphicx}
\graphicspath{{./Figures/}{./A_graphs/}{./Beta_graphs/}{./T_graphs/}}
\usepackage{caption}
\usepackage{multirow}

\usepackage{tabularx}
\newcolumntype{Y}{>{\centering\arraybackslash}X}

\usepackage[noindentafter]{titlesec} %Removes the indentation after new section
\counterwithout{section}{chapter} % Include if sections should be taken at the top level
\counterwithout{figure}{chapter} %Removes chapter number on figures
\counterwithout{table}{chapter} %Removes chapter number on figures
\numberwithin{equation}{section} % Sets numbering at the section level
\setsecnumdepth{subsection}

\titleformat{\section}{\centering \Large \bfseries \sffamily \color{blue!90!black} }{\thesection}{15pt}{}{}
\titlespacing{\section}{0pt}{15pt}{5pt}
\titleformat{\subsection}{\large \sffamily \color{blue!90!black} }{\thesubsection}{10pt}{}{}
\titlespacing{\subsection}{0pt}{10pt}{5pt}
\titleformat{\subsubsection}{\large \itshape \color{blue!70!black} }{\thesubsection}{10pt}{}{}
\titlespacing{\subsubsection}{0pt}{10pt}{0pt}

\makepagestyle{myruled}
\makeoddhead{myruled}{\leftmark}{}{\thepage}
\makeheadrule{myruled}{\textwidth}{\normalrulethickness}
\makepsmarks{myruled}{
	\nouppercaseheads
	\createmark{chapter}{both}{shownumber}{\ }{. \ }
	\createmark{section}{both}{shownumber}{}{. \ }
}
\maxtocdepth{subsection}
\makeatletter
\renewcommand*{\@tocmaketitle}{}
\makeatother

\definecolor{rossoCP3}{cmyk}{0,.88,.77,.40}
\definecolor{blue}{rgb}{0,0.396,0.741}
\colorlet{blueRef}{blue!80!black}
\hypersetup{
	colorlinks, 
	bookmarksnumbered,
	citecolor=blueRef, 		%color of links to bibliography
	linkcolor=rossoCP3,	%color of internal links
	urlcolor=rossoCP3,			%color of external links 
}

\newcommand{\braces}[1]{\left\lbrace #1 \right\rbrace}

\newcommand{\brakets}[1]{\left\langle #1 \right\rangle}

\newcommand{\dd}{\mathop{}\!\mathrm{d}}
\renewcommand{\div}{\mathop{}\!\text{div}}

\newcommand{\Tr}[1]{\mathop{}\!\mathrm{Tr}\! \left[ #1 \right]}
\newcommand{\diag}{\mathop{}\!\mathrm{diag}}

\newcommand{\transpose}{^{\mathrm{T}}}

\newcommand{\andeq}{\quad \mathrm{and} \quad}

\newcommand{\U}{\mathrm{U}}
\newcommand{\SU}{\mathrm{SU}}

\newcommand{\SO}{\mathrm{SO}}

\newcommand{\udindices}[2]{\phantom{}^{#1}\phantom{}_{#2}}

\newcommand{\grafer}{\texttt{GRAFER}\xspace}
\newcommand{\gfe}{Osborn's equation\xspace}
\def\msbar{$\overline{\text{{\small MS}}}$\xspace}
\def\mom{${\text{{\footnotesize MOM}}}$\xspace}

\newcommand{\cofg}[2]{\mathfrak{g}^{(#1)}_{#2}\,}
\newcommand{\cofq}[2]{\mathfrak{q}^{(#1)}_{#2}\,}
\newcommand{\cofy}[2]{\mathfrak{y}^{(#1)}_{#2}\,}
\newcommand{\cofA}[2]{\mathcal{A}^{(#1)}_{#2}\,}
\newcommand{\cofT}[2]{\mathcal{T}^{(#1)}_{#2}\,}
\newcommand{\cofsf}[2]{\mathfrak{f}^{(#1)}_{#2}\,}
\newcommand{\cofss}[2]{\mathfrak{s}^{(#1)}_{#2}\,}
\newcommand{\cg}[1]{[C_2(G)]_{#1}}
\newcommand{\tf}[1]{[S_2(F)]_{#1}}
\newcommand{\ts}[1]{[S_2(S)]_{#1}}
\newcommand{\cf}{C_2(F)}
\newcommand{\cft}{\tilde{C}_2(F)}
\renewcommand{\cs}[1]{[C_2(S)]_{#1}}
\newcommand{\yf}{Y_2(F)}
\newcommand{\yft}{\tilde{Y}_2(F)}
\newcommand{\ys}[1]{[Y_2(S)]_{#1}}
\newcommand{\cfe}[1]{C_2(F,\,#1)}

\newcommand{\cse}[2]{[C_2(S,\, #1)]_{#2}}
\newcommand{\tfe}[2]{[S_2(F, \, #1)]_{#2}}
\newcommand{\tse}[2]{[S_2(S, \, #1)]_{#2}}
\newcommand{\yfe}[1]{Y_2(F,\,#1)}

\newcommand{\yse}[2]{[Y_2(S,\, #1)]_{#2}}
\newcommand{\yyf}{Y_4(F)}
\newcommand{\yys}[1]{[Y_4(S)]_{#1}}

\newcommand{\nn}{\nonumber\\}
\DeclareMathAlphabet{\mathpzc}{OT1}{pzc}{m}{it}

\begin{document}

% % % % Title % % % % 
\thispagestyle{empty}
{\small \hfill CP3-Origins-2019-23 DNRF90}
\vspace*{0.04\textheight}
\begin{center}
{{\huge \sffamily 
	Constraints on 3- and 4-loop $\beta$-functions in a general four-dimensional Quantum Field Theory
}}\\[-.5em]
\rule{\textwidth}{3pt}\par 
\vspace{.02\textheight}
{{\bfseries \Large Colin Poole\footnote{ \texttt{cpoole@cp3.sdu.dk}} {\normalsize and} Anders Eller Thomsen\footnote{ \texttt{aethomsen@cp3.sdu.dk}}}}\\[.02\textheight]
$\text{CP}^{3}$-Origins, University of Southern Denmark,\\ Campusvej 55, DK-5230 Odense M, Denmark\\
\end{center}
% % % % Title % % % % 

\begin{abstract}\vspace{-.02\textheight}
The $ \beta $-functions of marginal couplings are known to be closely related to the $ A $-function through \gfe, derived using the local renormalization group. It is possible to derive strong constraints on the $\beta$-functions by parametrizing the terms in \gfe as polynomials in the couplings, then eliminating unknown $\tilde{A}$ and $T_{IJ}$ coefficients. In this paper we extend this program to completely general gauge theories with arbitrarily many Abelian and non-Abelian factors. We detail the computational strategy used to extract consistency conditions on $ \beta $-functions, and discuss our automation of the procedure. Finally, we implement the procedure up to 4-, 3-, and 2-loops for the gauge, Yukawa and quartic couplings respectively, corresponding to the present forefront of general $ \beta $-function computations. We find an extensive collection of highly non-trivial constraints, and argue that they constitute an useful supplement to traditional perturbative computations; as a corollary, we present the complete 3-loop gauge $\beta$-function of a general QFT in the \msbar scheme, including kinetic mixing. 
\end{abstract}
\rule{\textwidth}{1pt}\par
{
	\hypersetup{linkcolor = black}
	\tableofcontents*
}

\section{Introduction}
This paper represents the culmination of a rather sporadic research program, investigating constraints on the $\beta$-functions of a general Quantum Field Theory (QFT), given their relation to a particular scalar function of the couplings. In our experience, attempting to explain how such constraints are derived has proven difficult, not least due to the use of an intricate diagrammatic representation for all quantities involved. Therefore, we have attempted to make our paper as comprehensive and pedagogical as possible, starting with the introduction itself, which we have split into two parts: the first part is shorter, and seeks to contextualize our results with applications in mind; the second part is the introduction proper, written from the underlying theoretical perspective.

\subsection{The short version}
The Renormalization Group (RG) formalism implies a change in the couplings of a renormalizable Lagrangian density as one varies the energy scale~\cite{Wilson:1973jj}. The change in a coupling, \emph{i.e.} its RG flow, is described by its $\beta$-function, thus $\beta$-functions are of fundamental importance in QFT. At a given order in perturbation theory, the $\beta$-functions are simply polynomials in the couplings, with the precise form depending on the type of interaction. Given a general Lagrangian with tensor couplings, each $\beta$-function can be parametrized in terms of Tensor Structures (TSs); that is, a basis of contractions between the various couplings and gauge generators, in the style of Machacek and Vaughn~\cite{Machacek:1983tz,Machacek:1983fi,Machacek:1984zw}. This method essentially ``factors out" the actual field content of the theory, so that the coefficient multiplying each TS depends only on the renormalization scheme, allowing one to derive the $\beta$-functions for a completely general theory.

Once an adequate representation of the general $\beta$-functions has been found, investigating non-trivial constraints on possible RG flows becomes much more tractable; furthermore, by using a parametrization with TSs and unknown coefficients, constraints can be investigated in a manifestly scheme-independent manner. The constraints in which we are interested are those derived from \gfe (see Eq. \eqref{eq:GFE} below), which relates coefficients of each $\beta$-function at different loop orders. This is known as the ``3--2--1" ordering, since the coefficients of the gauge $\beta$-function are related to the Yukawa coefficients one loop lower, and the scalar coefficients two loops lower. Previous attempts at investigating these constraints have been limited in different ways: some analyses of \gfe have been done using the $\beta$-functions for a general scalar-fermion theory without gauge interactions \cite{Jack:2013sha}, or a general theory with a simple gauge group \cite{Jack:2014pua}; another analysis included the three gauge groups of the Standard Model (SM), but considered only the top Yukawa and Higgs self-interaction couplings, and limited its investigations to constraints that do not involve higher-order corrections \cite{Antipin:2013sga}.

The power of \gfe comes with the realization that one can investigate the associated constraints algebraically, without actually knowing the $\beta$-function coefficients. Phrased this way, \gfe becomes a way to constrain (or even derive) some higher-order coefficients using only lower-order coefficients. In this paper, we first extend the Machacek and Vaughn construction to manifestly include multiple gauge groups and kinetic mixing, thus creating a representation of the general gauge $\beta$-function in line with the Yukawa and quartic $\beta$-functions. We then perform a complete analysis of \gfe, deriving constraints up to order 4--3--2; not only does this allow us to determine, and for the first time present, the completely general 3-loop gauge $\beta$-function with kinetic mixing, it also gives us the first predictions for certain coefficients of the general 4-loop gauge $\beta$-function, and fixes the known ambiguities that arise at four loops when dimensionally-regularizing integrals that involve $\gamma_{5}$~\cite{Bednyakov:2015ooa,Zoller:2015tha,Poole:2019txl}.

\subsection{The long version}
In 1974, shortly after the formulation of the Renormalization Group (RG) itself \cite{Wilson:1973jj}, Wallace and Zia suggested that the RG flow of a general QFT could in fact be a gradient flow, with the $\beta$-functions of the theory related to derivatives of some scalar quantity via a positive-definite metric \citep{Wallace:1974dx,Wallace:1974dy}. They were able to show that such a gradient formula holds for the $\beta$-function of a general $\phi^{4}$ theory up to two loops, but that modifications at three loops spoil the manifest positive-definiteness of their metric. In addition, they remarked that allowing additional interactions creates similar issues, as a single term in the scalar function may then contribute to multiple $\beta$-functions at different loop orders. Eventually, as a consequence of work done on the Local Renormalization Group (LRG) by Jack and Osborn \cite{Osborn:1989td,Jack:1990eb,Osborn:1991gm}, these features were subsumed into a substantially more general framework, valid for completely general theories.

The LRG is a powerful extension of usual RG methods: not only does it allow one to include the effects of composite operators in a straightforward manner, but it also leads to additional restrictions on the allowed RG behavior of general renormalizable QFTs. The LRG was used in \cite{Osborn:1989td,Jack:1990eb,Osborn:1991gm} to re-derive the $c$-theorem \cite{Zamolodchikov:1986gt}, and to provide a perturbative proof of Cardy's conjectured $a$-theorem \cite{Cardy:1988cwa}. While a weaker formulation has been shown to hold non-perturbatively \cite{Komargodski:2011vj} subject to certain assumptions \cite{Shore:2016xor}, the stronger statement, concerning the existence of a function of the couplings that satisfies a gradient-flow equation and decreases monotonically under RG flow, is still lacking a non-perturbative proof. Nevertheless, for even-dimensional QFTs, there exists a function $\tilde{A}(g)$ of the couplings that satisfies a gradient-flow-like equation, and reduces to the coefficient $ a= \tilde{A}(g^{\ast}) $ of the Euler density in the trace anomaly at RG fixed points, $ g^{\ast} $.

By considering general QFTs as points on a manifold, specified by position-dependent couplings $g^{I}(x)$, one may derive relations between RG functions by imposing the Abelian nature of Weyl rescalings. These relations are known as Weyl Consistency Conditions (Weyl CCs), and include \gfe
	\begin{equation}\label{eq:GFE}
	\partial_{I}\tilde{A} \equiv \frac{\partial \tilde{A}}{\partial g^{I}} = T_{IJ}\beta^{J},
	\end{equation}
which forms the basis of our analysis\footnote{The strong $a$-theorem is equivalent to the existence of a choice of local counter terms, such that the symmetric part $G_{IJ}$ of the tensor $T_{IJ}$ appearing in \eqref{eq:GFE} is positive-definite. An even stronger version is that there exists a choice of local counter terms such that $ T_{IJ} \sim \hat{G}_{IJ}$ is itself symmetric and positive-definite; $\hat{G}_{IJ}$ would then act as a metric on the space of couplings, and the $\beta$-functions would satisfy the gradient-flow equation $\beta^{I} = \partial^{I}\tilde{A} \equiv \hat{G}^{IJ}\partial_{J}\tilde{A}$.}. If the kinetic terms of the theory possess a global symmetry, then $\beta^{I}$ in \eqref{eq:GFE} is replaced by a generalized function $B^{I}$, which begins to differ at three loops~\cite{Jack:2013sha}, and is related to the possible existence of theories with limit cycles \cite{Fortin:2012hn}. While the effects of this shift have been included for the simple case of six-dimensional $\phi^{3}$ theory \cite{Gracey:2015fia} and the more complicated four-dimensional scalar-fermion theory~\cite{Jack:2013sha}, equivalent analyses for general gauge theories have not yet reached the point where considering this shift is necessary \cite{Jack:2014pua}.

All quantities in \gfe are given as functions of the couplings and group structures of the theory. Thus, it is possible to parametrize Eq. \eqref{eq:GFE} in terms of a finite basis of TSs loop order by loop order. The existence of a function $\tilde{A}$ satisfying Eq. \eqref{eq:GFE} then implies relations between the coefficients of the $\beta$-functions of the theory at different loop orders\footnote{The order at which the gauge-, Yukawa- and quartic-$\beta$ functions are related is known as the ``3--2--1" ordering; for topological reasons that will later become clear, this ordering is manifestly preserved as one calculates at higher loops.}. These relations may be formulated in a scheme-independent manner, and are sufficient to predict some higher-order $\beta$-function coefficients, the precise magnitude of the shift $\beta\rightarrow B$ \cite{Gracey:2015fia}, and even the singular parts of certain Feynman integrals \cite{Jack:2016utw,Gracey:2016tuh}. Most recently, it has been shown that \eqref{eq:GFE} implies the existence of a constraint on the 4-loop strong-coupling $\beta$-function of the SM, involving Feynman integrals rendered ambiguous by the treatment of $\gamma_{5}$ in dimensional regularization; the constraint fixes the ambiguity by relating it to unambiguous terms in the general 3-loop Yukawa $\beta$-function \cite{Poole:2019txl}. In these papers, the construction of $\tilde{A}$ was carried out by hand, making use of a diagrammatic representation of the tensor couplings in the corresponding theory. This method was sufficient to render the first few loop orders tractable, but as with any perturbative calculation, the increase in complexity at higher loop orders limits its reliability and practicality. Consequently, the construction of $\tilde{A}$ for a general gauge theory (with a simple gauge group) has only been carried out to four loops, falling one loop short of being able to derive predictive constraints on cutting-edge perturbative calculations.

In order to overcome this restriction, we have endeavored to automate the diagrammatic construction of $\tilde{A}$, plus extraction of the associated consistency conditions (CCs), through the use of a custom Mathematica code, \grafer\footnote{As with all programs in theoretical physics, we too needed a punchy-sounding name with an associated acronym. We have settled on `GRAdient Flow Equation Reducer', which we think is marginally better than the original name, `GRAdient Flow Equation Rrrrrrrsomething'.}. We have also extended the previous diagrammatic representation of the gauge sector to allow for multiple gauge groups and kinetic mixing of the Abelian factors; combined with a parametrization of the $\beta^{I}\rightarrow B^{I}$ modification, \grafer allows us to extract the CCs for arbitrary renormalizable four-dimensional QFTs, based on a compact gauge group with any number of Abelian factors.
Since $\beta$-functions determine the change in couplings at different energy scales (hence controlling the strength of particle interactions), the ability to deduce or relate progressively-higher-order $\beta$-function coefficients has obvious phenomenological relevance. Whether through direct employment or as an independent cross-check of standard perturbative methods, we believe that our approach can be useful in the program of perturbative $ \beta $-function computations.

A key motivation for our work is attempting to demonstrate just how restrictive Weyl CCs can be. Ever since the development of the RG and subsequent discovery of asymptotic freedom in perturbative QFTs \cite{Gross:1973id,Politzer:1973fx}, $\beta$-function computations have been carried out for a huge range of models, with applications ranging from critical exponents, phase transitions and universality, to the existence of asymptotically-safe theories described by interacting RG fixed points \cite{Litim:2014uca}. Despite this ubiquity, comparatively little attention has been given to the idea of calculating $\beta$-functions for completely general theories; whereas the QCD $\beta$-function is now known to five loops \cite{Herzog:2017ohr}, and the $\beta$-functions of the full Standard Model are known to three loops \cite{Mihaila:2012pz,Bednyakov:2013cpa,Bednyakov:2014pia}, the $\beta$-functions for completely general theories have been stuck at two loops for over 35 years \cite{Machacek:1983tz,Machacek:1983fi,Machacek:1984zw} (modulo an extension in the early '00s to theories with kinetic mixing \cite{Luo:2002iq}). The highest-loop general $\beta$-function of which we are aware is the 3-loop gauge $\beta$-function for a general theory with a simple gauge group \cite{Pickering:2001aq}; we shall show that one can combine this result with the constraints imposed by Weyl CCs to infer the 3-loop gauge $\beta$-function for any theory with a compact gauge group. 

The paper is organized as follows: in section \ref{sec:background}, we give a brief review of the LRG, the derivation of \gfe (plus modifications), and the associated tensors; we also outline a conjecture, in which we speculate that it may always be possible to enforce symmetry of the tensor $T_{IJ}$ in \gfe by adding suitable local counter terms. In section \ref{sec:formalism}, we formulate a general, renormalizable, four-dimensional QFT with an arbitrary compact gauge group, indicating how all fields and couplings of a particular type can be assembled into field multiplets and tensor couplings. In section \ref{sec:strategy}, we introduce the graphical notation employed in \cite{Jack:2014pua}, extend the gauge sector to the general case, and describe the overall strategy for deriving constraints on the $\beta$-functions as employed in \grafer. In section \ref{sec:results}, we then present our main results: an extension to the general case of previously-derived CCs at orders 2--1--0 and 3--2--1, followed by a sample and discussion of the conditions at order 4--3--2\footnote{Given the sheer quantity of CCs, it would be impractical to include them all in the main text; instead, we have provided \href{https://arxiv.org/abs/1906.04625}{ancillary files} detailing the \grafer results with all tensors and CCs.}, chosen to highlight certain non-trivial features of \gfe at sufficiently-high order. Section \ref{sec:conclusion} concludes the paper, with comments on how our program can be extended to derive analogous constraints for theories in different spacetime dimensions, and some indications of previous results for supersymmetric theories. The appendix includes the fully general gauge, Yukawa, and quartic $\beta$-functions for a general four-dimensional QFT, in the \msbar scheme, up to order 3--2--2. To facilitate understanding of the diagrammatic notation, we have also included the explicit tensor contractions represented by each diagram in the $\beta$-functions, and diagrams of all $\tilde{A}$ and $T_{IJ}$ contributions at order 3--2--1. We have also included direct comparison with previous work on kinetic mixing and Weyl CCs.

Finally, we would like to comment on the results presented in section \ref{sec:results}: due to the ever-increasing complexity of constraints at higher orders, we have opted to give only \msbar-specific results, thus losing the manifest scheme-independence of the complete constraints derived from \gfe, but gaining \msbar-specific predictive power. Throughout the paper, we have attempted to comment on such scheme-dependent matters when relevant, and will repeatedly emphasize how the underlying method of deriving CCs is completely scheme-independent.

\section{Local Renormalization Group}
\label{sec:background}
Most of the material in this section is essentially yet another retelling of the LRG, the consequences of Weyl CCs, and the derivation of \gfe, based on the papers by Jack and Osborn~\cite{Jack:1990eb,Osborn:1991gm,Jack:2013sha}; the uninterested reader may choose to skip to section \ref{sec:formalism}. We have attempted to keep only those features relevant for the eventual extraction of CCs between different $\beta$-functions in a general four-dimensional theory; for a substantially more in-depth and pedagogical discussion of the LRG in various dimensions, including (but not limited to) its relation to the proof of the two-dimensional $c$-theorem and four-dimensional weak $a$-theorem, the question of scale- vs. conformal-invariance, and the existence of QFTs with limit cycles, we highly recommend  two reviews of the subject, given by Shore \cite{Shore:2016xor} and Nakayama \cite{Nakayama:2013is}.

\subsection{Weyl consistency conditions and Osborn's equation} \label{sec:LRG}

\gfe, \eqref{eq:GFE}, is derived from the LRG, which posits that the RG flow of a general QFT on a curved spacetime background can be reinterpreted as the response of a theory with ``local couplings" to a local Weyl rescaling $g_{\mu\nu}\rightarrow \Omega^{2}(x)g_{\mu\nu} = e^{2\sigma(x)}g_{\mu\nu}$, accompanied by the appropriate field transformations. By promoting the couplings $g^{I} $ to spacetime-dependent fields $g^{I}(x)$ with associated $\beta$-functions $\beta^{I} = \tfrac{\dd g^{I}}{\dd \ln \mu}$, one can define an action of the form
	\begin{equation} \label{eq:action}
	S = S_{\text{con}} + \int \dd^{d}x\,\sqrt{|g|}\left( g^{I}\mathcal{O}_{I} + \mathfrak{b}^{\alpha} \mathscr{R}_{\alpha} \right),
	\end{equation}
where $S_{\text{con}}$ is classically conformally-invariant (in our case it is a sum over conformally-coupled, non-interacting fields on curved spacetime), $\mathcal{O}_{I}$ is a set of marginal operators, $\mathscr{R}_{\alpha}$ is the appropriate set of all $d$-dimensional operators containing curvature terms and spacetime-derivatives of the couplings, and $\mathfrak{b}^{\alpha}(g^{I}) $ is the set of associated coefficients defined on the space of couplings; $\mathfrak{b}^{\alpha}\mathscr{R}_{\alpha}$ may simply be interpreted as the additional counter terms required to ensure finiteness of the associated quantum theory.

The response of the theory to a Weyl rescaling can be offset by a corresponding change of the couplings along their RG trajectory, leaving only an anomaly stemming from the added counter terms. We define the functional differential operator
	\begin{equation} \label{eq:Del_sigma}
	\Delta_{\sigma} = \int \dd^{d}x\,\sqrt{|g|}\left\{\left( \sigma(x)g_{\mu\nu}\right)\frac{2}{\sqrt{|g|}}\frac{\delta}{\delta g_{\mu\nu}} \right\} + \int \dd^{d}x \,\sqrt{|g|} \left\{ \left( \sigma(x)\beta^{I}\right)\frac{1}{\sqrt{|g|}}\frac{\delta}{\delta g^{I}}\right\}
	\end{equation}
that will therefore act on a suitably-renormalized vacuum generating functional\footnote{In general, the Schwinger Action Principle implies that $n$-point correlation functions of a local QFT in a general curved spacetime are obtained by corresponding functional derivatives of the vacuum expectation value, so that
	\begin{equation*}
	\left<T_{\mu\nu}\right> = \frac{2}{\sqrt{|g|}} \frac{\delta W}{\delta g^{\mu\nu}}, \quad \left<\mathcal{O}_{I}\right> = \frac{1}{\sqrt{|g|}}\frac{\delta W}{\delta g^{I}} \quad \implies \left<\mathcal{O}_{I}(x)\mathcal{O}_{J}(y)\right> = \frac{1}{\sqrt{|g_{x}|}} \frac{1}{\sqrt{|g_{y}|}}\frac{\delta^{2} W}{\delta g^{I}(x)\delta g^{J}(y)}
	\end{equation*}
and so on. The phrase ``suitably renormalized" indicates the notion that all such $n$-point functions are finite; indeed, this finiteness criterion is what first led to the introduction of additional terms containing spacetime-derivatives of the couplings.} $W$ according to
	\begin{equation}
	\Delta_\sigma W = \int \dd^{d}x\,\sqrt{|g|}\mathcal{A}_{\partial}.
	\label{eq:LRG}
	\end{equation}
$\mathcal{A}_{\partial} \equiv \mathcal{A}_{R} + \mathcal{A}_{\partial g} + \mathcal{A}_{\partial\sigma}$ is again some combination of curvature scalars, coupling-derivatives, and $\sigma$-derivatives. This is nothing more than an integrated version of the Trace anomaly with additional contributions from local couplings,
	\begin{equation}
	\left< T\udindices{\mu}{\mu} \right> = \beta^{I}\left<\mathcal{O}_{I}\right> + \mathcal{A}_{R} + \mathcal{A}_{\partial g}
	\label{eq:anomaly}
	\end{equation}
where $\mathcal{A}_{R}$ is the standard gravitational contribution\footnote{While all labels are ultimately arbitrary, we have opted for the notation $a_{2}$ in the two-dimensional Trace anomaly rather than the more conventional central-charge notation $c$, as we wish to emphasize that $a_{2}$ can be interpreted as the coefficient of $R = E_{2}$, the two-dimensional Euler density. Using this interpretation, the idea that a general even-dimensional analogue of the $a$-theorem is related to the coefficient of the associated Euler density becomes much less mysterious, and was indeed argued to be the case in \cite{Grinstein:2013cka,Stergiou:2016uqq}.},
	\begin{equation}
	\mathcal{A}_{R} = \begin{cases}
	\quad\quad\quad\quad\quad\quad\;\frac{1}{4\pi}a_{2}R, &d = 2\\
	\frac{1}{(4\pi)^{2}}\left(c_{4}\,C_{\mu\nu\rho\sigma}C^{\mu\nu\rho\sigma} - a_{4}E_{4} + b_{4}R^{2}\right), &d = 4\\
	\quad\quad\quad\quad\quad\quad\quad\;\vdots &d = 2k,\; k \in \mathbb{N}\,\backslash\left\{ 0 \right\}	\end{cases},
	\end{equation}
with coefficients
	\begin{equation}
	a_{2} = \frac{1}{3}(n_{s} + \tfrac{1}{4}n_{f}), \quad c_{4} = \frac{1}{120}(n_{s} + 3n_{f} + 12n_{v}), \quad a_{4} = \frac{1}{360}(n_{s} + \tfrac{11}{2}n_{f} + 62n_{v})
	\end{equation}
for a theory with $n_{s}$ real scalars, $n_{f}$ Weyl fermions, and $n_{v}$ gauge fields\footnote{The coefficient $b_{4}$, while non-zero for general theories, must vanish at RG fixed points \cite{Bonora:1983ff}. There are additional four-dimensional contributions to $\mathcal{A}_{R}$ involving $\square R$ and the Pontryagin density: the former can be removed by addition of a local $R^{2}$ counter term; the latter we shall address briefly at the end of Section \ref{sec:mods}.}. Since $g^{I}$ and $\mathfrak{b}^{\alpha}$ are marginal, one can simply define an analogous $\beta$-function, $\mathscr{B}^{\alpha}$, for the local counter terms, $\mathfrak{b}^{\alpha}$, and rewrite \eqref{eq:LRG} with a global Weyl rescaling $\sigma$ as a standard RG equation,
	\begin{equation}
	\left(\mu\frac{\partial}{\partial \mu} + \int \dd^{d}x\,\sqrt{|g|}\, \beta^{I}\frac{1}{\sqrt{|g|}}\frac{\delta}{\delta g^{I}} + \int \dd^{d}x\,\sqrt{|g|}\,\mathscr{B}^{\alpha} \frac{1}{\sqrt{|g|}} \frac{\delta}{\delta \mathfrak{b}^{\alpha}} \right) W = 0,
	\end{equation}
subject to the constraint
	\begin{equation}
	\left( \mu\frac{\partial}{\partial\mu} + \int \dd^{d}x\,\sqrt{|g|}\, g^{\mu\nu}\frac{2}{\sqrt{|g|}}\frac{\delta}{\delta g^{\mu\nu}}  \right)W \equiv \left(\mu\frac{\partial}{\partial\mu} + \Omega\frac{\partial}{\partial\Omega}\right)W = 0,
	\label{eq:RGscale}
	\end{equation}
\emph{i.e.} $\Omega = e^{\sigma}$ plays the r\^ole of an RG scale. For this reason, \eqref{eq:LRG} is taken to be the definition of an RG equation for a theory in curved spacetime with local couplings. 

Having maintained the renormalizability of the theory with local couplings (thus establishing consistency of the LRG), Osborn showed that one can derive restrictions on the form of some of the terms that appear in \eqref{eq:LRG} by applying the Wess--Zumino CCs for the Trace anomaly. Since a local Weyl rescaling is generated by the $ \Delta_\sigma $ operator \eqref{eq:Del_sigma}, the Abelian nature of Weyl rescalings dictates that the commutator of two such operators, acting on the renormalized action, must vanish:
	\begin{equation}
	[\Delta_{\sigma},\, \Delta_{\sigma'}] W = 0.
	\label{eq:WCC}
	\end{equation}
This criterion leads to a vast number of intricate identities relating different tensors; such constraints are referred to as Weyl CCs. In four dimensions, we may parametrize the possible terms according to
	\begin{align}
	\mathscr{B}^{\alpha}\mathscr{R}_{\alpha} &= C\,C_{\mu\nu\rho\sigma}C^{\mu\nu\rho\sigma} - AE_{4} + BR^{2} + \tfrac12 A_{IJ}\nabla^{2}g^{I}\nabla^{2}g^{J} \nn
	& \quad + \tfrac12 B_{IJK}\partial_{\mu}g^{I}\partial^{\mu}g^{J}\nabla^{2}g^{K} + \tfrac12 C_{IJKL}\partial_{\mu}g^{I}\partial^{\mu}g^{J}\partial_{\nu}g^{K}\partial^{\nu}g^{L} \nn
	& \quad + \tfrac13 E_{I}\partial_{\mu}R\partial^{\mu}g^{I} + \tfrac16 F_{IJ}R\partial_{\mu}g^{I}\partial^{\mu}g^{J} + \tfrac12 G_{IJ}G^{\mu\nu}\partial_{\mu}g^{I}\partial_{\nu}g^{J},
	\label{eq:4dbr}
	\end{align}
where $C_{\mu\nu\rho\sigma}$, $E_{4}$, and $G^{\mu\nu}$ are the Weyl curvature tensor, Euler density, and Einstein tensor respectively. Eq. \eqref{eq:LRG} then takes the form
	\begin{equation}
	\Delta_\sigma W = \int \dd^{d}x\sqrt{|g|} \Big[ \sigma\, \mathscr{B}^{\alpha}\mathscr{R}_{\alpha} + \partial_{\mu}\sigma\left( G^{\mu\nu}W_{I}\partial_{\nu}g^{I} + RH_{I}\partial^{\mu}g^{I} + \mathscr{Z}^{\mu}\right) + \nabla^{2}\sigma\left( DR + \mathscr{Y} \right)\Big],
	\label{eq:weylX}
	\end{equation}
with 
	\begin{align}
	\mathscr{Z}^{\mu} &= S_{IJ}\partial^{\mu}g^{I}\nabla^{2}g^{J} + T_{IJK}\partial^{\mu} g^{I}\partial_{\nu}g^{J}\partial^{\nu}g^{K}, \nn
	\mathscr{Y} &= U_{I}\nabla^{2}g^{I} + V_{IJ}\partial_{\mu}g^{I}\partial^{\mu}g^{J}.
	\end{align}
Imposing \eqref{eq:WCC} for arbitrary Weyl rescalings $\sigma, \sigma'$ then requires that the contributions proportional to $\sigma'\partial_{\mu}\sigma - \sigma\partial_{\mu}\sigma'$, $\partial_{\mu}\sigma' \partial_{\nu}\sigma - \partial_{\nu}\sigma' \partial_{\mu}\sigma$, and $\partial_{\mu}\sigma'\nabla^{2}\sigma - \nabla^{2}\sigma'\partial_{\mu}\sigma$ must vanish separately; furthermore, requiring that these identities hold for arbitrary position-dependent couplings eventually leads to the two crucial conditions
	\begin{align}
	8\partial_{I}A &= G_{IJ}\beta^{J} - \left( \beta^{J}\partial_{J}W_{I} + \partial_{I}\beta^{J}W_{J} \right), \\
	G_{IJ} + 2A_{IJ} + 2\partial_{I}\beta^{K}A_{KJ} + \beta^{K}B_{IJK} &= \partial_{I}\beta^{K}S_{KJ} + \partial_{J}\beta^{K}S_{IK} + \beta^{K}\partial_{K}S_{IJ}.
	\end{align}
The first condition shows that the function $\tilde{A} = 8A + W_{I}\beta^{I}$ satisfies \eqref{eq:GFE},
	\begin{equation}
	\partial_{I}\tilde{A} = T_{IJ}\beta^{J}, \quad T_{IJ} = G_{IJ} + 2 \partial_{[I}W_{J]},
	\end{equation}
while the second relates the symmetric tensor $G_{IJ} = T_{(IJ)}$ to two- and three-point functions of $\mathcal{O}_{I}$ \cite{Jack:1990eb}. Since $\tilde{A}$ reduces to the Euler coefficient $a_{4}$ of the Trace anomaly at RG fixed points ($\beta^{I} = 0$) and contracting \eqref{eq:GFE} with $\beta^{I}$ gives
	\begin{equation}
	\frac{\dd \tilde{A}}{\dd \ln \mu} = \beta^{I}\partial_{I}\tilde{A} = \beta^{I}G_{IJ}\beta^{J},
	\end{equation}
the function $\tilde{A}$ would satisfy the strong $a$-theorem if $G_{IJ}$ were positive-definite. Unfortunately, since there is no manifest positivity constraint on $\left<\mathcal{O}_{I}\mathcal{O}_{J}\mathcal{O}_{K}\right>$, the second condition is not enough to guarantee positive-definiteness \cite{Jack:1990eb}; nevertheless, the existence of the function $\tilde{A}$ for a general four-dimensional QFT is sufficient to imply CCs on the $\beta$-functions, and it is these conditions that we wish to derive in this work. By a slight abuse of terminology, the $\beta$-function conditions are \emph{also} referred to as Weyl CCs, as they are ultimately derived from \gfe.

\subsection{Modifications of \gfe}
\label{sec:mods}
There are two necessary modifications that must be made to the derivation of \eqref{eq:GFE}, one of which is highly non-trivial. The first is simply the observation that we may always add additional local counter terms to the theory, and that these counter terms can take the same generic form as \eqref{eq:4dbr}; that is, we are free to change the finite part of the counter terms in Eq. \eqref{eq:action} by $ \mathfrak{b}^{\alpha} \to \mathfrak{b}^{\alpha} +  \left\{ a, b, c, a_{IJ}, b_{IJK}, \ldots \right\}$, \emph{i.e.} terms whose coefficients are lower-case versions of the tensors in \eqref{eq:4dbr}. Doing so induces various shifts in the terms of $\mathscr{B}^{\alpha}$, the most immediately relevant of which are
	\begin{align}
	A &\sim A + \mathcal{L}_{\beta}a = A + \beta^{K}\partial_{K}a,\nn
	G_{IJ} &\sim G_{IJ} + \mathcal{L}_{\beta}g_{IJ} = G_{IJ} + \beta^{K}\partial_{K}g_{IJ} + (\partial_{I}\beta^{K})g_{KJ} + (\partial_{J}\beta^{K})g_{IK}, \nn
	W_{I} &\sim W_{I} - \partial_{I}a + g_{IJ}\beta^{J}.
	\label{eq:shifts1}
	\end{align}
From this, it is easy to see that
	\begin{align}
	\tilde{A} \sim \tilde{A} + \beta^{I}g_{IJ}\beta^{J}, \quad T_{IJ} \sim T_{IJ} + \mathcal{L}_\beta g_{IJ} + 2 \partial_{[I} (g_{J]K} \beta^{K}),
	\end{align}
and so \gfe is invariant under the addition of local counter terms---away from RG fixed points, $\tilde{A}$ therefore possesses an arbitrariness that is quadratic in $\beta^{I}$. 

The second modification is based on the inclusion of relevant operators, in addition to the usual marginal operators. For a four-dimensional theory containing scalars and fermions, one can include vector operators, \emph{i.e.} spin-1 composite operators with dimension 3. The effect of these operators on the LRG can be investigated by exploiting the continuous part, $G_{K}$, of the global symmetry present in the kinetic terms of a general theory with multiple scalars or fermions\footnote{This continuous part of the global symmetry is simply the freedom to rotate the fields, hence the associated global symmetry is $\SO(n_{\phi})$ for the scalars, and $\U(n_{\psi})$ for the Weyl fermions.}. If a generic field $\phi$ has a global symmetry transformation $\delta_{\omega}\phi = -\omega\phi$, then one must include a compensating transformation $\delta_{\omega}g^{I}(x) = -\omega^{I}_{\;\;J}g^{J}(x)$ for the local couplings, such that the action remains invariant. 
The classical action then has an associated current $J^{\mu}(x)$, being the aforementioned spin-1 operator; if
$ J^{\mu} $ is sourced by a background gauge field $ a_\mu(x) $ associated to the symmetry, renormalizability of the quantized theory is guaranteed. 
Thus, all partial derivatives $\partial_{\mu}$ are replaced by gauge-covariant derivatives $D_{\mu} = \partial_{\mu} + a_{\mu}$, and the following transformation laws are obeyed:
	\begin{align}
	\delta_{\omega}g^{I}(x) &= -\omega^{I}_{\;\;J}(x)g^{J}(x) \equiv -(\omega g)^{I}(x), \nn
	\delta_{\omega}a_{\mu}(x) &= \partial_{\mu}\omega(x) + [a_{\mu}(x),\omega(x)] \equiv D_{\mu}\omega(x),
	\end{align}
for local transformation $ \omega(x) \in \mathfrak{g}_K $, where $ \mathfrak{g}_K $ is the Lie algebra of $ G_K $. The background gauge field then behaves as a source term for the local vector current of the corresponding quantum theory,
	\begin{equation}
	\left<J^{\mu}\right> = -\frac{1}{\sqrt{|g|}}\frac{\delta W}{\delta a_{\mu}},
	\end{equation}
allowing the inclusion of dimension-3 operators. If one introduces an operator that generates these local $G_{K}$ transformations,
	\begin{equation}
	\Delta_{\omega} = \int \dd^{d}x\,\sqrt{|g|}\left( D_{\mu}\omega \cdot \frac{1}{\sqrt{|g|}}\frac{\delta}{\delta a_{\mu}} - (\omega g)^{I} \dfrac{\delta}{\delta g^{I}} \right),
	\end{equation}
with ``$\,\cdot\,$'' denoting the appropriate inner product on $\mathfrak{g}_{K}$, manifest background gauge invariance immediately implies the corresponding Ward identity
	\begin{equation}
	\Delta_{\omega}W = 0 \quad \implies \quad \omega\cdot D_{\mu}\left<J^{\mu}\right> + (\omega g)^{I} \left< \mathcal{O}_{I}\right> = 0,
	\label{eq:gkward}
	\end{equation}
similar to how the Weyl rescaling generates the Trace anomaly. Eq. \eqref{eq:LRG} is now extended by replacing the partial derivatives of local couplings $\partial_{\mu}g^{I}$ with gauge-covariant derivatives $D_{\mu}g^{I} = \partial_{\mu}g^{I} + (a_{\mu}g)^{I}$, and by introducing the new Weyl variation component
	\begin{equation}
	\Delta^{a}_{\sigma} = \int \dd^{d}x\sqrt{|g|}\left\{ \sigma(x)\rho_{I}D_{\mu}g^{I}\frac{1}{\sqrt{|g|}}\frac{\delta}{\delta a_{\mu}} - \partial_{\mu}\sigma S \frac{1}{\sqrt{|g|}}\frac{\delta }{\delta a_{\mu}} \right\}, \quad \rho_{I}, S \in \mathfrak{g}_{K}
	\label{eq:dasigma}
	\end{equation}
with the assumption of manifest $G_{K}$-covariance,
	\begin{equation}
	[\Delta_{\omega},\, \Delta_{\sigma}] = [\Delta_{\omega},\, \Delta^{a}_{\sigma}] = 0
	\label{eq:NoGfAnomaly}
	\end{equation}
so that the Trace anomaly is still obtained from the action of a local Weyl rescaling on the vacuum generating functional,
	\begin{equation}
	\left( \Delta_{\sigma} - \Delta^{a}_{\sigma} \right)W = \int \dd^{d}x \sqrt{|g|}\, \mathcal{A}_{\partial,a}.
	\label{eq:LRG2}
	\end{equation}
Now though, the anomalous variation includes additional contributions related to the local current $\left<J^{\mu}\right>$ and the background field-strength tensor $f_{\mu\nu} = \partial_{\mu}a_{\nu} - \partial_{\nu}a_{\mu} + [a_{\mu},a_{\nu}]$, stemming from new counter terms involving the background gauge field. Since the $\beta$-functions are now only determined up to $\mathfrak{g}_{K}$ transformations, there is an inherent ambiguity in the Trace anomaly as defined by \eqref{eq:LRG2}---this ambiguity may be removed by using the Ward identity \eqref{eq:gkward} to express \eqref{eq:dasigma} in terms of $G_{K}$-invariant RG quantities
	\begin{equation}
	B^{I} = \beta^{I} - (Sg)^{I}, \quad \tilde{\rho}_{I} = \rho_{I} + \partial_{I}S,
	\label{eq:Sdef}
	\end{equation}
using which one can rewrite the Trace anomaly derived from Eq. \eqref{eq:anomaly2} in the unambiguous, $G_{K}$-invariant form
	\begin{equation}
	\left<T\udindices{\mu}{\mu} \right> = B^{I}\left<O_{I}\right> + B^{a}_{\mu}\left<J^{\mu}\right> + \mathcal{A}_{R} + \mathcal{A}_{\partial g} + \mathcal{A}_{a},
	\label{eq:anomaly2}
	\end{equation}
with additional contributions
	\begin{equation}
	B^{a}_{\mu} = \tilde{\rho}_{I}D_{\mu}g^{I}, \quad \mathcal{A}_{a} = -\frac14 f_{\mu\nu}\cdot\kappa\cdot f^{\mu\nu}.
	\end{equation}
This alone is a rather striking result, as it demonstrates that conformal invariance of a theory is decided by the vanishing of $B$, not $\beta$, even for a standard QFT in flat spacetime with global couplings. However, even more interesting is what happens if one re-derives the Weyl CCs using the new Weyl rescaling operator in \eqref{eq:LRG2}. The action of this new operator on the vacuum functional is
	\begin{align}
	[\Delta_{\sigma} - \Delta^{a}_{\sigma}]W &= \Delta_{\sigma} W \Big\rvert_{\partial_{\mu}\rightarrow D_{\mu}} + \int \dd^{d}x\sqrt{|g|}\sigma(x)\left\{ -\frac14 f_{\mu\nu}\cdot\kappa\cdot f^{\mu\nu} + f^{\mu\nu}\cdot P_{IJ}D_{\mu}g^{I}D_{\nu}g^{J} \right\}\nn
	&\qquad\qquad\qquad\quad + \int \dd^{d}x\sqrt{|g|}\partial_{\mu}\sigma(x)\left\{ f^{\mu\nu}\cdot Q_{I}D_{\nu}g^{I} \right\} + \mathcal{O}(\nabla^{2}\sigma),
	\label{eq:LRG2b}
	\end{align}
\emph{i.e.} \eqref{eq:weylX} with partial derivatives replaced by covariant derivatives, plus new terms related to the background field-strength, as expected. Imposing
	\begin{equation}
	[\Delta_{\sigma} - \Delta^{a}_{\sigma},\, \Delta_{\sigma'} - \Delta^{a}_{\sigma'}] W = 0,
	\end{equation}
then leads to generalized versions of the Weyl CCs, including
	\begin{equation}
	8\partial_{I}A = G_{IJ}B^{J} - \mathcal{L}_{B,\tilde{\rho}}W_{I} \equiv G_{IJ}B^{J} - \mathcal{L}_{B}W_{I} - (\tilde{\rho}_{I}g)^{J}W_{J}
	\end{equation}
subject to the constraint
	\begin{equation}
	\tilde{\rho}_{I}B^{I} = 0.
	\label{eq:RhoB}
	\end{equation}
Hence, $\tilde{A} = 8A + W_{I}B^{I}$ satisfies an equation of the same form as \eqref{eq:GFE}, with $\beta$ replaced by $B$, and the tensor $T_{IJ}$ extended by an additional antisymmetric component:
	\begin{equation} \label{eq:full_TIJ}
	\partial_{I}\tilde{A} = T_{IJ}B^{J}, \quad T_{IJ} = G_{IJ} + 2\partial_{[I}W_{J]} + 2 \tilde{\rho}_{[I} \cdot Q_{J]} .
	\end{equation}
By adding appropriate local counter terms, one can again verify that the new Weyl CCs are still invariant under such additions, and that $\tilde{A}$ has the expected arbitrariness quadratic in $ B^{I} $. The new function $\tilde{A}$, valid for theories with additional vector operators, therefore satisfies
	\begin{equation}
	B^{I}\partial_{I}\tilde{A} = B^{I}G_{IJ}B^{J}
	\end{equation}
thus establishing the perturbative $a$-theorem for leading-order positive-definite $ G_{IJ} $. For RG flows that include limit cycles, the interpretation of this is that $G_{IJ}$ has the potential to act as a metric on the quotient space $\mathcal{M} = \left\{ g^{I} \right\} /\,G_{K}$, where couplings related by $G_{K}$ transformations are identified, such that $B^{I} = 0$ define fixed points on $\mathcal{M}$, and the associated QFT is conformal \cite{Jack:2013sha,Shore:2016xor}. 
A final extension of the LRG formalism includes the introduction of dimension-2 operators. However this bears no influence on the $ \beta $-function constraints that we wish to explore in the present paper, so we will not address this extension further. 

The quantity $S$, appearing in the definition of $B$, stands out as the only symbolic modification to \gfe that is of relevance to us. As shown in \cite{Fortin:2012hn}, it effectively parametrizes any possible difference between a scale-invariant theory ($\beta^{I} = 0$) and a conformally-invariant theory ($B^{I} = 0$). Since $S$ is an element of the Lie algebra $\mathfrak{g}_{K}$, it is necessarily an antisymmetric tensor, and is directly related to the additional counter terms required when one renormalizes field multiplets with position-dependent couplings\footnote{Since the wave-function-renormalization matrices $Z$ are functions of $g^{I}(x)$, $\partial_{\mu}Z$ is non-trivial, thus one requires new counter terms. For example, expanding a scalar kinetic term $\tfrac12 \partial_{\mu}\Phi^{T}_{0}\partial^{\mu}\Phi_{0}$ with $\Phi_{0} = \sqrt{Z}\Phi$ leads to a counter term of the form $\partial_{\mu}\Phi^{T} N_{I} \Phi \partial^{\mu}g^{I}$.}. 
The spacetime-dependent couplings of the theory necessitate a new kind of counter term, 
	\begin{equation}
	N_{I} \cdot (j^{\mu} \partial_\mu g^{I}),
	\end{equation}
where $ j_\mu \in \mathfrak{g}_K $ is the current associated with the global symmetry. $S$ is related to the simple pole of the counter term coefficient, $ N_{I} \in \mathfrak{g}_K $, by $ S = -N^{(1)}_{I}g^{I} $.
$ S $ can therefore be calculated by considering self-energy diagrams with local couplings playing the r\^ole of spectator fields, then simply extracting any possible antisymmetric combinations of tensors.

Strictly speaking, the derivation and modifications outlined above make use of two non-trivial assumptions. First, the LRG formalism itself is predicated on the Trace anomaly conserving parity, thus \eqref{eq:4dbr} contains no $\epsilon$-tensor contributions such as the Pontryagin density, although the coefficient of this term typically vanishes.\footnote{Indeed, it is still unknown whether any unitary QFT can actually produce the Pontryagin term in its associated Trace anomaly: see \cite{Nakayama:2012gu,Bonora:2017gzz,Bastianelli:2018osv} and many references therein.}. Second, by Eq. \eqref{eq:NoGfAnomaly}, the Ward identities for any global flavor symmetries are assumed to be non-anomalous, and in particular that there is no associated chiral anomaly. These assumptions were relaxed in \cite{Keren-Zur:2014sva}, where it was shown that any such anomalous behavior can be ``consistently factorized out" of the LRG formalism; the inclusion of $\epsilon$-tensors in \eqref{eq:4dbr} leads to no additional constraints on the RG flow, and the effects of a chiral anomaly can be absorbed into a further redefinition of $B$. The end result is that \gfe, in its $\beta\rightarrow B$ form, is effectively unaltered by the presence of these anomalies.

\subsection{Imposing symmetry on $T_{IJ}$} \label{sec:sym}
Beyond the modifications of the formalism discussed in the previous section, we would also like to comment on an additional feature of \gfe, which has (so far) been shown to hold in all perturbative calculations, but which we have not seen explicitly addressed in the current literature; namely, whether one can always choose local counter terms to make $ T_{IJ} $ symmetric. As we have seen, the addition of local counter terms reveals an arbitrariness in $\tilde{A}$, but we have not explicitly discussed their effect on $T_{IJ}$. 

In the case where one includes the effects of relevant operators, we have already seen that $T_{IJ}$ acquires an additional antisymmetric contribution $ 2 \tilde{\rho}_{[I} \cdot Q_{J]} $ (again neglecting the effects of dimension-2 operators). Additionally, there exist new local counter terms, defined by lower-case versions of the tensors in \eqref{eq:LRG2b}, so the corresponding shifts of the anomalies in \eqref{eq:shifts1} become more involved. Making use of various non-trivial RG identities implied by manifest $G_{K}$-covariance, the end result is that the transformations of Eq. \eqref{eq:shifts1} are simply extended to their $G_{K}$-covariant form with additional transformations for the other new tensors. Since only $Q_{I}$ appears in $T_{IJ}$, we only consider the shifts
	\begin{align}
	A &\sim A + \mathscr{L}_{B,\tilde{\rho}}\,a & G_{IJ} &\sim G_{IJ} + \mathscr{L}_{B,\tilde{\rho}}\,g_{IJ} \nn
	W_{I} &\sim W_{I} - \partial_{I}a + g_{IJ}B^{J} & Q_{I} &\sim Q_{I} + p_{IJ}B^{J}
	\label{eq:shifts2}
	\end{align}
with the local counter term coefficient $p_{IJ} = -p_{JI} \in \mathfrak{g}_K $ defined via Eq. \eqref{eq:LRG2b}. The total shift of the full $ T_{IJ} $, as specified in Eq. \eqref{eq:full_TIJ}, under addition of the finite counter terms is thus 
	\begin{equation}
	T_{IJ} \sim T_{IJ} + \mathcal{L}_{B,\tilde{\rho}} g_{IJ} + 2 \partial_{[I}(g_{J]K} B^{K}) + 2 \tilde{\rho}_{[I} \cdot (p_{J]K}B^{K}). 
	\end{equation}
	
We can now raise the question of whether there exist local counter terms that make the anti-symmetric part of $ T_{IJ} $ vanish. A sufficient condition for this to be the case, is if there exist a symmetric $ x_{IJ} $, two scalars $ z, t $, and an antisymmetric $ s_{IJ} \in \mathfrak{g}_K $, such that 
	\begin{equation}
	\begin{split}
	W_{I} &= x_{IJ} B^{J} + \partial_I z \andeq  \\
	Q_I &= s_{IJ} B^{J} + \tilde{\rho}_{I} \, t.
	\end{split}
	\end{equation}
In this case one can simply add counter terms $ g_{IJ} = -x_{IJ} $ and $ p_{IJ} = -s_{IJ} $, to remove the anti-symmetric part of $ T_{IJ} $. Up to two loops, it is possible to show directly that such a decomposition of $ W_{I} $ and $ Q_I $ exists, and that $T_{IJ}$ is indeed symmetric, but it is unclear how the program proceeds at higher loop orders. This is perhaps not surprising, as it has already been shown that the first off-diagonal contributions to $T_{IJ}$ in a general theory occur at three loops \cite{Jack:2014pua}.

When deriving constraints on the $\beta(B)$-functions, imposing symmetry on $ T_{IJ} $ would correspond to identifying the coefficients of each pair of tensors related by the interchange of open indices $ I $ and $ J $, and setting them equal---in six-dimensional $\phi^{3}$ theory at sufficiently high loop order, it has been shown that this does indeed force an additional CC, which (rather amazingly) turn out to be satisfied by the \msbar coefficients, \emph{and} renormalization-scheme independent~\cite{Gracey:2015fia}. Effectively, we conjecture that it is always possible to choose local counter terms such that $ T_{IJ} $ is symmetric, and that an analogous construction can be made in the extension to all even dimensions.

Finally, before concluding our discussion of the LRG, we should draw attention to the fact that \gfe has been generalized to an arbitrary even-dimensional spacetime \cite{Grinstein:2013cka}, hence the procedure of using Weyl CCs to derive constraints on the $\beta$-functions will always work, as long as there exists a corresponding even-dimensional, renormalizable QFT with marginal interactions. For example, such constraints have been thoroughly investigated for the case of a general six-dimensional $\phi^{3}$ theory \cite{Gracey:2015fia}, where $\tilde{A}$ was constructed up to five loops, and all non-trivial features (a non-zero shift $\beta \rightarrow B$, the ability to impose symmetry of $T_{IJ}$) were shown to hold for arbitrary renormalization schemes.

\section{Formalism} \label{sec:formalism}
This section introduces a compact notation for general four-dimensional renormalizable theories, and sketches the perturbative computation of $ \beta $-functions and $ T_{IJ} $ in such theories. We have used the terminology ``compact gauge group" throughout: given a general QFT, renormalizability constrains the gauge group to be compact, hence it can be any product of simple and Abelian Lie groups.

\subsection{General QFT in four dimensions}
We begin with a completely general, four-dimensional, renormalizable QFT, allowing for all marginal operators.
None of the relevant couplings in the form of mass terms or trilinear scalar couplings impact the $ \beta $-functions of the marginal couplings, so we can safely ignore them for our purpose.  
With this in mind, the Lagrangian is typically given on the form\footnote{Matching one's favorite Lagrangian onto the general form used here can be surprisingly difficult. Explicit representations of the couplings and generators for a single SM generation are given in appendix D of \cite{Mihaila:2012pz}. More generally, one can derive the expressions in any theory, by using the ``structure-delta" method of \cite{Molgaard:2014hpa}.}:
	\begin{align}\label{eq:traditional_Lagrangian}
	\mathcal{L} &= -\tfrac{1}{4} \sum_{u} F_{u,\mu\nu}^{A_u} F^{A_u \mu\nu}_{u} + \tfrac{1}{2} (D_\mu \phi)_{a} (D^{\mu} \phi)_a + i \psi^{\dagger}_i \bar{\sigma}^\mu (D_\mu \psi)_i \nn
	& \quad - \tfrac{1}{2} \left(Y_{aij} \psi_i \psi_j + \mathrm{h.c.} \right) \phi_a - \tfrac{1}{24} \lambda_{abcd}\phi_a \phi_b \phi_c \phi_d.
	\end{align}
All scalars and fermions have been gathered into two multiplets, which may generally be reducible representations of the gauge group. We work with real scalars $ \phi_a $ and Weyl spinors $ \psi^i $, as any complex scalar or Dirac fermion can be put into this form. For the gauge symmetry, we consider a compact gauge group with any number of Abelian and non-Abelian factors, $ \mathcal{G} = \times_u \mathcal{G}_u $. Since $ \mathcal{G} $ may contain more than one Abelian factor, we must allow for kinetic mixing terms in the Lagrangian; we shall return to this complication shortly. As per usual, the covariant derivatives are taken to be
	\begin{align}
	D_\mu \phi_a &= \partial_\mu \phi_a -i\sum_u g_u \,V_{u,\mu}^{A_u}  (T_{\phi,u}^{A_u})_{ab} \phi_b \quad \andeq \\
	D_\mu \psi_i &= \partial_\mu \psi_i -i\sum_u g_u \,V_{u,\mu}^{A_u}  (T_{\psi,u}^{A_u})_{ij} \psi_j
	\end{align}
respectively. For Abelian factors the generators $ T_{(\phi, \psi), u} $ are just charge matrices---diagonal for the complex Weyl spinors and antisymmetric for the real scalar fields. 

With an eye on future simplifications, we will take several steps to make the notation more compact, in the sense that all fields of a given spin are gathered into just one multiplet. At the risk of adding an additional layer of abstraction, this will help us keep our expressions more manageable later on. Following \cite{Jack:2014pua}, we define a new multiplet $\Psi_{i}$ of Majorana spinors $ \Psi = \binom{\psi }{\psi^{\dagger} } $, taking advantage of the unitary equivalence between Weyl and Majorana fermions\footnote{An in-depth exposition of the connections between Dirac, Majorana and Weyl spinors in two- and four-component notation may be found in \cite{Dreiner:2008tw}: our conventions match those spelt out in Appendix G.}. This allows us to assemble the Yukawa couplings and fermionic generators into larger matrices,
	\begin{equation}
	\quad y_a = \begin{pmatrix}	Y_a & 0 \\ 0 & Y_{a}^{\ast }\end{pmatrix}, \andeq T^{A_u}_u = \begin{pmatrix} T_{\psi,u}^{A_u} & 0 \\ 0 & -(T_{\psi,u}^{A_u})^\ast \end{pmatrix},  
	\end{equation} 
employing matrix notation for the fermion indices to reduce the clutter. The spinor part of the Lagrangian thus reads
	\begin{equation}
	i \psi_i^{\dagger} \bar{\sigma}^\mu (D_\mu \psi)_i - \tfrac{1}{2} \left(Y_{aij} \psi_i \psi_j + \mathrm{h.c.} \right) \phi_a = \tfrac{i}{2} \Psi\transpose \begin{pmatrix} 0 & \sigma^\mu \\ \bar{\sigma}^\mu & 0 \end{pmatrix} D_\mu \Psi - \tfrac{1}{2} \Psi^{T} y_{a} \Psi \phi_{a}. 
	\end{equation}
Gathering the fermions like this removes the distinction between left- and right-handed fermions (in the sense of their representations under the Lorentz group), $ \psi $ and $ \psi^\dagger $; that is, for any left-handed fermion loop appearing in a given Feynman diagram there is an equivalent diagram with right-handed fermion running in that loop. By using $ y_a $ and $ T^{A_u}_{u} $, both contributions are automatically included in any quantity constructed from the couplings. Of course, this construction neglects the chiral nature of theories such as the SM, where there \emph{is} a distinction between left-handed and right-handed fermions. The effects of such a difference do not appear in any $\beta$-functions up to order 3--2--1, but at order 4--3--2 one encounters the first non-trivial occurrences of $ \gamma_5 $. In such diagrams, left- and right-handed fermions contribute with opposite rather than same signs, thus we must slightly extend our tensor notation to accommodate this. We will return to the question of $ \gamma_5 $ as it becomes relevant for our discussion in section \ref{sec:gamma5}.

For the gauge interactions, we absorb the gauge couplings into the gauge field $ g_u V_{u,\mu}^{A_u} \to V_{u,\mu}^{A_u} $, so that the gauge kinetic term becomes $ - \sum_{u} \tfrac{1}{4g_u^2} F_{u,\mu\nu}^{A_u} F^{A_u \mu\nu}_{u} $ and the covariant derivatives come without any gauge coupling. 
Rather than dealing with every product of the gauge group individually, we again employ a compact notation to account for all of them at the same time. From a computational standpoint, all gauge field have identical propagators and Lorentz structure; only the group factors differ depending on the group. Thus, all gauge fields can be incorporated in each diagram using sufficiently general couplings in the vertices, similar to the way all scalar and fermion fields are incorporated into just one multiplet each.

We gather the gauge fields of the separate product groups into one multiplet $ A^{A}_\mu $ taking values in the Lie Algebra $ \mathfrak{g} = \oplus_u \mathfrak{g}_u $. The group generators and structure constants are correspondingly gathered by introducing a general adjoint index taking values in
	\begin{equation}
	A \in \braces{(u, A_u): \; A_u \leq d(\mathcal{G}_u) } \quad \text{with summation convention} \quad \sum_{A} = \sum_{u} \sum_{A_u =1}^{d(\mathcal{G}_u)}.
	\end{equation}
The gauge couplings are arranged in a diagonal matrix,
	\begin{equation}
	G_{AB}^2 = g^2_u \delta_{uv} \delta^{A_u B_v},
	\end{equation}
in this notation. It is a positive definite matrix and works as a metric on $ \mathfrak{g} $; it is still subject to gauge symmetry, so it stays diagonal and is proportional to the identity on each subalgebra $ \mathfrak{g}_u $. Gauge invariance ensures that this remains the case along the RG flow. For the group structure, we similarly define the generators 
	\begin{equation}
	T^{A} = T_u^{A_u} \andeq f^{ABC} = \delta_{uv} \delta_{u w} f_u^{A_u B_v C_w}.
	\end{equation} 
A simple consequence of this construction is that the structure constants satisfy $ G^{2}_{AD} f^{DBC}  = G^{2}_{BD} f^{ADC} $. The benefit of this construction is that it allows us to ignore much of the inconvenience of having multiple gauge groups in the theory. We may treat all gauge couplings simultaneously by considering $ G^2_{AB} $ directly.

When including multiple Abelian factors in the gauge group, there is an extra complication in the form of kinetic mixing~\cite{Holdom:1985ag}. Invariance of the Abelian field-strength tensors under gauge transformations implies that terms such as 
	\begin{equation}
	- \tfrac{1}{4} h_{uv}^{-2} F_{u, \mu\nu} F_{v}^{\mu\nu},
	\end{equation}
for Abelian gauge groups $ u,v $ are allowed in the Lagrangian by symmetry. In fact, they are needed to renormalize the theory and are generated by the RG evolution, even if set to vanish at a particular scale. The inclusion of these terms in the general 2-loop \msbar $ \beta $-function has been discussed in Refs. \cite{Luo:2002iq,Fonseca:2013bua}. We wish to emphasize that the kinetic mixing of multiple Abelian groups fit into the formalism employed here in a very natural way. The gauge couplings have been relegated to the coupling matrix $ G^{2}_{AB} $ in the kinetic term, thus kinetic mixing is simply the observation that when several of the product groups are Abelian, gauge invariance allow for non-diagonal terms in $ G^{2}_{AB} $ between the corresponding groups. 

To summarize, we consider the general case where the gauge group is taken to be 
	\begin{equation}
	\mathcal{G} = \times_u \mathcal{G}_u = \U(1)_1 \times \ldots \times \U(1)_n \times \mathcal{G}_{n+1} \times \ldots \, ,
	\end{equation}
such that the factors $ \mathcal{G}_u $ are non-Abelian for $ u > n $. The first $ n $ instances of the adjoint index, $ A $, thus label the $ \U(1) $ groups, while $ A >n $ takes values in the adjoint indices of the non-Abelian groups.
For the coupling matrix in particular we have
	\begin{equation}
	G_{AB}^2 = \begin{cases}h_{uv}^2 & \text{for } A, B\leq n \\
	g_u^2 \delta_{uv} \delta^{A_u B_v} & \text{for } A > n \end{cases},
	\end{equation}
where $ h^2_{uv} $ is a symmetric $ n\times n $ coupling matrix. The off-diagonal terms in $ h^2_{uv} $ ensure that the renormalized Lagrangian contain all possible kinetic terms allowed by gauge symmetry. The generators of the individual $ \U(1) $ groups are still just the associated charges and require no change. 

Altogether, in our notation, the Lagrangian \eqref{eq:traditional_Lagrangian} becomes
	\begin{align} \label{eq:general_Lagrangian}
	\mathcal{L} &= -\tfrac{1}{4} G^{-2}_{AB} F^{A}_{\mu\nu} F^{B\mu\nu} + \tfrac{1}{2} (D_\mu \phi)_a (D^\mu \phi)_a + \tfrac{i}{2} \Psi\transpose \begin{pmatrix} 0 & \sigma^\mu \\ \bar{\sigma}^\mu & 0 \end{pmatrix} D_\mu \Psi \nn
	&\quad - \tfrac{1}{2} \phi_{a}\, \Psi\transpose  y_{a} \Psi -\tfrac{1}{24} \lambda_{abcd} \phi_a \phi_b \phi_c \phi_d,
	\end{align}
and the couplings of the theory are $ g^{I} = \braces{G^2_{AB},\, y_{aij},\, \lambda_{abcd}} $. Note that $ G^2 $ rather than $ G $ is used for the gauge coupling in the general framework. This is done for two reasons: first, only $ G^2 $ appear in the Lagrangian; second, for the purpose of including multiple Abelian couplings, it is more convenient to work with $ G^2 $ for the counter terms, as $ G^{2}_{AB} $ will pick up off-diagonal elements. We have verified that our formalism is indeed capable of reproducing the previously known 2-loop results for the kinetic mixing of the Abelian factors~\cite{Luo:2002iq}---for details of the comparison and matching to previously used notation see App. \ref{app:kin_mixing}. Later, we will deduce the full 3-loop gauge $ \beta $-function for the general theory \eqref{eq:general_Lagrangian}, including kinetic mixing; although the result for a semi-simple gauge group has in fact been calculated (but not published) previously~\cite{Molgaard:xxx}, as far as we know, this will be the first time that the 3-loop running of kinetic mixing terms has been presented.

\subsection{Perturbative expansion of quantities in \gfe}
Presently we will not concern ourselves with the finer details of perturbative $\beta$-function computation; on the contrary, we aim to demonstrate (and emphasize) that constraints derived from \eqref{eq:GFE} require no knowledge of the exact perturbative results in any scheme. Instead, we shall work with a parametrization of all the TSs that \emph{may} appear in each $\beta$-function at a particular loop order, where each possible tensor carries an associated coefficient, the value of which then depends on the chosen renormalization scheme\footnote{This method of parameterizing the $\beta$-functions is essentially implicit in the classic 1-loop calculations of Cheng \emph{et al.} \cite{Cheng:1973nv}, and the 2-loop calculations of \citet{Machacek:1983fi}, and Jack and Osborn \cite{Jack:1984vj}.}. The possible TSs that can appear in each $\beta$-function are completely determined by the form of the interactions in the Lagrangian density, with the Feynman rules providing an exact correspondence between the Feynman diagrams employed in perturbation theory and the associated tensor-coupling contractions. Gauge invariance of a theory with additional Yukawa and scalar interactions requires that certain non-trivial identities involving tensor couplings and gauge generators must hold, which then lead to relations between certain TSs, reducing the number of unique terms. In certain renormalization schemes such as \msbar, the number of contributing TSs will be further reduced on general grounds. We will take advantage of this simplification, but we wish to stress that the derivation of CCs works in arbitrary schemes if one includes the additional TSs not present in \msbar.

In dimensional regularization the bare couplings are renormalized by setting
	\begin{align}
	\begin{split}
	y_{0,aij} &= \mu^{\epsilon/2} (Z_\phi^{-1/2})_{ab} (Z_\Psi^{-1/2})_{ik} (Z_\Psi^{-1/2})_{j\ell} \left(y_{bk \ell} + \delta y_{bk \ell} \right),\\
	\lambda_{0,abcd} &= \mu^{\epsilon} (Z_\phi^{-1/2})_{ae} (Z_\phi^{-1/2})_{bf} (Z_\phi^{-1/2})_{cg} (Z_\phi^{-1/2})_{dh}  \left( \lambda_{efgh} + \delta \lambda_{efgh} \right).
	\end{split}
	\end{align}
$ \delta y^{aij} $ and $ \delta \lambda_{abcd} $ are counter terms necessary for removing the overall divergences of 1PI vertex functions. In \msbar they are singular in $ \epsilon = 4-d \to 0 $ and contain no finite part. The field strength normalizations, $ (Z_\Psi)_{ij} $ and $ (Z_\phi)_{ab} $, are needed to renormalize 1PI 2-point functions of the fermion and scalar fields respectively. They are given by $ (Z_\Psi)_{ij} = \delta_{ij} + \text{singular terms} $ and similarly for the scalars. 

The $ \beta $-function associated with a coupling $ g^{I} $ is then given by 
	\begin{equation} \label{eq:generic_beta_function}
	\beta^I = \dfrac{\dd g^{I}}{\dd \ln\mu} = \bigg(-\rho_I + \sum_{j} \rho_J g^{J} \partial_J \bigg) \left[ \mu^{-\rho_I \epsilon} g_0^{I} \right]_{1/\epsilon},
	\end{equation}
with $ \rho_{J} $ being the dimension of the bare coupling in $ 4-\epsilon $ dimensions. The simple pole of the bare coupling must come from either the vertex counter term or from the wave function renormalization of one of the external fields. Hence, the only TSs that can appear in the $ \beta $-functions are 1PI vertex corrections, and 1PI 2-point structures attached to each leg of a vertex.

The situation for the gauge couplings is \emph{a priori} a little less straightforward, as multiple interaction vertices involve the gauge field. However, using the background field method, gauge invariance is preserved through renormalization~\cite{DeWitt:1967ub,Abbott:1981ke}, and so only one counter term is needed: 
	\begin{equation}
	G_{0,AB}^{-2} = \mu^{-\epsilon} \left(G_{AB}^{-2} + \delta G^{-2}_{AB} \right),
	\end{equation}
from which one finds
	\begin{equation} \label{eq:G2_ct}
	G^{2}_{0,AB} = \mu^{\epsilon} \left(G^2 - G^2 \delta G^{-2}\, G^2 + \ldots \right)_{AB},
	\end{equation}
where the neglected terms only contribute to higher order poles. The counter term is in turn determined by the divergence of the 2-point function of the \emph{background} field and is given by  
	\begin{equation} \label{eq:gauge_renorm_condition}
	\delta G_{AB}^{-2} = \div\left( \Pi_{AB} (p^2 = 0) \right).
	\end{equation}
Thus the only TSs that can appear in $ \beta_G = \tfrac{\dd G^2}{\dd \ln\mu }$ are those needed to renormalize the amputated 1PI 2-point function of the gauge field. Whereas one must take ghost fields and gauge fixing into account when doing the perturbative computation, we need not concern ourselves with them here: ghost fields introduce no new couplings, and the $ \beta $-functions are independent of gauge-fixing parameters. 

We will also need to understand how the tensor $ T_{IJ} $ in \gfe comes about in perturbation theory. 
At leading order, $T_{IJ}$ is related to the counter terms of 2-point functions between marginal operators, $ \brakets{\mathcal{O}_I \mathcal{O}_J} $~\cite{Jack:1990eb}. Thus, it contains the insertions of the couplings found in $ \mathcal{O}_I $. With the couplings $ g^{I} = \braces{G^2_{AB}, \, y_{aij}, \, \lambda_{abcd}} $ of the general theory, Eq. \eqref{eq:general_Lagrangian}, we have 
	\begin{equation}
	\begin{split}
	\mathcal{O}_{AB} &= \dfrac{\delta S_0}{\delta G^2_{AB}} = \dfrac{1}{4} G^{-2}_{AC} G^{-2}_{BD}  F^{C}_{\mu\nu} F^{D\mu\nu} + \ldots, \\
	\mathcal{O}_{aij} &= \dfrac{\delta S_0}{\delta y_{aij}} = - \dfrac{1}{2} \phi_a \Psi_i \Psi_j + \ldots, \\
	\mathcal{O}_{abcd} &= \dfrac{\delta S_0}{\delta \lambda_{abcd}} = - \dfrac{1}{24} \phi_a \phi_b \phi_c \phi_d +\ldots,
	\end{split}
	\end{equation}
where the ellipses denote higher-order contributions arising as a consequence of renormalization. Each open index in $ T_{IJ} $ thus has an insertion of one of the tensors (we resolve the collective indices $ I,J $ down to their constituent indices, depending on the coupling in question)
	\begin{equation} \label{eq:open_index_tensors}
	G^{-2}_{I_A J_C} G^{-2}_{I_B J_D}, \qquad \delta_{I_a J_b} \delta_{I_i J_k} \delta_{I_j J_\ell}, \qquad \delta_{I_a J_e} \delta_{I_b J_f} \delta_{I_c J_g} \delta_{I_d J_h}, 
	\end{equation}
depending on whether the index corresponds to a gauge, a Yukawa, or a quartic coupling respectively; this also holds at higher orders. Borrowing terminology from Feynman diagrams, the TSs appearing in $ T_{IJ} $ are then simply all possible 1PI contractions with two such open indices and no tadpole substructures (to be precise, the tensors should be 1PI only after contracting e.g. couplings on the two open indices). 

Having specified the form of $\beta^{I}$ and $T_{IJ}$, we will need to include the change from $ \beta^{I} $ to $ B^{I} $, c.f. Eq. \eqref{eq:Sdef}, albeit $ \beta $ is our ultimate interest in the present work. Thus, we need to determine the structure of $ (Sg)^{I} $. Given the continuous symmetry $ G_K $ of the scalar and fermion kinetic terms, $ S $ takes values in the associated Lie algebra, $\mathfrak{g}_{K}$. Since the scalars and fermions are charged under this symmetry, it acts non-trivially on scalar and fermion indices. Separating out the scalar and fermion groups $ G_K = G_\phi \times G_\Psi $ and Lie algebras $ \mathfrak{g}_K = \mathfrak{g}_\phi \oplus \mathfrak{g}_{\Psi} $ we may write $ S = S_{ab} \oplus S_{ij} $. The real scalars being in a real representation of $ G_\phi $ implies that $ S_{ab} $ can be written in terms of anti-symmetric TSs with two open scalar indices. Similarly, the collective fermion index $ i $ running over both left- and right-handed fermions transform as if in a real representation and $ S_{ij} $ is parameterizable in terms of TSs antisymmetric in two open fermion indices. Consequently, to identify possible contributions to $S$, one needs only consider the two-point TSs at a given loop order, determine which are not symmetric under exchange of indices, and define an associated $S$ term as an antisymmetric combination of the tensor and its transpose. As seen in \cite{Fortin:2012hn} and \cite{Jack:2013sha}, the first such antisymmetric TSs may appear at 3-loop order, giving non-zero contributions to the 3-loop Yukawa $ B $-function; when expanding \gfe to construct the $A$-function, failing to allow for these shifts leads to an inconsistent system of equations.

For the last component of \gfe, we can infer the form of $ \tilde{A} $ directly, since it has no open indices in coupling space. Hence, it must be parameterizable by fully contracted 1PI TSs. There is one more constraint on the TSs that may appear in $ \tilde{A} $, following from \gfe: no TSs that can be split into two disconnected parts by removing any one tensor (coupling or generator) will appear in $ \tilde{A} $. Suppose for contradiction that one or more such TS appears in $ \tilde{A} $: $ \partial_I \tilde{A} $ will contain at least one TS with disconnected parts. It would then follow from \gfe that the same TS appears in $ T_{IJ} B^{J} $, but both $ T_{IJ} $ and $ B^{J} $ are connected, and so is their contraction. Hence, $ \tilde{A} $ does not contain any TSs that can be disconnected by eliminating just one tensor.

The discussion in this section mostly apply to other renormalization schemes, but some modifications may be necessary. In more general renormalization schemes, one is free to define the finite part of each renormalization constant however one wishes, hence there would exist schemes in which the restrictions above no longer apply (although the restriction of $\beta_{G}$ contributions to 2-point functions will always hold, as it follows from renormalizability). If such a scheme is chosen, one need only include extra TSs representing any additional non-zero contributions to each $\beta$-function. In addition, the relation between the $\beta$-functions in two different schemes may be deduced by calculating the effects of a coupling-redefinition. Given a redefinition $g^{I} \rightarrow \tilde{g}^{I} = g^{I} + \delta g^{I}$, there exist two relations between the original function $\beta^{I}$ of couplings $g^{I}$ and the new function $\tilde{\beta}^{I}$ of couplings $\tilde{g}^{I}$:
	\begin{align}
	\tilde{\beta}^{I}(\tilde{g}) &\equiv \mu\frac{\partial \tilde{g}^{I}}{\partial \mu} = \beta^{I} + \beta^{J} \frac{\partial}{\partial g^{J}}\delta g^{I}\nn
	&= \tilde{\beta}^{I}(g + \delta g) = \left[ 1 + \delta g^{K}\frac{\partial}{\partial g^{K}} + \frac{1}{2} \, \delta g^{J} \, \delta g^{K} \frac{\partial^2 }{\partial g^{J} \,\partial g^{K}} + \ldots\right]\tilde{\beta}^{I}(g).
	\end{align}
Expanding $\delta g^{I} = (\delta g^{I})^{(1)} + (\delta g^{I})^{(2)} + \ldots $ by loop order, the change $\delta \beta^{I} \equiv \tilde{\beta}^{I}(g) - \beta^{I}(g)$ may then be inferred order-by-order, giving
	\begin{align}
	(\delta \beta^{I})^{(1)} &= 0\nn
	(\delta \beta^{I})^{(2)} &= (\beta^{J})^{(1)}\frac{\partial}{\partial g^{J}} (\delta g^{I})^{(1)} - (\delta g^{J})^{(1)} \frac{\partial}{\partial g^{J}}(\beta^{I})^{(1)},\nn
	(\delta \beta^{I})^{(3)} &= (\beta^{J})^{(1)} \frac{\partial}{\partial g^{J}} (\delta g^{I})^{(2)} + (\beta^{J})^{(2)} \frac{\partial}{\partial g^{J}} (\delta g^{I})^{(1)} - (\delta g^{J})^{(2)} \frac{\partial}{\partial g^{J}} (\beta^{J})^{(1)},\nn
	&\quad\; - (\delta g^{J})^{(1)} \frac{\partial}{\partial g^{J}}(\beta^{I})^{(2)} - \dfrac{1}{2} (\delta g^{J})^{(1)} (\delta g^{K})^{(1)} \frac{\partial^2 }{\partial g^{J} \,\partial g^{K}} (\beta^{I})^{(1)} \nn
	&\quad\; - (\delta g^{J})^{(1)} \frac{\partial}{\partial g^{J}} \left[ (\beta^{K})^{(1)}\frac{\partial}{\partial g^{K}}(\delta g^{I})^{(1)} - (\delta g^{K})^{(1)}\frac{\partial}{\partial g^{K}} (\beta^{I})^{(1)} \right],
	\end{align}
and so on. $\delta g^{I}$ may then be parametrized as a sum over a gauge basis of \emph{all} TSs, both 1PI and 1PR, each with its own coefficient. One will then find that the changes in some of the $\beta$-function coefficients are zero, establishing scheme-independence; furthermore, one can easily see whether a given scheme requires 1PR contributions, as the necessary choice of $\delta g^{I}$ will lead to non-zero 1PR coefficients. By extension, if one simply imposes the desired relations between $\delta g^{I}$ coefficients, it is possible to define a class of \msbar-like renormalization schemes in which 1PR contributions are absent\footnote{It turns out that there is a subtlety in these calculations, which only occurs for general theories containing field multiplets. Attempting to ensure that all 1PR terms cancel forces certain relations between the coefficients of a general redefinition $\delta g^{I}$, but these relations may leave behind certain contributions, formed by attaching antisymmetric combinations of 1PR 2-point TSs to a tensor coupling, similar to the terms generated by $S$ in the $B$-functions. In \cite{Gracey:2015fia}, the coupling redefinitions required to relate \msbar to momentum subtraction (\mom) produced non-zero coefficients for these 1PR terms, despite the explicit \mom calculation giving zero as expected. The resolution to this apparent paradox is described in \cite{Jack:2016tpp}, whereby such 1PR terms can be absorbed into a redefinition of the otherwise-arbitrary antisymmetric part of the anomalous dimension matrix for the scalar multiplet.}. By calculating the effects of the most general possible coupling redefinition, it can also explicitly be shown that constraints derived using \gfe are scheme-independent, as all associated variations in the CCs cancel.

\section{Procedure for extracting $\beta$-function constraints} \label{sec:strategy}
We will now explain the procedures employed in \grafer to identify unique TSs and to determine the CCs on the coefficients of the $ \beta $-functions.

\subsection{Tensor-graph identification} \label{sec:graph_notation}
Part of constructing the Weyl CCs comes down to identifying the unique ways of contracting the indices of the couplings. For suitably large combinations of the couplings, it will quickly become unmanageable to manually (or automatically) identify unique combinations using regular dummy-index notation. One way of dealing with this problem is to map the TSs into graphs, in which case identifying identical TSs becomes a problem of identifying graph isomorphisms. 

The rules for mapping between TSs and graphs are a further development of the notation developed in Ref. \cite{Jack:2014pua}; they essentially come down to associating a graph with the tensor part of a Feynman diagram. The starting point is that the edges of the graph come in three varieties,
	\begin{equation}
	\vcenter{\hbox{\includegraphics{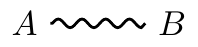}}} = G^2_{AB}, \qquad 
	\vcenter{\hbox{\includegraphics{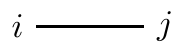}}} = \delta_{ij}, \andeq 
	\vcenter{\hbox{\includegraphics{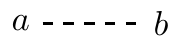}}} = \delta_{ab},
	\end{equation}
and are to be understood as contractions of field indices of the corresponding kind: gauge (adjoint), fermion, and scalar respectively. Remember here that all \emph{internal} gauge edges carry the gauge coupling matrix by definition. 

There are five kinds of vertices in the graph notation, each corresponding to a coupling or a group generator:
	\begin{equation} \label{eq:graph_vertices}
	\begin{gathered}
	\vcenter{\hbox{\includegraphics{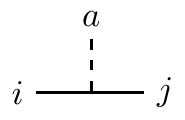}}} = y_{aij}, \qquad \vcenter{\hbox{\includegraphics{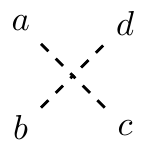}}} = \lambda_{abcd}, \\
	\vcenter{\hbox{\includegraphics{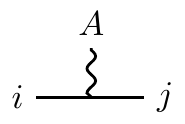}}} = (T^{A})_{ij}, \qquad \vcenter{\hbox{\includegraphics{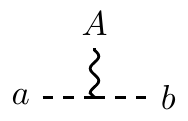}}} = (T^{A}_\phi)_{ab}, \andeq 
	\vcenter{\hbox{\includegraphics{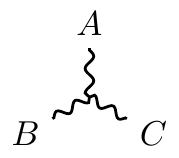}}} = G^{-2}_{AD} f^{DBC}.
	\end{gathered}
	\end{equation}
As discussed previously, the $G^{-2}_{AB}$ factor may be ``commuted through" the triple-gauge vertex. The graph vertices roughly correspond to the interaction vertices of a QFT, except for the interactions between two scalars and two gauge bosons, or between four gauge bosons. The couplings associated to those interactions can be reduced to combinations of the vertices above.

The Feynman rules for the fermions in the Lagrangian \eqref{eq:general_Lagrangian} introduce a $ \sigma_1 $ in the $ 2\times 2 $ subspace spanned by $ \psi $ and $ \psi^{\dagger} $ on every fermion propagator and in any gauge-fermion vertex\footnote{The difference between $ \sigma_\mu $ and $ \bar{\sigma}_\mu $ vanishes in every fermion loop, with the exception of the $ \gamma_5 $ contributions discussed in Sec. \ref{sec:gamma5}.}. We introduce the notation
	\begin{equation} \label{eq:tilde_quantities}
	\tilde{y}_a = \sigma_1 y_a \sigma_1 = \begin{pmatrix}
	Y_a^{\ast} &0 \\ 0 & Y_a
	\end{pmatrix} \andeq \tilde{T}^{A} = \sigma_1 T^{A} \sigma_1 = \begin{pmatrix} -(T_{\psi}^{A})^\ast & 0 \\ 0 & T_{\psi}^{A}  \end{pmatrix}
	\end{equation}
to account for the corresponding chirality flips.
To see how this works out on a fermion line, consider the following example:
	\begin{equation}
	\vcenter{\hbox{\includegraphics{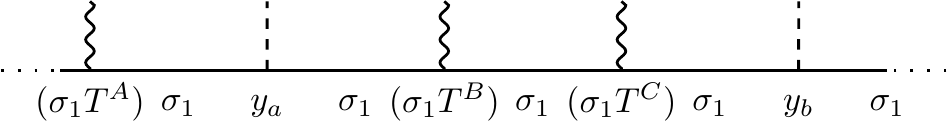}}} = \ldots \tilde{T}^{A} y_a T^{B} T^{C} \tilde{y}_b \ldots 
	\end{equation}   
In practice, it becomes unnecessary to explicitly worry about including the $ \sigma_1 $s on the fermion lines: the effect of them is merely the unique choice that ensures that $ \psi $ is always contracted with $ \psi^\dagger $ between two vertices. At the level of the tensors it is sufficient to remember the following rules, reading left to right along a fermion line:
	\begin{enumerate} \setlength\itemsep{0.2em}
	\item $ y $ is always followed by a $ \tilde{y} $ regardless of the number of gauge generators between the two.
	\item All generators following $ y $ and before $ \tilde{y} $ are of the kind $ T $.
	\item All generators before $ y $ and following $ \tilde{y} $ are of the kind $ \tilde{T} $.
	\end{enumerate}  
Any fermion line satisfying these rules makes for a valid tensor. On the other hand, the first rule cannot be satisfied for a fermion loop with an odd number of Yukawa insertions. These are invalid because a fermion loop must have an even number of chirality flips. 

As an example of how the graph representation works in practice, take the 1-loop gauge $ \beta $-function. In graph notation this is given by 
	\begin{align}
	\dfrac{\dd G^{2}_{AB}}{\dd \ln \mu} &= \dfrac{G^2_{AC}}{(4\pi)^2} \left( -\dfrac{22}{3} \vcenter{\hbox{\includegraphics[]{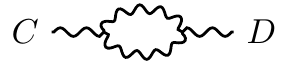}}} + \dfrac{2}{3} \vcenter{\hbox{\includegraphics[]{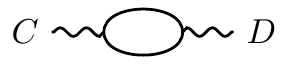}}} + \dfrac{1}{3} \vcenter{\hbox{\includegraphics[]{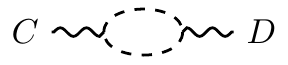}}} \right) G^2_{DB} \nn
	&= \dfrac{G^2_{AC}}{(4\pi)^2} \left( -\dfrac{22}{3} f^{CEF} f^{EFD}+ \dfrac{2}{3} \Tr{T^{C} T^{D}} + \dfrac{1}{3} (T_\phi^{C} T_\phi^{D})_{aa} \right) G^2_{DB}.
	\end{align}
This can be compared to the traditional notation for the case of a simple gauge group~\cite{Luo:2002ti}:
	\begin{equation}
	\dfrac{\dd g^2}{\dd \ln\mu} =\dfrac{g^4}{(4\pi)^2} \left(- \dfrac{22}{3} C_2(G) + \dfrac{4}{3} S_2(F) + \dfrac{1}{3} S_2(S) \right),
	\end{equation}
where $ C_2(G) $ is the quadratic Casimir of the adjoint and $ S_2(F[S]) $ denotes the trace normalization of the fermion [scalar] representation. The graph notation accurately incorporates the appropriate group invariants for any choice of indices $ A=B $ selecting a specific gauge group. The factor of two difference in the coefficient of the second term between our notation and the traditional result is due to our notation automatically including the complex conjugate of every fermion loop. In this simple---but well known---example, our notation unfortunately comes across as slightly cumbersome, which is the price to pay for complete generality. At higher loop orders or in theories with more gauge groups, however, the notation comes into its own. As a more advanced example of the notation, consider the following TS from the 3-loop gauge $ \beta $-function:
	\begin{equation}
	\vcenter{\hbox{\includegraphics{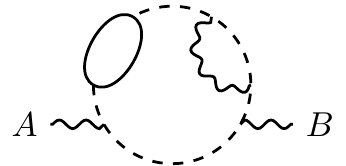}}} = (T_\phi^{A} T_\phi^{B} T_\phi^{C} T_\phi^{D})_{ab} \Tr{y_b \tilde{y}_a} G^{2}_{CD}.
	\end{equation}
Note the factor $ G^{2} $ coming from the internal gauge line. This line contains contributions from all the gauge groups, and will give a sum of the quadratic Casimirs weighted with the respective gauge couplings. 

There is a potential pitfall with the diagram notation, which must be addressed. Since the gauge generators are antisymmetric, there is an ambiguity in which leg is associated with each index. For scalar generators, we will impose the rule that all generators on the same scalar line (beginning and ending at Yukawa or quartic interactions or open indices) have the same orientation, that is, they join in combinations such as $ (T^{A}_\phi T^{B}_\phi)_{ab} $. For an even number of generators on a line, as is almost exclusively the case in the TSs we work with, this resolves the ambiguity. In other cases the orientation will have to be specified explicitly. On fermion loops the ambiguity comes down to what way one should go around the loop. This results in a sign ambiguity when there is an odd number of gauge generators on said loop, but that is not the case for any of the TSs in our basis. 

A final point regarding the graph notation is that we will use a small blob to denote an open index, $ I $, such as in e.g. $ T_{IJ} $, with corresponding tensors
	\begin{equation}
	\begin{gathered}
	\vcenter{\hbox{\includegraphics{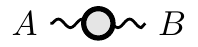}}} = G^{-2}_{I_A A} G^{-2}_{I_B B}, \qquad \vcenter{\hbox{\includegraphics{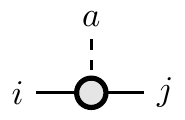}}} = \delta_{I_aa} \delta_{I_ii} \delta_{I_jj}, \\
	\vcenter{\hbox{\includegraphics{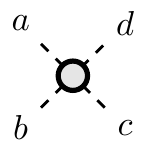}}} = \delta_{I_aa} \delta_{I_bb} \delta_{I_cc}, \delta_{I_dd}, 
	\end{gathered}
	\end{equation}
as per Eq. \eqref{eq:open_index_tensors}. To distinguish between different open indices we will use a diamond instead where it is required, though the interpretation wrt. tensors is the same. 

At this point, we can finally address the underlying topological reasoning for the 3--2--1 ordering. By construction, $\beta_{AB}$, $\beta_{aij}$ and $\beta_{abcd}$ are sums over two-, three-, and four-point TSs respectively. Similarly then, the open indices in $ T_{IJ} $ involve two, three, and four internal lines. It then becomes clear that the leading contribution to $ T_{I(AB)} $, $ T_{I(aij)} $, and $ T_{I(abcd)} $ occur at loop orders $ 1 $, $ 2 $, and $ 3 $ respectively. This is basically a consequence of the characteristic formula for graphs: $ \ell = I - V +1 $, where $ \ell $ is the number of loops, $ I $ the number of internal lines, and $ V $ the number of vertices. Since in the loop expansion of \gfe the number of loops of the $ \beta $-function and $ T_{IJ} $ must add to that of $ \tilde{A} $, it follows that $ \beta_{abcd} $ always enter at one loop order less than $ \beta_{aij} $, which itself is one order less than $ \beta_{AB} $. Hence the ordering $\tilde{A}^{(4)}$ receives contributions from the 3-loop gauge, 2-loop Yukawa and 1-loop quartic $\beta$-functions, and relates the various coefficients via \eqref{eq:GFE}.  Clearly, this pattern extends to all higher orders where $\tilde{A}^{(5)}$ simply relates the $\beta$-functions at 4-, 3-, and 2-loops.

The graph notation developed in this section is similar to that of Feynman diagrams. This is no accident and the similarities makes it easy to visualize a typical Feynman diagram with a similar contraction of the couplings. However, the graphs are \emph{not} to be thought of as diagrams, as they contain no information about loops or momentum structure, and several Feynman diagrams can share coupling dependence due to gauge symmetry.

\subsection{Gauge basis}
In the Lagrangian \eqref{eq:general_Lagrangian}, the tensor couplings $ y_{aij} $ and $ \lambda_{abcd} $ are contracted with field multiplets, which have non-trivial gauge transformations. Clearly, if \eqref{eq:general_Lagrangian} is to be gauge-invariant, there must be non-trivial relations involving the tensor couplings and the gauge generators; by demanding invariance under an infinitesimal gauge variation, one finds that
	\begin{align}
	0 &= -\tilde{T}^{A} y_a + y_a T^{A} + y_b (T_{\phi}^{A})_{ba},\\
	0 &= (T^{A}_\phi)_{ae} \lambda_{ebcd} +(T^{A}_\phi)_{be} \lambda_{aecd} + (T^{A}_\phi)_{ce} \lambda_{abed} +(T^{A}_\phi)_{ad} \lambda_{abce}.
	\end{align}
In addition to these identities, the gauge generators still satisfy the Lie algebra, which in our notation reads the usual way:
	\begin{align}
	0 &= [T^A, T^{B}] - i f^{ABC} T^{C},\\
	0 &= [T^A_\phi, T^{B}_\phi]_{ab} - i f^{ABC} (T^{C}_\phi)_{ab},\\
	0 &= f^{ABE} f^{CDE} - f^{ACE} f^{BDE} + f^{ADE} f^{BCE}.
	\end{align} 
This follows from the structure constants vanishing whenever two of the gauge indices belong to distinct product groups. 

The gauge identities in the tensor space introduce redundancy between all the unique graphs that can be constructed. Since \gfe involves derivatives of the TSs associated with each graph, there will exist degeneracies in the resulting system of equations, unless one attempts to express all TSs in terms of a basis; one must therefore account for this arbitrariness before trying to extract the Weyl CCs. At higher loop orders, removing this redundancy also greatly reduces the number of arbitrary coefficients introduced into the system. For instance, \grafer produces 2890 unique graphs in the 4-loop order gauge $ \beta $-function, but after accounting for the gauge identities we are left with a basis consisting of just 198 TSs.

To construct a suitable gauge basis of TSs for each $\beta$-function, we follow the steps outlined here, using the $ \ell $-order contribution to the Yukawa $ \beta $-function, $ \beta_{aij}^{(\ell)} $ as an example:
\begin{enumerate}
	\item Determine all unique $ \ell $-loop graphs with one scalar and two fermion legs (indices) open, removing the isomorphic graphs.
	\item Construct the full set of all gauge relations involving the graphs.
	\item In \msbar, $\beta_{aij}^{(\ell)} $ only receives contributions from the 1PI leg and 1PI vertex graphs. Starting from this set, use the gauge identities to reduce the set of $\ell$-loop graphs to a basis. 
	\item If possible, choose a basis that maximizes the number of group Casimirs and Dynkin indices.
\end{enumerate}
The last step may seem a bit strange and deserves a comment. In principle, one is free to construct $\tilde{A}$ using any of the large number of equivalent gauge bases. However, for practical purposes it is convenient to work with TSs that, in so far as is possible, can be written in terms of group invariants, as explicit $\beta$-function calculations for gauge theories are usually expressed in terms of such invariants. Such TSs have very useful commutativity properties, and the correspondence between our graph notation and explicit calculations becomes more straightforward.

What we find using this procedure is a basis of TSs that can appear in \msbar, plus any scheme related to \msbar by a coupling redefinition that preserves the form of the TSs\footnote{If one wishes to work with a completely general scheme, one need only remove the \msbar-like restriction in step 3 above and repeat the procedure, taking all possible graphs into account.}. Once the Feynman diagrams have been evaluated, it may turn out that some coefficients vanish, as was found in previous general calculations. As there is no \emph{a priori} way to know if a particular coefficient will be zero, or whether such a zero is (at sufficiently-high order) merely an artefact of using \msbar, the tensor should still be included.

\subsection{Overall strategy}
To extract the CCs from \gfe, one must first parameterize each of the quantities in terms of a basis of TSs with unknown coefficients. The external indices are then contracted in the combination prescribed by the equation, before the two sides are set equal TS by TS in the basis. The result is a system of equations between all the unknown coefficients.

It is typically convenient to think of \gfe in differential form,
	\begin{equation}
	\dd \tilde{A} = \dd g^{I}\, \partial_I \tilde{A} = \dd g^{I} \, T_{IJ} B^{J}.
	\end{equation} 
In this form the differentiated $ A $-function can be constructed by marking couplings in each graph one by one, and tallying up the result. Working to loop order $ \ell $, \grafer performs the following steps for each side of \gfe:

\noindent\emph{For the LHS:} 
\begin{enumerate}
	\item Generate all unique closed graphs at $ \leq \ell $'th order, using the five kinds of vertices of Eq. \eqref{eq:graph_vertices}. Each graph corresponds to a TS.
	\item Use the gauge identities to determine all relations between the TSs. Eliminate tensors using linear dependence to achieve a basis. These are the unique TSs parameterizing the $ A $-function, and each is given an unknown coefficient.
	\item The differential operator is applied to the $ A $-function, acting on each TS. In each TS new TSs are made by marking a Yukawa or quartic vertex or gauge edge signifying a derivative, and equivalent TSs are gathered and matched to a new basis using gauge identities (counting multiplicity). The unique, marked TSs are all the terms that can show on the LHS. 
\end{enumerate}

\noindent\emph{For the RHS:} 
\begin{enumerate}
	\item For each coupling, the TSs showing up in the $ \beta $-function are determined from the unique closed contractions with a corresponding coupling removed, viz. removing a vertex in the case of Yukawa or quartic $ \beta $-functions, or breaking a gauge line for the gauge $ \beta $-function. 
	\item All TSs that do not correspond to 1PI vertex or field-strength graphs are eliminated, as they do not appear in \msbar-like schemes. The remaining TSs are thinned to a basis using gauge identities and each is given a coefficient\footnote{The diagrams chosen for the basis itself are not necessarily 1PI.}. At sufficiently-high loop order, the shift $\beta\rightarrow B$ is taken into account by identifying any possible antisymmetric combinations of $2$-point TSs, attaching to a leg, and assigning a new arbitrary coefficient.
	\item The TSs in the $ T_{IJ} $ tensor are constructed from all unique closed contractions appearing in the $ A $-function at order $ <\ell $. In each structure two couplings are marked, one as a derivative and the other as a point where a $ \beta $-function can be inserted. Each unique TS is given a coefficient. 
	\item The $\ell_{1}$-loop $\beta$-function tensors are inserted into the $\ell_{2}$-loop $T_{IJ}$ tensors at the relevant positions, such that the total loop order is $ \ell_{1} + \ell_{2} \leq \ell $. The resulting TSs are then matched onto the same basis as the LHS.
\end{enumerate}

Once both sides of \gfe are constructed and written in terms of the same gauge basis, equality must apply for the coefficients of each TS individually. In the resulting set of equations, all the unknown coefficients from $ T_{IJ} $ and $ \tilde{A} $ are eliminated. This leaves a system of equations involving only the coefficients of the $ \beta $-functions, which we refer to as the CCs. 

Before proceeding to our analysis we would like to make a brief comment on the inclusion of multiple Abelian groups in the computations. 
Whereas the $ \beta $-function TSs automatically include the all contributions from multiple Abelian factors once they are set up to work with a semi-simple gauge group, this is not the case for $ T_{IJ} $.
The reason for this being that in the semi-simple case all 2-point subtensors on a gauge line are diagonal in the gauge indices. Thus they commute with e.g. $ \dd G^2_{AB} $ or gauge $ \beta $-functions that could be contracted into the open indices. This is not the case with multiple Abelian groups where matter loops, e.g. fermion-loops like $ \Tr{T^{A} T^{B}} $, do not have to be diagonal. If one were to perform the analysis with multiple Abelian factors, it would therefore require more TSs in the basis of $ T_{IJ} $ and in the final basis for \gfe. Whereas the former introduce more unknowns into the system, the latter provides additional constraints as equality must hold for each TS in the basis. What we found was that the inclusion of multiple Abelian factors in the analysis did not contribute extra CCs, nor did it produce fewer (which would have been a red flag, as all the semi-simple information is included in the analysis with multiple Abelian factors). Since no extra information was gained by the extra complication, we present the analysis of the next section in the case with at most one Abelian factor. We stress that the CCs are nevertheless valid for $ \beta $-functions of theories with multiple such factors.

\section{Weyl consistency conditions} \label{sec:results}
We have now---finally---arrived at the point where we can present the constraints on the $ \beta $-functions. In this section we will discuss the CCs order by order in the loop expansion and discuss new complications as they occur.

\subsection{Tensor basis for the $ \beta $-functions}
Before discussing the CCs we will begin by expanding the $ \beta $-functions by loop order, as 
	\begin{equation} \label{eq:beta_parametrization}
	\begin{gathered}
	\beta_{AB} = \dfrac{\dd G^{2}_{AB}}{\dd \ln\mu} = \dfrac{1}{2} \sum_{\mathrm{perm}} \sum_{\ell} G^{2}_{AC} \dfrac{ \beta^{(\ell)}_{CD} }{(4 \pi)^{2\ell}} G^{2}_{DB}, \\
	\beta_{aij} = \dfrac{\dd y_{aij}}{\dd \ln \mu} = \dfrac{1}{2} \sum_{\mathrm{perm}} \sum_{\ell} \dfrac{\beta^{(\ell)}_{aij}}{(4\pi)^{2\ell}}, \andeq \beta_{abcd} = \dfrac{\dd \lambda_{abcd}}{\dd \ln \mu} = \dfrac{1}{24} \sum_{\mathrm{perm}} \sum_{\ell} \dfrac{\beta^{ (\ell) }_{abcd} }{(4 \pi)^{2 \ell}}.
	\end{gathered}
	\end{equation}
The sum over permutations are over the 2 permutations of the gauge indices in $ \beta_{AB} $, the 2 permutations of the fermion indices in $ \beta^{(\ell)}_{aij} $, and the 24 permutations of the scalar indices in $ \beta^{(\ell)}_{abcd} $. Symmetrizing the $ \beta $-functions in this way ensures that they respect the same symmetries as the couplings themselves; in the case of the gauge $ \beta $-function, symmetrization is only required in the event of multiple Abelian factors. The symmetrization is performed at the level of the $ \beta $-functions rather than for individual TSs, to keep notation to a minimum beyond this point.

At each loop order $ \beta^{(\ell)} $ is in turn  parametrized in terms of a basis of TSs, so we write 
	\begin{equation}
	\begin{gathered}
	\beta^{(\ell)}_{AB} = \sum_n \cofg{\ell}{n} \!\! \vcenter{\hbox{\includegraphics{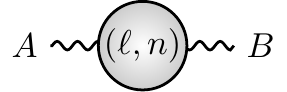}}} , \\
	\beta^{(\ell)}_{aij} = \sum_n \cofy{\ell}{n} \!\! \vcenter{\hbox{\includegraphics{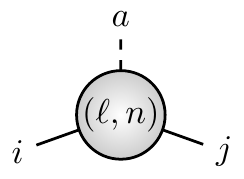}}}, \andeq \beta^{(\ell)}_{abcd} = \sum_n \cofq{\ell}{n} \!\! \vcenter{\hbox{\includegraphics{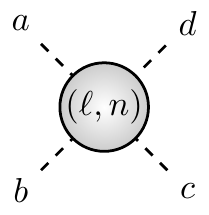}}},
	\end{gathered}
	\end{equation}
where the blobs denote TSs with the appropriate external indices and the sums run over the full basis at each loop order. The gauge $ \beta $-function stands out, as we have factored out two common gauge-coupling matrices. In this way the TSs in $ \beta^{(\ell)}_{AB} $ correspond to 2-point structures of the background field, c.f. Eq. \eqref{eq:gauge_renorm_condition}. 

We can now present a basis of TSs for the general \msbar $ \beta $-functions, or indeed any scheme in which 1PR contributions do not appear. Our choice of basis defines the coefficients that will eventually be constrained by the CCs. The gauge basis we have used for the $ \beta $-function is chosen such that there are as few gauge generators as possible that appear outside (possibly nested) 2-point subgraphs of group invariants. Here we have presented the TSs by mapping them onto graphs using the notation of Sec. \ref{sec:graph_notation}, as we believe this to be the most intuitive way of representing them, in the same manner that Feynman diagrams are usually the easiest way of thinking about loop amplitudes. If the reader is unsure as to the interpretation of the graphs we refer to App. \ref{app:beta_tensors}, where TSs are written out explicitly.  

\begin{figure}
	\centering
	\includegraphics[width=.75\textwidth]{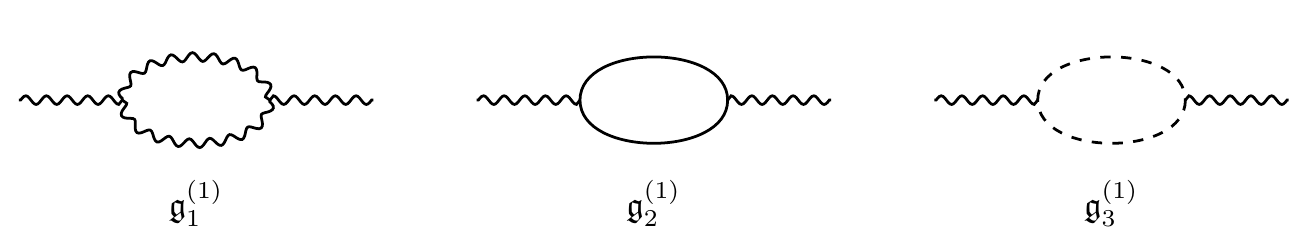}
	\includegraphics[width=\textwidth]{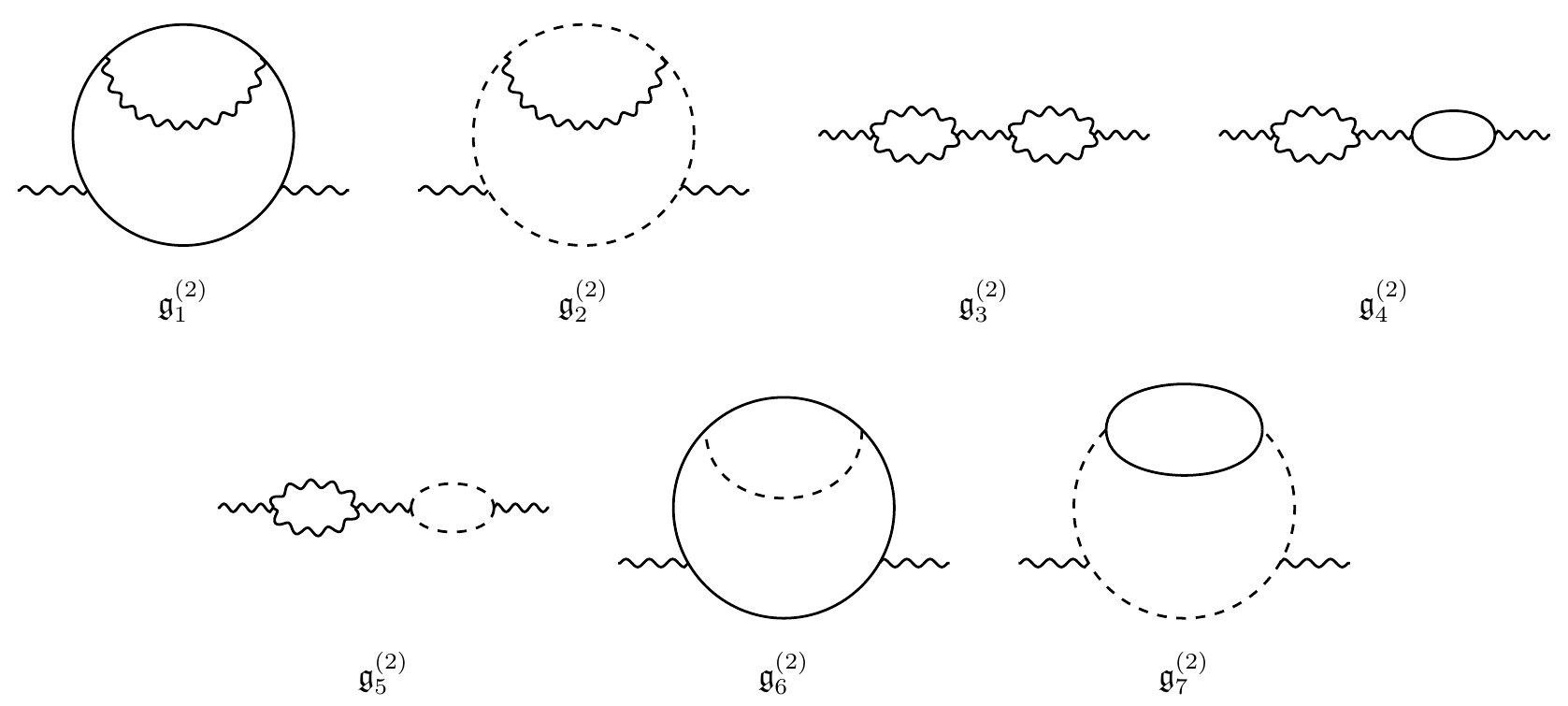}
	\includegraphics[width=\textwidth]{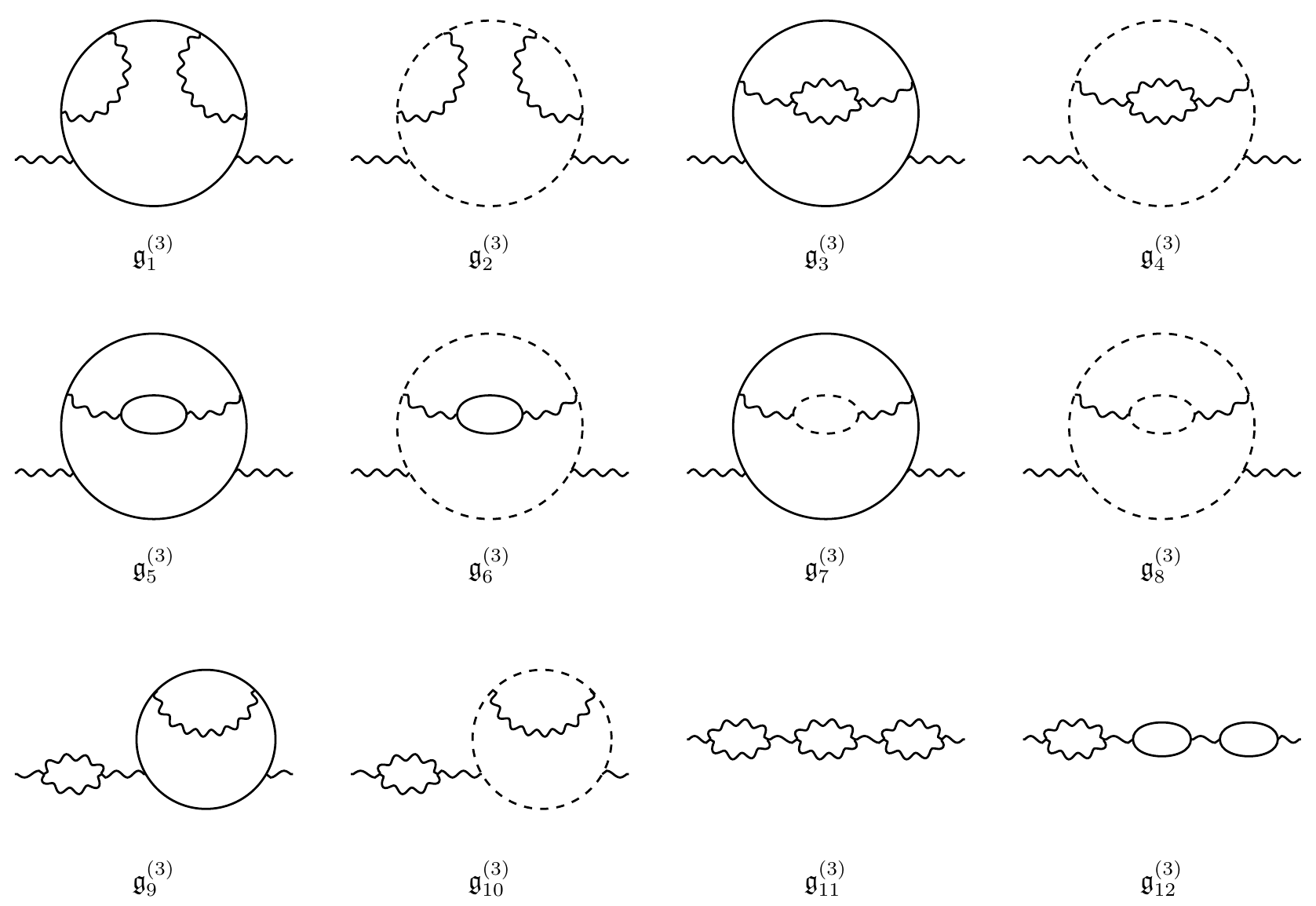}
	\caption{Graph representation of the TSs appearing in the gauge $ \beta $-function, $ \beta^{(\ell)}_{AB} $, at 1 to 3 loop orders. Recall that the external lines do not carry powers of the gauge coupling.}
	\label{fig:betag_1}
\end{figure}

\begin{figure}
	\centering
	\includegraphics[width=\textwidth]{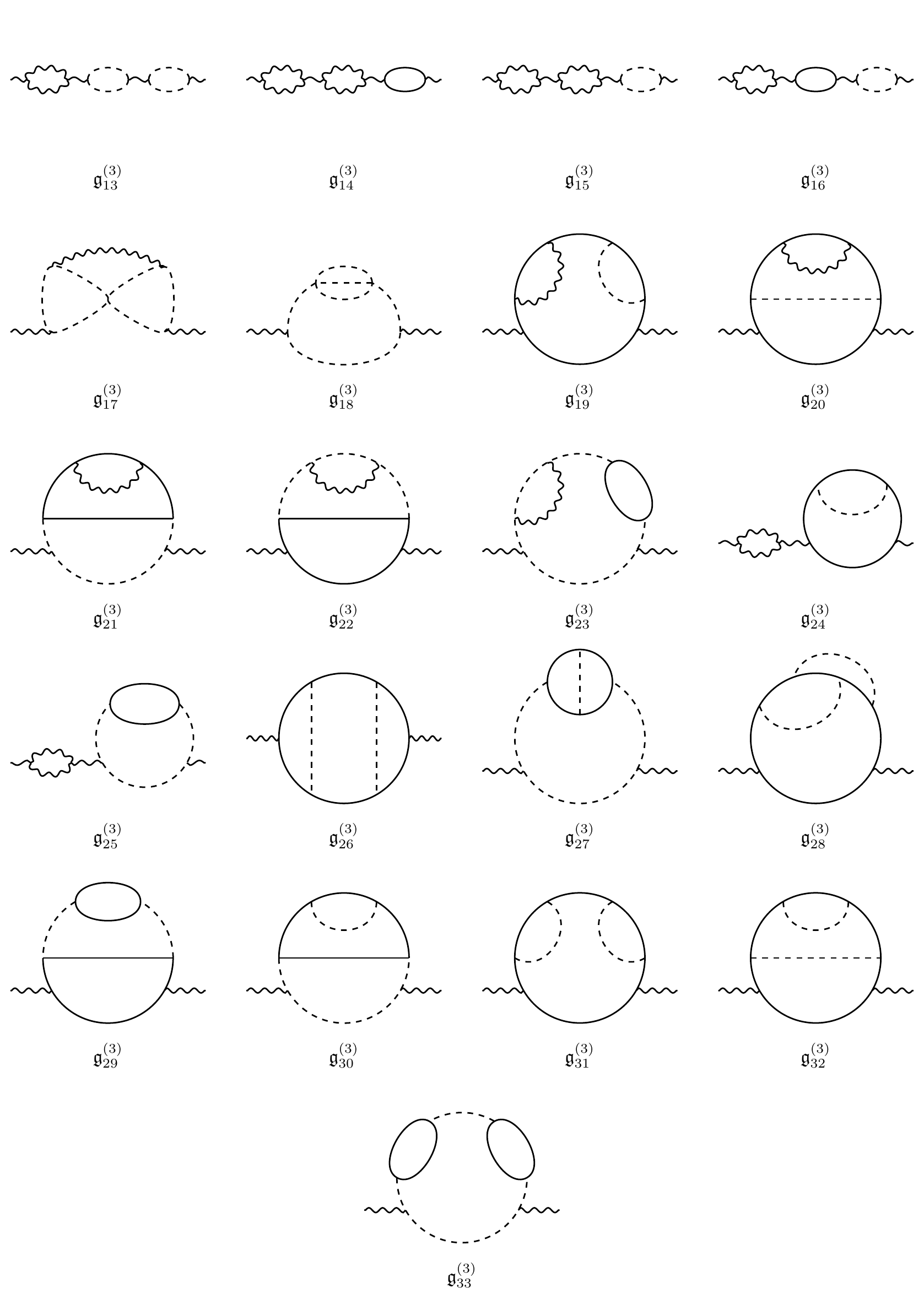}
	\caption{(Continued) Graph representation of the TSs appearing in the gauge $ \beta $-function, $ \beta^{(\ell)}_{AB} $, at 1- to 3-loop orders.}
	\label{fig:betag_2}
\end{figure}

\begin{figure}
	\centering
	\includegraphics[width=\textwidth]{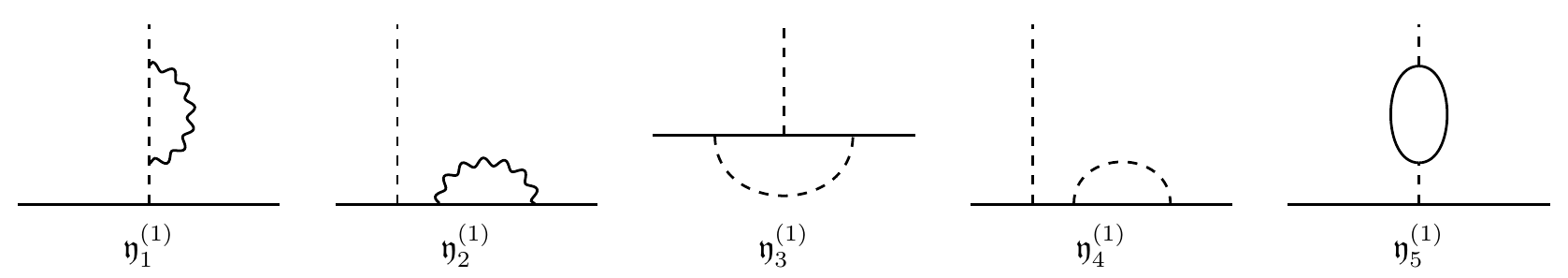}
	\includegraphics[width=\textwidth]{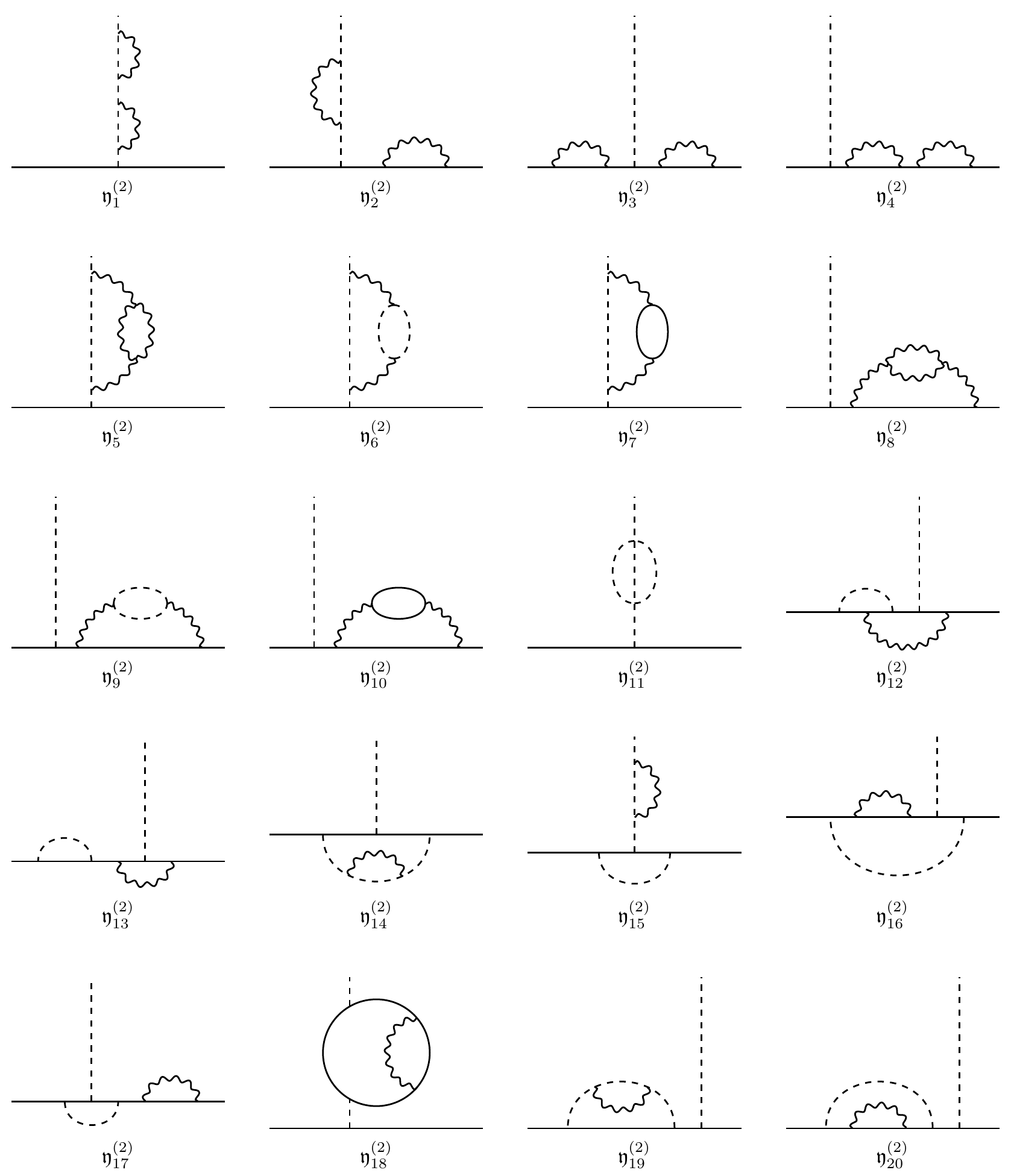}
	\caption{Graph representation of the TSs appearing in the Yukawa $ \beta $-function, $ \beta^{(\ell)}_{aij} $, at 1- and 2-loop orders.}
	\label{fig:betay_1}
\end{figure}

\begin{figure}
	\centering
	\includegraphics[width=\textwidth]{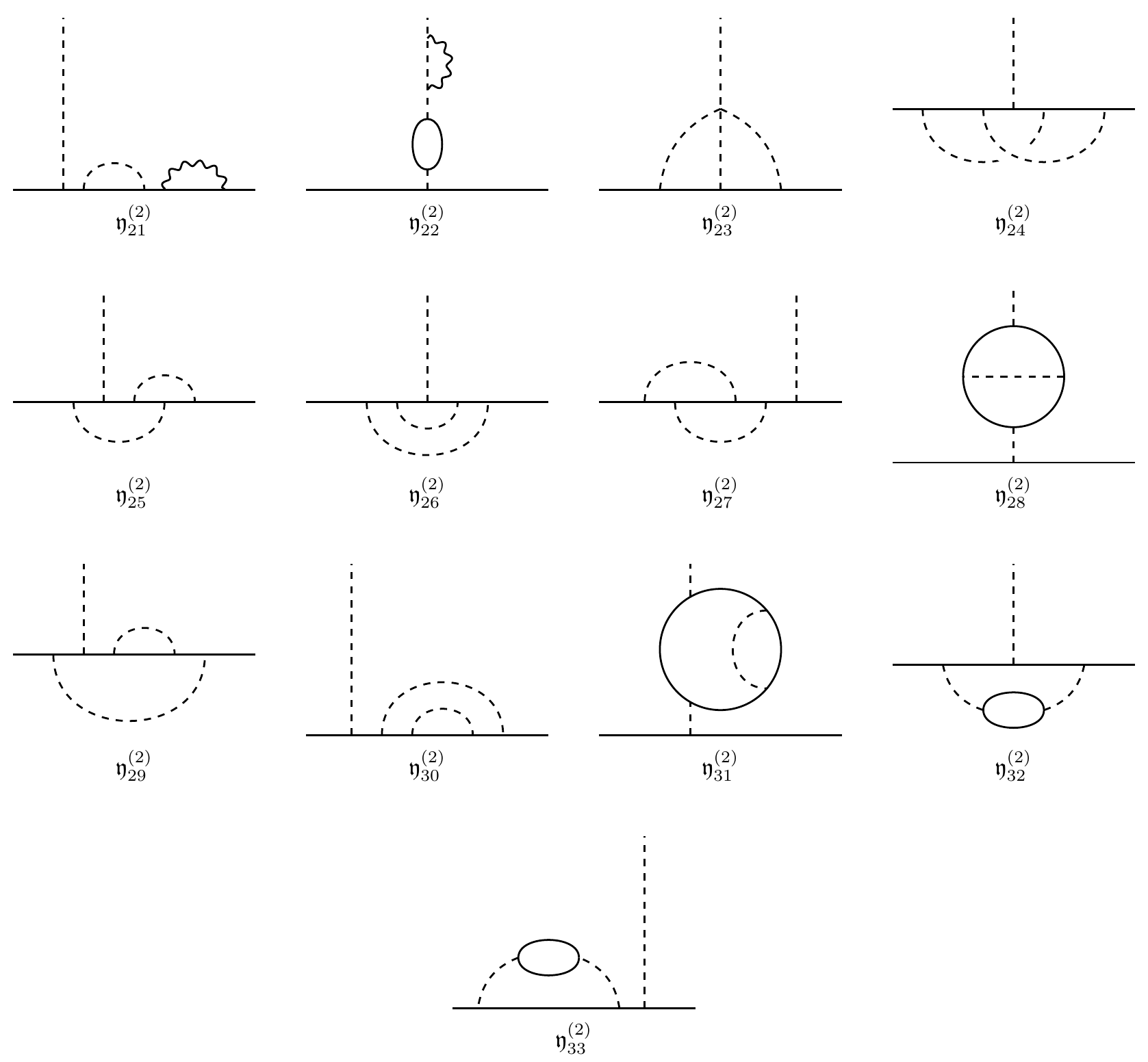}
	\includegraphics[width=\textwidth]{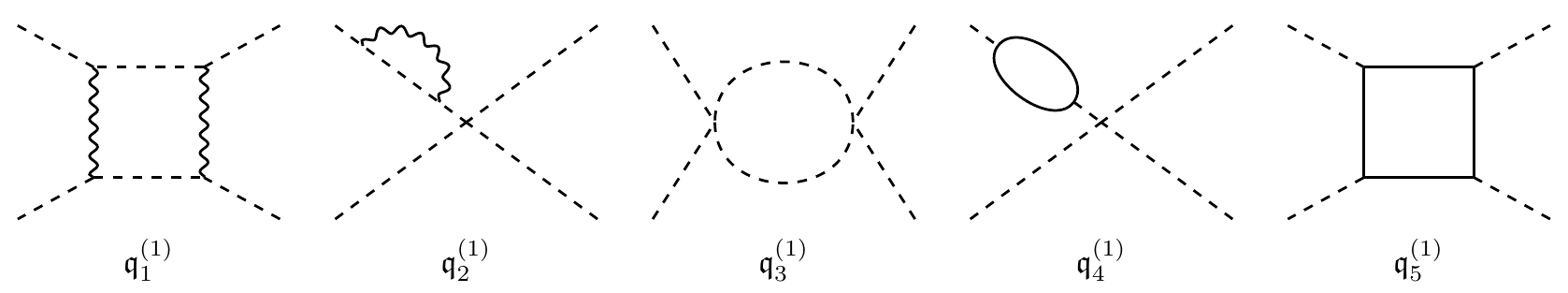}
	\caption{(Continued) Graph representation of the TSs appearing in the Yukawa $ \beta $-function, $ \beta^{(\ell)}_{aij} $, at 2-loop order and the quartic $ \beta $-function, $ \beta^{(\ell)}_{abcd} $, at 1-loop order.}
	\label{fig:betay_2}
\end{figure}

The basis of TSs for the general gauge $\beta$-function, up to three loops, is shown in Figs. \ref{fig:betag_1} and \ref{fig:betag_2}. The 1-loop basis consists of the three well-known contributions from fermion, scalar, and gauge/ghost loops respectively. At 2-loop order there are another 7 TSs, whereas typical \msbar result quote just 6 contributions~\cite{Machacek:1983tz,Machacek:1983fi,Machacek:1984zw,Luo:2002ti}. The discrepancy is due to $ \cofg{2}{7} = 0 $, which can only be determined from computation of the contributing Feynman diagrams. At 3-loop order we find another 33 TSs in $ \beta_{AB}^{(3)} $. In the limiting case where the gauge group is simple, the tensors with coefficients $ \cofg{3}{3} $ and $ \cofg{3}{9} $ coincide, as do those with coefficients $ \cofg{3}{4} $ and $ \cofg{3}{10} $ (a point which we will elaborate further on in Sec. \ref{sec:3--2--1}), reducing our basis to the 31 TSs of \citet{Pickering:2001aq}. For the general Yukawa $ \beta $-function, the Tensor basis include 5 and 33 TSs for 1- and 2-loop orders respectively, shown in Figs. \ref{fig:betay_1} and \ref{fig:betay_2}. This basis is more extensive than the one of \citet{Luo:2002ti}, as 6 of the coefficients vanish, c.f. appendix \ref{app:coefficients}. In Fig. \ref{fig:betay_2} we also include the 5 TSs parameterizing the general 1-loop quartic $\beta$-function; Figs. \ref{fig:betag_1}-\ref{fig:betay_2} therefore comprise the full list of TSs contributing to $\beta^{I}$ in \gfe, up to $A^{(4)}$. 

As has been mentioned, all of the \msbar coefficients multiplying these TSs have been computed previously, though the determination of $\cofg{3}{3, 4, 9, 10}$ is unpublished. We list the coefficients in appendix~\ref{app:coefficients}; in section \ref{sec:3--2--1}, we detail how Weyl CCs augment the results of \cite{Pickering:2001aq} to fix the remaining four coefficients. Since the 2-loop quartic coefficients are also known, we have included the \msbar results for $\beta^{(2)}_{abcd}$, along with explicit representations of our basis of TSs. Appendices \ref{app:beta_tensors} and \ref{app:coefficients} therefore contains the full \msbar $\beta$-functions for a general QFT based on a compact gauge group up to order 3--2--2.

\subsection{Consistency condition at order 2--1--0}
Now, using tensor notation, we shall begin using \gfe to parametrize $\tilde{A}$ loop order by loop order. The first-order calculation of $\tilde{A}$ is precisely the trace anomaly, thus contributions to $\tilde{A}^{(1)}$ are simply in one-to-one correspondence with the field content of the theory. These terms are effectively the constant of integration left undetermined by \eqref{eq:GFE}, and hence are not related to the $\beta$-functions.

\begin{figure}
	\includegraphics[width=\textwidth]{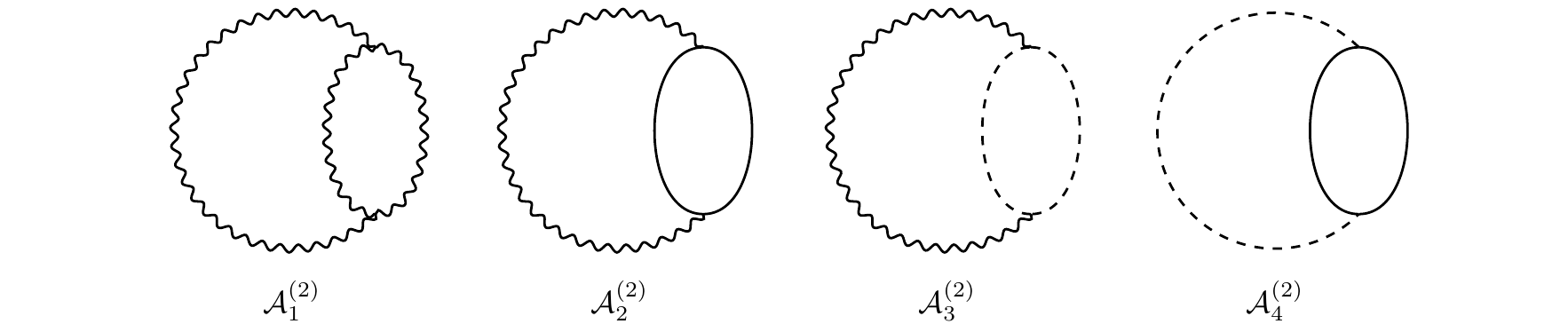}
	\includegraphics[width=\textwidth]{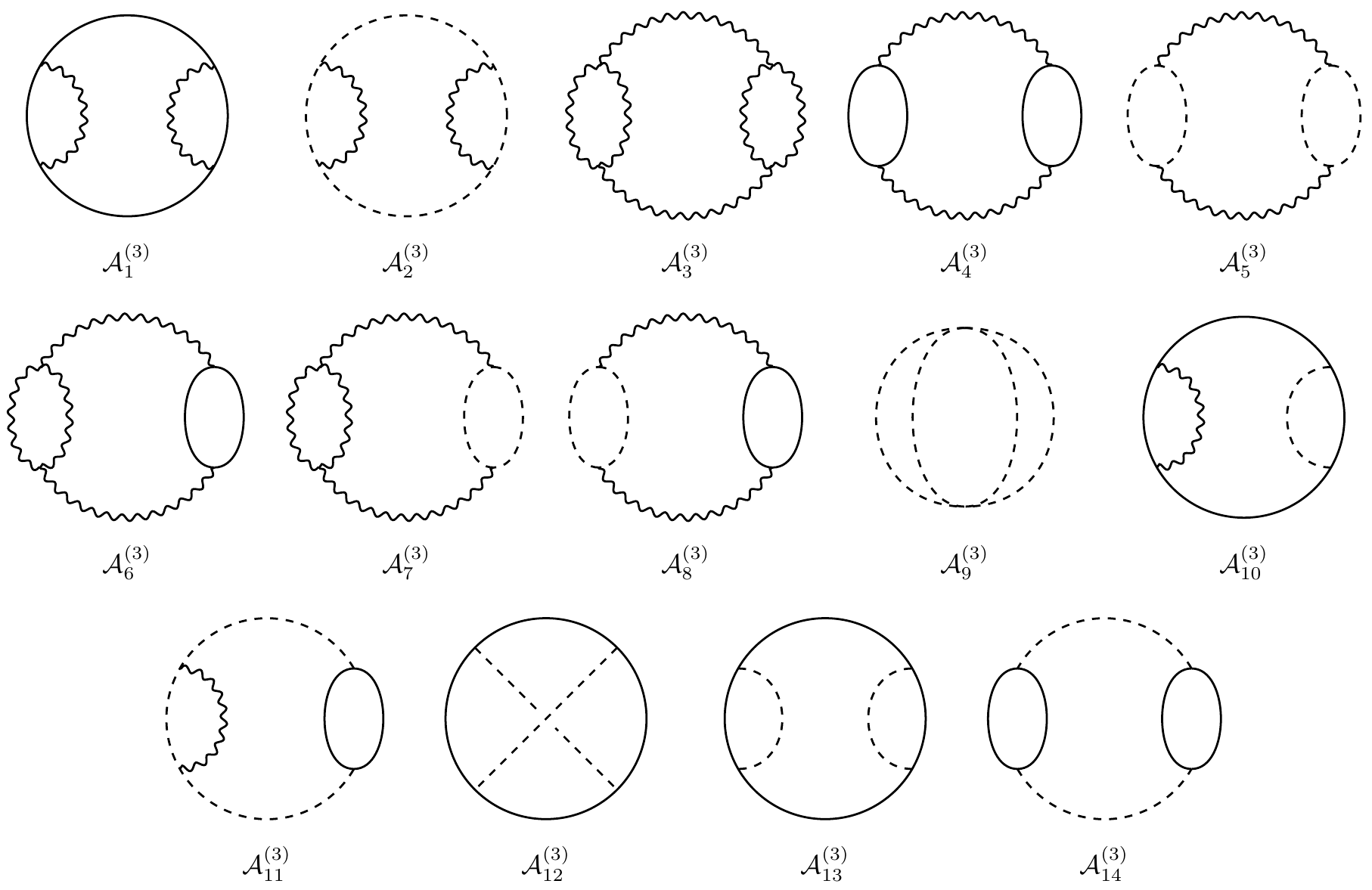}
	\caption{Graph representation of the 2- and 3-loop contributions to the $A$-function. Note that the scalar lines in $\mathcal{A}_{12}^{(3)}$ do not intersect.}
	\label{fig:a_3-loop}
\end{figure}

The first order at which $\tilde{A}$ is related to the $\beta$-functions is order 1--0--0, corresponding to the 2-loop $ A $-function. $ \tilde{A}^{(2)} $ contains 4 fully contracted TSs and thus 4 unknown coefficients shown in Fig. \ref{fig:a_3-loop}. In addition there is a single coefficient in $ T^{(1)}_{IJ} $ and 3 in $ \beta^{(1)}_{AB} $. \gfe reduces down to just 4 equations, and it would be impossible to extract information on $ \beta^{(1)}_{AB} $ without knowing terms in $T_{IJ}$ and/or $\tilde{A}$ (though one finds that $ \cofA{2}{4} = 0 $). This is in some sense due to the TSs of $ \tilde{A}^{(2)} $ being so simple that there are the same number of TSs in the basis before and after differentiation wrt. the couplings. 

The first non-trivial CC in a general theory occurs at loop order 2--1--0, or equivalently $ \tilde{A}^{(3)} $, and was in fact missed by \citet{Jack:2014pua}, as they substituted in all (scheme-independent) 1-loop coefficients. The main difference in our approach is that we have generalized the notation to allow for any compact gauge group. Nevertheless, at this order, the structure of the calculation is identical, hence we do not obtain any additional CCs beyond this one.

\begin{figure}
	\includegraphics[width=\textwidth]{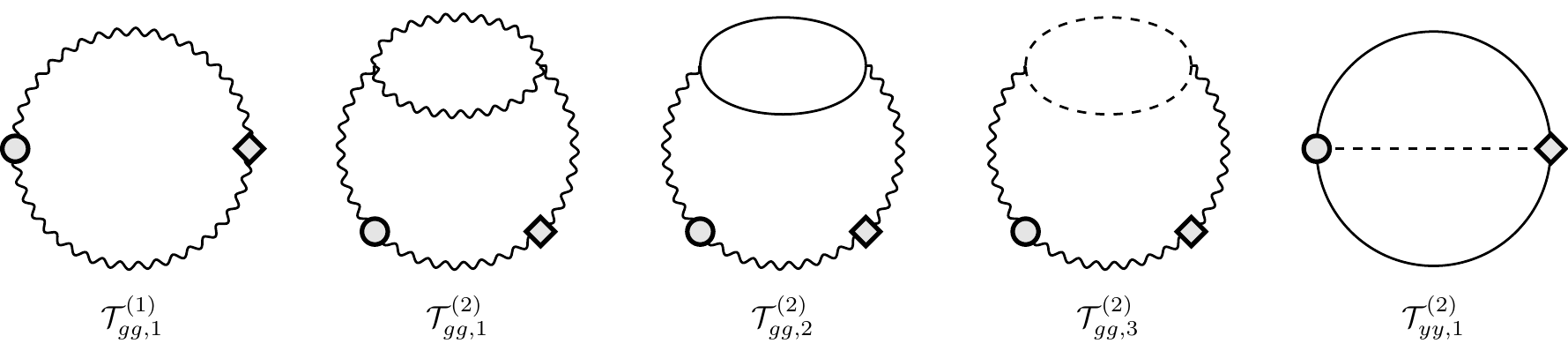}
	\caption{Graph representation of the 1- and 2-loop contributions to $ T_{IJ} $. A circle indicates an open index $ I $ and a diamond an index $ J $.}
	\label{fig:T_2-loop}
\end{figure}

At order 2--1--0 there are 14 new TSs in the $A$-function, illustrated in Fig. \ref{fig:a_3-loop}, and 4 new TSs in $ T_{IJ}^{(2)} $. The non-trivial constraints on the $\beta$-function coefficients stem from the terms in the $A$-function with coefficients $ \cofA{3}{10} $ and $ \cofA{3}{11} $. When differentiating the corresponding TSs wrt. the couplings to obtain the LHS of \gfe, the derivative can act either on a gauge or a Yukawa coupling, giving rise to two distinct TSs. The part of \gfe involving these terms read 
	\begin{align}
	\cofA{3}{10} &= \cofT{1}{gg,1} \cofg{2}{6}, & \cofA{3}{11} &= \cofT{1}{gg,1} \cofg{2}{7}, \nn 
	2 \cofA{3}{10} &= \cofT{2}{yy,1} \cofy{1}{2}, &	2 \cofA{3}{11} &= \cofT{2}{yy,1} \cofy{1}{1}.
	\end{align}
The leading-order coefficients of $ T_{IJ} $---in the sense that there are no contributions involving the respective coupling at lower order---were calculated in Ref. \cite{Jack:1990eb}, and are given by
	\begin{equation}
	\cofT{1}{gg,1} = \tfrac{1}{2}, \quad \cofT{2}{yy,1} = \tfrac{1}{6}, \quad \cofT{3}{\lambda\lambda,1} = \tfrac{1}{144}.
	\label{eq:Tcoeffs}
	\end{equation}
The knowledge that $ \cofT{1}{gg,1}, \cofT{2}{yy,1} \neq 0$ is sufficient to eliminate the $ T_{IJ} $ coefficients from the resulting system of equations, leaving the CC 
	\begin{equation}
	\cofg{2}{7} \cofy{1}{2} = \cofg{2}{6} \cofy{1}{1}.
	\end{equation}
This condition is indeed satisfied, as it is well known that $ \cofg{2}{7} = \cofy{1}{1} = 0 $. No other CCs occur at this order, since no other term in the $A$-function gives multiple unique TSs under coupling differentiation. By considering the contributions to $T_{IJ}$ listed in Fig.~\ref{fig:T_2-loop}, our graphical notation should also make it obvious that symmetry of $T_{IJ}$ at this order is in fact trivial, and hence imposes no extra CCs.

It is worth noting that this CC can be used in another way: if we \emph{did not know} $ \cofT{1}{gg,1}, \cofT{2}{yy,1} $, the same set of equations would imply that
	\begin{equation}
	\frac{2\cofT{1}{gg,1}}{\cofT{2}{yy,1}} = \frac{\cofy{1}{2}}{\cofg{2}{6}} = \frac{\cofy{1}{1}}{\cofg{2}{7}},
	\end{equation}
assuming that at least one of the ratios were well-defined. Since we know that 
	\begin{equation}
	\cofy{1}{2} = -6, \quad \cofg{2}{6} = -1,
	\end{equation}
Eq. \eqref{eq:GFE} would automatically imply the ratio
	\begin{equation}
	\cofT{2}{gg,1} = 3\cofT{1}{yy,1}
	\label{eq:ratio1}
	\end{equation}
between the leading-order coefficients, revealing a hitherto-unknown restriction on $T_{IJ}$. While the impact of this relation is again somewhat diminished by the explicit calculations, we hope it is clear that similar relations will exist at higher orders, where the corresponding coefficients are not known.

At the next loop-order we find that not only does the number of CCs increase dramatically, but so too does the ratio between the number of CCs and the number of $ \beta $-function coefficients, making the whole endeavour much more worthwhile.

\subsection{Consistency conditions at order 3--2--1} \label{sec:3--2--1}
Proceeding to \gfe at order 3--2--1, or $ \tilde{A}^{(4)} $, the list of possible TSs starts becoming overwhelming: the number of $A$-function tensors alone rises from 14 to 49 (down from 147 before eliminating redundant TSs with gauge identities), and most receive multiple contributions from the various $\beta$-functions and $T_{IJ}$ tensors at each loop order. All of these are shown in App. \ref{app:A4}, as we will not make any further use of them here. At this order we also introduce 71 new $ \beta $-function coefficients. The upshot of the increased complexity is that we obtain 26 new CCs, compared to the single CC at order 2--1--0.

At this loop-order, once one starts eliminating the coefficients of $ \tilde{A} $ and $ T_{IJ} $, the CCs in full generality start becoming rather involved. To keep things tractable and (almost) linear, we input the known coefficients at order 2--1--0\footnote{Recall that the gauge $\beta$-function is scheme-independent to two loops, and the Yukawa $\beta$-function to one loop, thus these conditions are still fully general, at least with respect to schemes preserving the structure of the \msbar $\beta$-function.}. Expressed this way, the CCs reflect only the new information gained at the highest loop order. We find 
\begin{subequations}\label{eq:321}
	\begin{align}
	& \cofg{3}{5} + \cofg{3}{8} & &= & &\tfrac{1}{2} \cofg{3}{6} + 2 \cofg{3}{7}\label{eq:X1}\\
	& 4 \cofg{3}{3} +44 \cofg{3}{5} + \cofg{3}{10} & &= & & 2\cofg{3}{4} + 22 \cofg{3}{6} + 2 \cofg{3}{9}\label{eq:X2}\\
	&\cofg{3}{23} && = && \tfrac{1}{3} \cofy{2}{1}\\
	&3 \cofg{3}{19} + \cofy{2}{3} && = && 3 \cofg{3}{20} + \cofy{2}{4}\\
	&22 \cofy{2}{6} + \cofy{2}{5} && =&& 3 \cofg{3}{25}\label{eq:X5}\\
	& \cofy{2}{7} && = && 2\cofy{2}{6} \label{eq:X6}\\
	& 3 \cofg{3}{3} + 66 \cofg{3}{7} + \cofy{2}{8} + 22 \cofy{2}{9} &&=&& \tfrac{3}{2} \cofg{3}{9} + 3 \cofg{3}{24} \label{eq:X7}\\
	& 3 \cofg{3}{5} + \cofy{2}{10} &&=&& 6 \cofg{3}{7} + 2 \cofy{2}{9}\label{eq:X8}\\
	& \cofg{3}{26} &&=&& \tfrac{1}{12} \cofy{2}{12}\\
	& \tfrac{1}{3} \cofy{2}{2} &&=&& 2 \cofg{3}{21} + 12 \cofg{3}{30} + \cofy{2}{13} \\
	& 2 \cofg{3}{22} + \cofy{2}{14} &&=&& 2 \cofg{3}{21} + 6 \cofg{3}{27} \\
	& \cofy{2}{15} + \tfrac{1}{3} \cofy{2}{2} &&=&&  2 \cofg{3}{21} + 6 \cofg{3}{27}\\
	& \cofg{3}{21} + 6 \cofg{3}{30} &&= && \cofg{3}{22} +  \cofy{2}{19}\\
	&12 \cofg{3}{31} + \cofy{2}{20} &&=&& 12 \cofg{3}{32} + \cofy{2}{21}\\
	& 2\cofg{3}{21} + 24 \cofg{3}{33} &&=&& \cofg{3}{22} + \tfrac{1}{6} \cofy{2}{2}+ 2 \cofy{2}{22}\\
	& 4 \cofg{3}{20} + 6 \cofg{3}{28} + \cofy{2}{16} + 3 \cofy{2}{26} &&=&& \tfrac{4}{3} \cofy{2}{3} +2 \cofy{2}{17} + \tfrac{3}{2} \cofy{2}{25}\label{eq:X16}\\
	& 2 \cofg{3}{20} + 2 \cofg{3}{21} +3 \cofg{3}{28} + 12 \cofg{3}{30} + 2 \cofy{2}{20} + 3 \cofy{2}{27} &&=&& \tfrac{1}{3} \cofy{2}{2} + \tfrac{2}{3} \cofy{2}{3} + \tfrac{1}{2} \cofy{2}{17}\label{eq:X17}\\
	& \cofy{2}{16} &&=&& \cofy{2}{17} + 4 \cofy{2}{18} + 12 \cofy{2}{28}\\
	& 2 \cofg{3}{20} + 2 \cofg{3}{21}+ 12 \cofg{3}{30} +12 \cofg{3}{32} + \tfrac{1}{2} \cofy{2}{16} + 3 \cofy{2}{29} &&=&& \tfrac{1}{3} \cofy{2}{2} + \tfrac{2}{3} \cofy{2}{3} + \tfrac{1}{2} \cofy{2}{17} \label{eq:X19}\\
	& \cofg{3}{20} + 2 \cofg{3}{21} + 12 \cofg{3}{30} + 6 \cofg{3}{32} + \cofy{2}{20} + 6\cofy{2}{30} &&=&& \tfrac{1}{3} \cofy{2}{2} + \tfrac{1}{3} \cofy{2}{3} \label{eq:X20}\\
	& \cofy{2}{17} + 4 \cofy{2}{18} + 24 \cofy{2}{31}  &&=&& \cofy{2}{16} + 6 \cofy{2}{29}\\
	& \tfrac{1}{3} \cofg{3}{20} + \cofg{3}{28} + \tfrac{1}{12} \cofy{2}{16} &&=&& 2 \cofg{3}{29} + \tfrac{1}{9} \cofy{2}{3} + \tfrac{1}{4} \cofy{2}{17} + \cofy{2}{32} \\
	& \cofg{3}{20} + \cofg{3}{21} + 6 \cofg{3}{29} + 6 \cofg{3}{30} + 6 \cofy{2}{33} + \cofy{2}{20} &&=&& \tfrac{1}{6} \cofy{2}{2} + \tfrac{1}{3} \cofy{2}{3} \\
	& \cofq{1}{2} \cofg{3}{17} &&=&& 4 \cofq{1}{1} \cofg{3}{18}\label{eq:X24}\\
	& 3 \cofq{1}{4} \cofg{3}{17} &&=&& 2 \cofq{1}{1} \cofy{2}{11}\label{eq:X25}\\
	& 6 \cofq{1}{5} \cofg{3}{17} &&=&& \cofq{1}{1} \cofy{2}{23}\label{eq:X26}.
	\end{align} 
\end{subequations}
As remarked previously, the construction at order 3--2--1 has been done for a single gauge group by \citet{Jack:2014pua}. Our construction for a completely general theory has produced an additional 7 CCs, including some relations between tensors with different group Casimirs and Dynkin indices, which was not resolved in that work. Since our general construction must reduce to the special case, we would expect that the conditions obtained by Jack and Poole can be expressed as linear combinations of our new conditions, and we have indeed verified that this is the case (see App. \ref{app:321} for further details).

We may now demonstrate the power of these CCs, by showing how one can obtain the complete 3-loop gauge $\beta$-function for a general theory. In the case where the gauge group is simple, there is a one-to-one correspondence between 29 of the TSs in our expression for $ \beta^{(3)}_{AB} $ and  the TSs described in Ref.~\cite{Pickering:2001aq}. As mentioned in the previous subsection, the last 4 of our TSs, with $ \cofg{3}{3,4,9,10} $, reduce to the two remaining TSs of Pickering \emph{et al.}, as the TSs coincide in this simple case. Direct comparison with their result then translates into the requirements that 
	\begin{equation}
	\cofg{3}{3} + \cofg{3}{9} = \dfrac{205}{18} \andeq \cofg{3}{4} + \cofg{3}{10} = \dfrac{1129}{36}.
	\end{equation}
Meanwhile, utilizing the known results for all the remaining coefficients at order 3--2--1, the CCs translate into just the two conditions (from Eqs. \eqref{eq:X2} and \eqref{eq:X7})
	\begin{equation}
	2\cofg{3}{3} -\cofg{3}{9} = \dfrac{97}{9} \andeq  2\cofg{3}{4} -\cofg{3}{10} = \dfrac{227}{9}.
	\end{equation}
These four constraints are sufficient to uniquely determine the four coefficients, which are found to be 
	\begin{equation}
	\cofg{3}{3} = \dfrac{133}{18}, \qquad \cofg{3}{4} = \dfrac{679}{36}, \qquad \cofg{3}{9} = 4, \andeq \cofg{3}{10} = \dfrac{25}{2}.  
	\end{equation}
With this result, the full gauge $ \beta $-function is known to 3-loop order (c.f. the appendix). It is thus possible to verify that with our result, the general $ \beta $-function can reproduce the explicit 3-loop computation for the SM\footnote{Incidentally, it was pointed out to us by Esben Mølgaard~\cite{Molgaard:xxx} that comparison with the SM provides an alternative route to extract the semi-simple result, which in turn is structurally identical to the compact gauge group result (at least in our notation).}~\cite{Bednyakov:2014pia,Herren:2017uxn}.

Seeing as the difficulty of the perturbative computation increase rapidly with increasing loop order, an interesting prospect is seeing how much information can be extracted for the $ \beta^{(3)}_g $ coefficients given only the coefficients of $\beta^{(1)}_{y}$, $ \beta^{(2)}_y $, and $ \beta^{(1)}_\lambda $. Between the first 16 coefficients we are left with just 3 CCs:
	\begin{gather}
	\cofg{3}{5} = \tfrac{2}{3} + 2 \cofg{3}{7}, \qquad \cofg{3}{6} = \tfrac{4}{3} + 2 \cofg{3}{8}, \nonumber\\
	2\cofg{3}{3} +22 \cofg{3}{5} + \tfrac{1}{2} \cofg{3}{10} = \cofg{3}{4} + 11 \cofg{3}{6} + \cofg{3}{9}.
	\label{eq:purebetag3}
	\end{gather}
These 16 coefficients are those that determine $ \beta^{(3)}_g $ in a general gauge theory without Yukawa or quartic interactions. By itself, \eqref{eq:purebetag3} is insufficient to make headway in any computation, and would have to serve as a cross check for a regular perturbative calculation. This is not a surprising result; knowing the Yukawa and quartic $ \beta $-functions does not provide much information about the $ \beta $-function of a purely gauge theory. However, the picture changes dramatically once we consider the remaining 17 coefficients of $ \beta^{(3)}_g $, all of which involve Yukawa or quartic couplings. We see that the CCs with the known results for $ \beta_y^{(2)} $ and $ \beta_\lambda^{(1)} $ predict the remaining coefficients up to an ambiguity involving just two parameters:
\begin{align}
  	\cofg{3}{17} =& 1
&	\cofg{3}{18} =& -\tfrac{1}{12} 
&	\cofg{3}{19} =& -\tfrac{1}{2} - 6x
&	\cofg{3}{20} =& \tfrac{1}{2} - 6x \nonumber
\\	\cofg{3}{21} =& -\tfrac{5}{2} - 12 y
&	\cofg{3}{22} =& - 7 
&	\cofg{3}{23} =& - \tfrac{7}{2}
&	\cofg{3}{25} =& \tfrac{9}{4} \nonumber
\\	\cofg{3}{26} =& 1 
&	\cofg{3}{27} =& -\tfrac{1}{2} + 4 y 
&	\cofg{3}{28} =& 1 + 4x
&	\cofg{3}{29} =& \tfrac{3}{4} + x \nonumber
\\	\cofg{3}{30} =& \tfrac{3}{4} +2y
&	\cofg{3}{31} =& x 
&	\cofg{3}{32} =& \tfrac{1}{4} + x
&	\cofg{3}{33} =& y	\nonumber
\\ &&\mathclap{\cofg{3}{24} = -\tfrac{14}{3} + \cofg{3}{3}	+ 11 \cofg{3}{5} - \tfrac{1}{2} \cofg{3}{9}.} 
\end{align}  
Amongst these, $ \cofg{3}{24} $ stands out like a sore thumb, but it is predicted in terms of the purely gauge terms. What all of this shows is that Weyl CCs may be used to postdict the 3-loop $ \beta $-function in a general theory, if one knows the following:
	\begin{itemize} \setlength\itemsep{0.2em}
	\item[\emph{i})] The coefficients of $ \beta_{AB}^{(2)} $, $ \beta_{aij}^{(2)} $, and $ \beta_{abcd}^{(1)} $ 
	\item[\emph{ii})] The part of $ \beta^{(3)}_{AB} $ that does not involve Yukawa or quartic interactions
	\item[\emph{iii})] The coefficient of two of the remaining tensors, e.g. $ \cofg{3}{31} $ and $ \cofg{3}{33} $. 
	\end{itemize}
Having achieved such a large amount of additional information on the 3-loop gauge $\beta$-function, our hope is that constructing the $A$-function at the next order and repeating the approach will be sufficient to infer the 4-loop gauge $\beta$-function. As shown by the above analysis, this would almost certainly require the full 3-loop Yukawa $\beta$-function, and even then the purely-gauge terms are most likely under-determined. However, various purely-gauge calculations (such as $\SU(N)$ Yang-Mills with matter) have already been done at 4 loops \cite{vanRitbergen:1997va}, so it is plausible that the general 4-loop result can be almost completely determined by this approach. 

Finally, we shall comment on an augmentation of Eq.~\eqref{eq:ratio1} that follows from a similar subset of equations involving $\tilde{A}^{(4)}$. If one considers the $A$-function tensors with coefficients $\cofA{4}{32}$ and $\cofA{4}{41}$ (see App. \ref{app:A4}), one finds
	\begin{align}
		2 \cofA{4}{32} &= \cofT{2}{yy,1} \cofy{2}{11}, & 4 \cofA{4}{41} &= \cofT{2}{yy,1} \cofy{2}{23}, \nn
		2 \cofA{4}{32} &= \cofT{3}{\lambda\lambda,1} \cofq{1}{4}, & \cofA{4}{41} &= \cofT{3}{\lambda\lambda,1} \cofq{1}{5}.
	\end{align}
implying that
	\begin{equation}
	\frac{\cofT{2}{yy,1}}{\cofT{3}{\lambda\lambda,1}} = \frac{\cofq{1}{4}}{\cofy{2}{11}} = \frac{4\cofq{1}{5}}{\cofy{2}{23}}.
	\end{equation}
As with \eqref{eq:ratio1}, this can be used in two ways: eliminating the $T_{IJ}$ coefficients gives a linear combination of CCs \eqref{eq:X25} and \eqref{eq:X26}; alternatively, using the known $\beta$-function coefficients produces another (scheme-independent) ratio between $\cofT{2}{yy,1}$ and $\cofT{3}{\lambda\lambda,1}$. Consequently, combining this result with \eqref{eq:ratio1}, we find that \emph{all three} leading-order contributions to $T_{IJ}$ for a general theory must be proportional:
	\begin{equation}
	\cofT{1}{gg,1} = 3\cofT{2}{yy,1} = 72\cofT{3}{\lambda\lambda,1}.
	\label{eq:ratio2}
	\end{equation}
This relation is indeed satisfied by the known results, c.f. Eq.~\eqref{eq:Tcoeffs}. Again, despite being somewhat redundant due to the existence of explicit calculations, Eq.~\eqref{eq:ratio2} is a powerful, hitherto-unknown restriction on the allowed RG flow of a completely general QFT: the mere existence of \gfe automatically implies that leading-order positive-definiteness of $T_{IJ}$ (and thus a perturbative proof of the $a$-theorem) actually follows from leading-order positivity of $T_{IJ}$ for a pure-gauge ($\cofT{1}{gg,1}$), pure-Yukawa ($\cofT{2}{yy,1}$), or pure-scalar ($\cofT{3}{\lambda\lambda,1}$) theory.

At this order, we have not yet explicitly mentioned the conjecture regarding symmetry of $T_{IJ}$, nor any possible contributions to $S$. The new TSs appearing in $T_{IJ}$ at this order are all given in App \ref{app:A4}---it is again clear that all diagonal contributions are manifestly symmetric, thus we need only investigate the new mixed terms. Our conjecture requires that we deliberately set
	\begin{equation}
	\cofT{3}{gy,1} = \cofT{3}{yg,1} \quad\quad\text{and}\quad\quad \cofT{3}{gy,2} = \cofT{3}{yg,2}
	\end{equation}
then re-derive the CCs. As it happens, no additional CCs are obtained, and so imposing symmetry of $T_{IJ}$ is still trivial, at least in this sense. This clarifies earlier assertions made in two papers: in \cite{Jack:2013sha}, the authors speculated on the potential for imposing such symmetry, concluding that ``this need not be true in general renormalization schemes"; in \cite{Jack:2014pua}, the authors mistakenly specify values for two $T_{IJ}$ coefficients, as a consequence of using explicit \msbar coefficient for the rest of the computation.

\subsection{Treatment of $ \gamma_5 $} \label{sec:gamma5}
So far we have ignored  possible contributions to the $ \beta $-functions from Feynman diagrams with $ \gamma_5 $ insertions. To obtain the full $ \beta $-function at order 4--3--2 one must include such contributions, and their corresponding TSs. In dimensional regularization the treatment of $ \gamma_5 $, or equivalently $ \epsilon_{\mu\nu\rho\sigma} $, is ambiguous\footnote{This ambiguity has been known since the original 't Hooft--Veltman paper \cite{tHooft:1972tcz}. A full analysis of the difficulties encountered when attempting to treat $\gamma_{5}$ is given in \cite{Jegerlehner:2000dz}.}, leading to an undetermined parameter in the 4-loop $ \beta $-function of the SM strong coupling~\cite{Bednyakov:2015ooa,Zoller:2015tha}. This ambiguity is not present in the 3-loop Yukawa $ \beta $-function, where a ``semi-na\"ive" treatment of $\epsilon_{\mu\nu\rho\sigma}$ is sufficient \cite{Chetyrkin:2012rz,Bednyakov:2012en,Herren:2017uxn}. As we previously pointed out~\cite{Poole:2019txl}, the use of Weyl CCs relates the coefficients of the ambiguous terms in the 4-loop gauge $ \beta $-function to those of the unambiguous (and known) terms in the Yukawa $ \beta $-function, uniquely settling the treatment of $ \gamma_5 $ in the corresponding 4-loop diagrams. 

Contributions from Feynman diagrams with non-vanishing $ \gamma_5 $ contributions are treated separately in our framework, because $ \gamma_5 $ contributions in a fermion loop give opposite signs to each fermion chirality. TSs constructed just from $ y_a $ and $ T^{A} $ give same-sign contributions to the two chiralities by construction. Thus, we include $ \sigma_3 $ on fermion lines at the level of TSs to account for the sign difference. The introduction of $ \sigma_3 $ on the fermion lines introduces the need for a specific interpretation of the use of tensors with or without tilde, see Eq. \eqref{eq:tilde_quantities}. We use the graph notation 
	\begin{equation}
	\vcenter{\hbox{\includegraphics{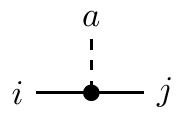}}} = (\sigma_3 y_{a})_{ij} \quad \mathrm{or} \quad  (\tilde{\sigma}_3 \tilde{y}_{a})_{ij},
	\end{equation} 
that is, a blob on a Yukawa vertex corresponds to a $ \sigma_3 $ on the fermion line. One should use $ \sigma_3 $ if the Yukawa is taken to be $ y_a $ and $ \tilde{\sigma}_3  = \sigma_1 \sigma_3 \sigma_1 = - \sigma_3 $  if the Yukawa is taken to be $ \tilde{y}_a $. In this way we preserve the symmetry of the fermion lines under replacements
	\begin{equation}
	(y_a,\, T^{A},\, \sigma_3) \quad \leftrightarrow \quad (\tilde{y}_a,\, \tilde{T}^{A},\, \tilde{\sigma}_3).
	\end{equation}

We should not blindly go ahead and try to build all possible TSs with $ \sigma_3 $ on the fermion lines, as there are only a few select Feynman diagrams that can give non-vanishing $ \gamma_5 $ contributions. For a fermion loop to give rise to an $ \epsilon_{\mu\nu\rho\sigma} $ contribution in the Feynman diagram there needs to be at least four independent Lorentz indices on the loop, which can either come from momenta or from Lorentz indices on the gauge lines. Denoting by $ (m,n) $ a fermion line with $ m $ gauge insertions and $ n $ Yukawa insertions, then for a line to contribute a Levi-Cevita symbol it must be of type at least $ (3,0) $, $ (2,1) $, $ (1,3) $, or $ (0, 5) $. Furthermore, the Feynman diagrams must have at least two distinct fermion lines contributing an $ \epsilon_{\mu\nu\rho\sigma} $ for the contribution to be non-vanishing, so there must be two such lines (recall that loops must have an even number of Yukawa couplings). First we note that lines of type $ (1,3) $ and $ (0, 5) $ cannot appear at the present loop-order. Additionally, a Feynman diagram with a line of type $ (3,0) $ leads to a $ \gamma_5 $ contribution containing
	\begin{equation}
	\Tr{\sigma_3 T^{A} T^{B} T^{C}} = \Tr{T_\psi^{A} \braces{T_\psi^{B}, T_\psi^{C}}} = 0.
	\end{equation}
Such terms vanish in sound gauge theories that are anomaly free; a similar argument eliminates diagrams with a $ (4,0) $ line. This leaves only TSs from $ \gamma_5 $ contributions stemming from diagrams with 2 fermion lines of type $ (2,1) $ or higher. 

\begin{figure}
	\centering
	\includegraphics[width=\textwidth]{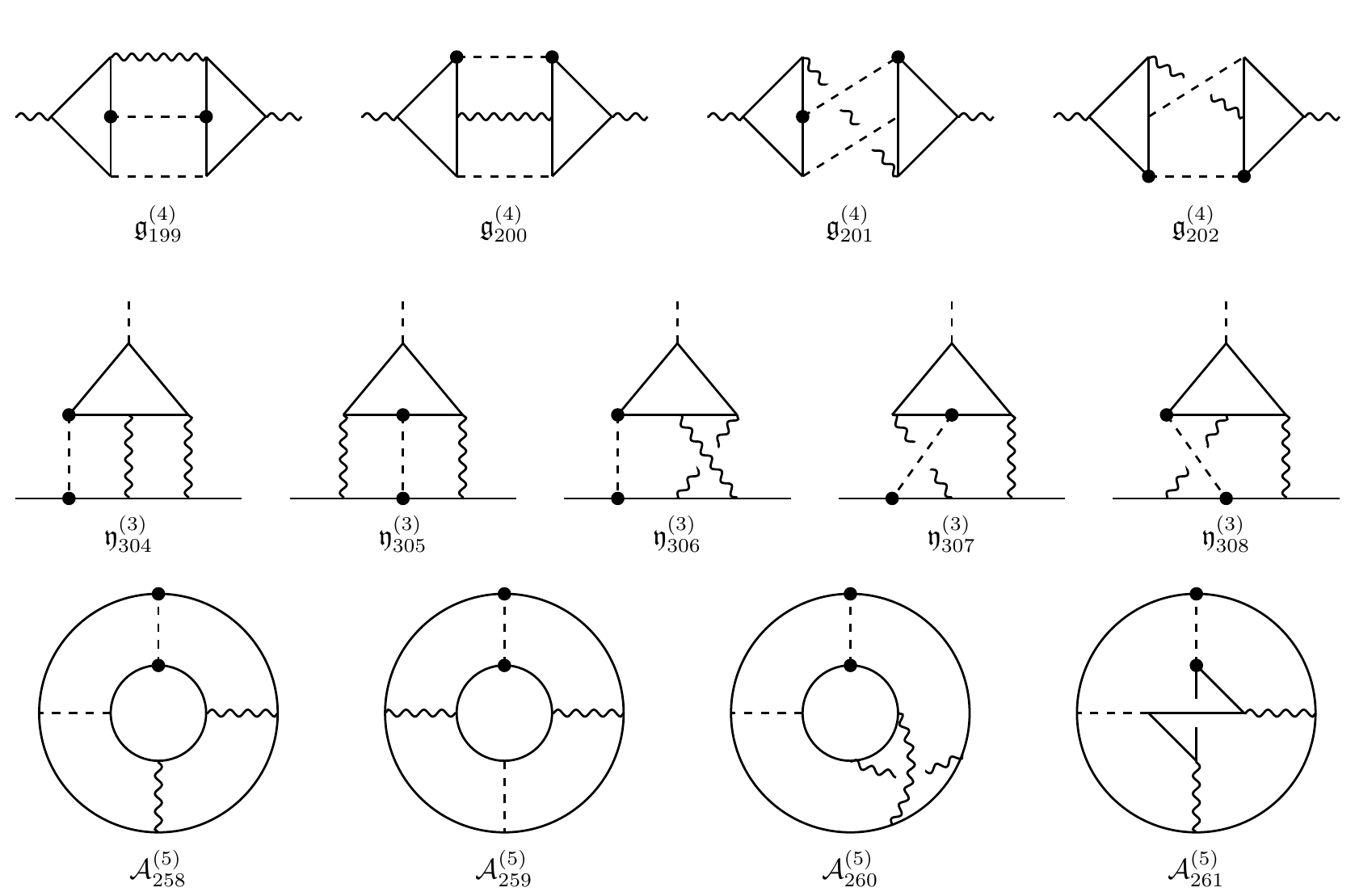}
	\caption{The first two rows contains the TSs from non-vanishing $ \gamma_5 $ contributions to $ \beta^{(4)}_{AB} $ and $ \beta^{(3)}_{aij} $, while the last row list the 4 corresponding TSs in $ \tilde{A}^{(5)} $. The blobs on the fermion lines symbolize $ \sigma_3 $ insertions.}
	\label{fig:gamma5_beta}
\end{figure}

From our considerations we arrive at just 4 kinds of Feynman diagrams with $ \gamma_5 $ insertions that can contribute to $ \beta^{(4)}_{AB} $, and 5 that can contribute to $ \beta^{(3)}_{aij} $. In neither case do gauge identities lead to a redundancy between the TSs resulting frome these diagrams. Though it is possible to rewrite the basis of $ \gamma_5 $-relevant TSs to one with more simple sub-tensors (such as 2-point functions and Casimirs), we have deviated slightly from our usual modus operandi here, and have kept the basis directly corresponding to the relevant Feynman diagrams\footnote{A $ \sigma_3 $ on a fermion line does not influence the gauge identities, as $ [\sigma_3, y_a] = [\sigma_3, T^{A}] = 0 $.}. In this basis, the correspondence between each Feynman diagram and its TS becomes much more explicit. The extra contributions to the $ \beta $-functions are shown in Fig. \ref{fig:gamma5_beta}.

Since our chosen basis for $\gamma_{5}$ terms involves only primitive graphs, we may use the topological argument indicated in \cite{Poole:2019txl}, which guarantees that the only relevant contributions to \gfe come from inserting these TSs into the leading-order $T_{IJ}$. Alternatively one can argue that no $ \sigma_3 $ terms can enter in the RHS of the \gfe without it either entering in in $ T_{IJ} $ or $ B^{I} $. No $T_{IJ}$ terms at present order can include $ \gamma_5 $ contributions, so we have identified all relevant terms below. Thus, the $ \gamma_5 $--relevant TSs with $ \sigma_3 $ on the fermion lines separate neatly into their own subsystem of \gfe. There are four new terms in the $ A $-function, shown in Fig. \ref{fig:gamma5_beta}. Plugging the new $ A $-function and $ \beta $-function terms into \gfe yields 9 equations, corresponding to the 9 distinct derivatives of $\cofA{5}{258-261}$:
	\begin{align}
	2 \cofA{5}{258} & = T^{(1)}_{gg,1} \cofg{4}{199}, & 2 \cofA{5}{259} & = T^{(1)}_{gg,1} \cofg{4}{200}, & 2 \cofA{5}{260} & = T^{(1)}_{gg,1} \cofg{4}{201}, & 2 \cofA{5}{261} & = T^{(1)}_{gg,1} \cofg{4}{202}, \nn
	4 \cofA{5}{258} &= T^{(2)}_{yy,1} \cofy{3}{304}, & 4 \cofA{5}{259} &= T^{(2)}_{yy,1} \cofy{3}{305}, &  4 \cofA{5}{260} &= T^{(2)}_{yy,1} \cofy{3}{306}, &  2 \cofA{5}{261} &= T^{(2)}_{yy,1} \cofy{3}{307}, \\
	&& && && 2 \cofA{5}{261} &= T^{(2)}_{yy,1} \cofy{3}{308}. \nonumber  
	\end{align}
Eliminating the coefficients $ \cofA{5}{i} $ from the system, yields
	\begin{equation}
	\begin{gathered}
	\cofg{4}{199} = \dfrac{\cofT{2}{yy,1}}{2\cofT{1}{gg,1}}\cofy{3}{304}, \qquad 	
	\cofg{4}{200} = \dfrac{\cofT{2}{yy,1}}{2\cofT{1}{gg,1}}\cofy{3}{305}, \qquad
	\cofg{4}{201} = \dfrac{\cofT{2}{yy,1}}{2\cofT{1}{gg,1}}\cofy{3}{306},\\
	\cofg{4}{202} = \dfrac{\cofT{2}{yy,1}}{\cofT{1}{gg,1}}\cofy{3}{307}, \andeq
	\cofy{3}{307} = \cofy{3}{308}.
	\end{gathered}
	\end{equation}
and since the ratio between $\cofT{1}{gg,1}$ and $\cofT{2}{yy,1}$ is fixed by \eqref{eq:ratio2}, we find\footnote{Through direct comparison with the SM results of \cite{Herren:2017uxn}, we extracted explicit values of the Yukawa coefficients:
	\[
	\cofy{3}{304} = -24,\quad \cofy{3}{305} = -12, \quad \cofy{3}{306} = \cofy{3}{307} = \cofy{3}{308} = 8 -24\zeta_3.
	\]
}
	\begin{equation}
	\cofg{4}{199} = \dfrac{1}{6}\cofy{3}{304}, \quad 	
	\cofg{4}{200} = \dfrac{1}{6}\cofy{3}{305}, \quad
	\cofg{4}{201} = \dfrac{1}{6}\cofy{3}{306}, \quad
	\cofg{4}{202} = \dfrac{1}{3}\cofy{3}{307}, \quad
	\cofy{3}{307} = \cofy{3}{308}.
	\end{equation}
These conditions are what allowed us to uniquely determine the $ \gamma_5 $--relevant terms in $ \beta^{(4)}_{AB} $ in Ref. \cite{Poole:2019txl}.

\subsection{Consistency conditions at order 4--3--2}
We will now move on to discuss, for the first time, the Weyl Consistency Conditions for the $ \beta $-functions at loop order 4--3--2, or $\tilde{A}^{(5)}$. A summary of our results are shown in Table \ref{tab:results}. 

\begin{table} 
	\centering
	\begin{tabularx}{.85\textwidth}{|c| Y Y Y | Y Y Y | c c |}
		\hline \hline
		& \multicolumn{6}{c|}{No. of coefficients} & & \\ %\cline{2-7}
		$ \ell $ & $ \tilde{A}^{(\ell+1)} $ & $ T_{IJ}^{(\ell)} $ & $ S^{(\ell-1)} $ & $ \beta^{(\ell)}_{AB} $ & $ \beta^{(\ell-1)}_{aij} $ & $ \beta^{(\ell-2)}_{abcd} $ & TS basis & CCs \\ \hline
		1	& 4		& 1		&	& 3 	& 		&	& 4		&		\\
		2	& 14	& 4		&	& 7		& 5		&	& 16	& 1		\\
		3	& 49	& 27	&	& 33	& 33	& 5	& 91	& 26	\\
		4	& 257	& 260	& 9	& 198	& 303	& 33& 703	& 265	\\ 
		4 ($ \gamma_5 $) & 4 & & & 4 & 5 & & 9 & 5 \\
		\hline \hline
	\end{tabularx}
	\caption{The table gives an overview of the number of coefficients appearing at each loop-order, $ \ell $, in \gfe. It also includes the number of TSs in the basis of the LHS after differentiation; this is the number of individual equations relating all coefficients. Finally we list the number of consistency conditions relating the coefficients of the $ \beta $-functions at each order after eliminating the coefficients of $ \tilde{A} $, $ T $, and $ S $. The last line lists the number of new tensors occurring at the highest order when we allow for non-trivial contributions from $ \gamma_5 $.}
	\label{tab:results}
\end{table}

We have derived the CCs at order 4--3--2 using the \grafer code that we have developed for this purpose. This naturally begs the question of validation of the program to ensure the validity of the results. As described in the previous section, we have used the \grafer routines to derive the CCs at order 3--2--1, where all $\beta$-function coefficients are known. In that case all the conditions are fulfilled, providing a strong check. As a further point regarding validation, we will point to the fact that all the CCs we have recovered at order 4--3--2 are internally consistent. This would not a priori be the case if there were to be any errors in the program implementation. With the large number of overlapping equations, this is not a trivial check.

\begin{figure}
	\centering
	\includegraphics[width=\textwidth]{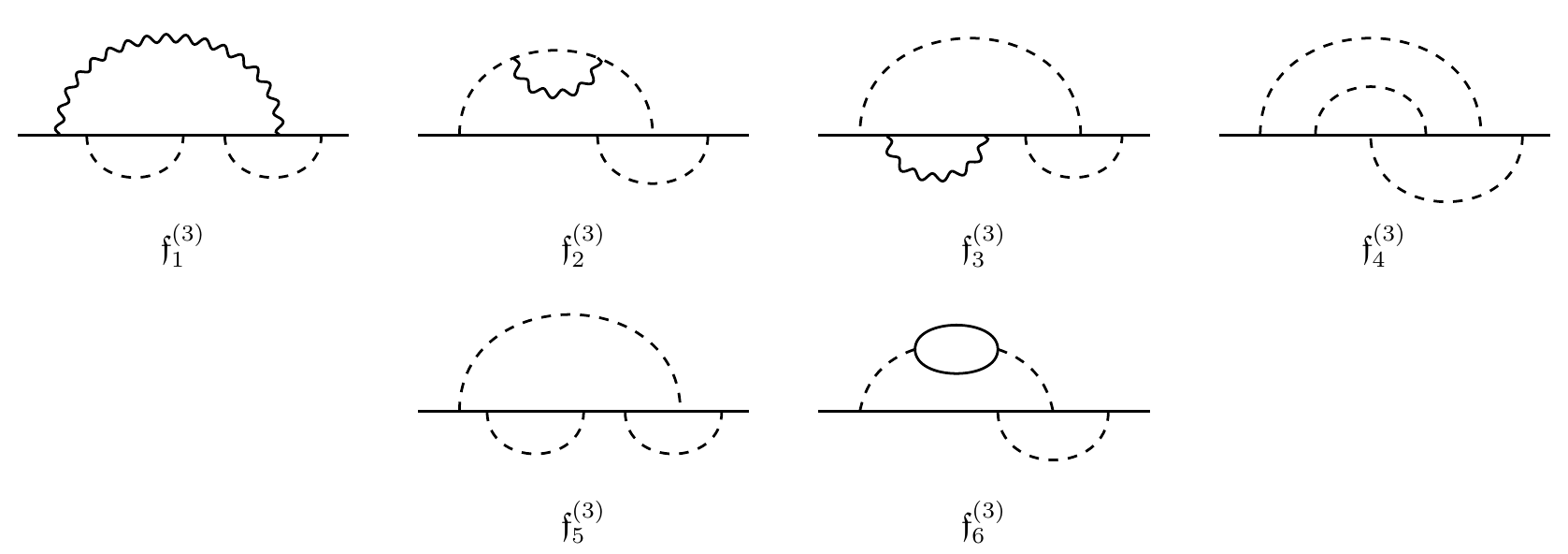}
	\includegraphics[width=\textwidth]{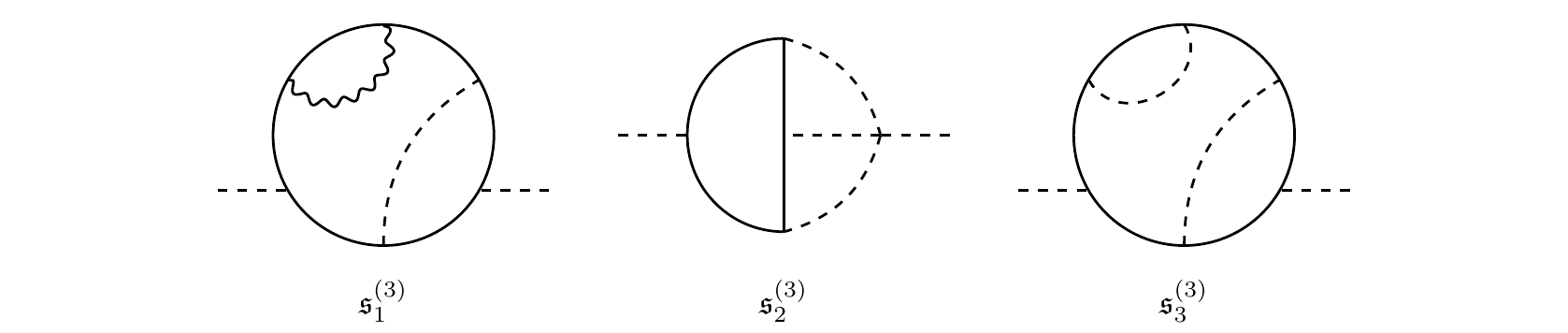}
	\caption{Graph representation of the TSs appearing in $ S^{(3)}_{ij} $ and $ S^{(3)}_{ab} $ respectively.}
	\label{fig:sFunction}
\end{figure}

Order 4--3--2 is the first occurrence of 2-point TSs for scalar and fermion lines that are not inherently symmetric. There are a total of 6 such structures with fermion indices, $ S^{(3)}_{ij} $, and 3 with scalar indices $ S^{(3)}_{ab} $. As discussed in Sec. \ref{sec:LRG}, these must be included in $ B^{I} = \beta^{I} -(Sg)^{I} $, otherwise the system of equations will be inconsistent. These extra terms are included in the 3-loop Yukawa $ \beta $-function, by setting 
	\begin{equation}
	B^{(3)}_{aij} = \beta^{(3)}_{aij} - 2S^{(3)}_{ik} \, y_{akj} - S^{(3)}_{ab} \,y_{bij}.
	\end{equation}
No changes are needed for the quartic $ \beta $-function, as this only enters at two loops. The $ S $ tensor is parametrized by 
	\begin{equation}
	\begin{split}
	S^{(\ell)}_{ij} &= \sum_{n} \cofsf{\ell}{n}\!\! \left(\vcenter{\hbox{\includegraphics{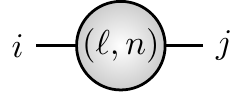}}} - (i\leftrightarrow j) \right), \\
	S^{(\ell)}_{ab} &= \sum_{n} \cofss{\ell}{n}\!\! \left(\vcenter{\hbox{\includegraphics{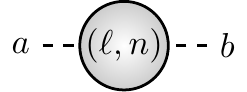}}} - (a \leftrightarrow b) \right),
	\end{split}
	\end{equation}
and the individual TSs are given in Fig. \ref{fig:sFunction}. Eliminating $ S $ from \gfe predict all the $ \cofss{3}{1-3} $ and $ \cofsf{3}{1,2} $ in terms of the standard $\beta$-function coefficients. Between the remaining four terms, $ \cofsf{3}{3-6} $, there is one undetermined parameter\footnote{Explicit results for the $ \cofsf{3}{} $ and $ \cofss{3}{} $ coefficients are included in the ancillary file.}. 

As the reader may already have seen in Table \ref{tab:results}, the number of TSs at order 4--3--2 increase dramatically compared to previous orders, as does the number of CCs that can be inferred from \gfe. For this reason we find ourselves unable to present the full result in the present paper. For interested parties we have instead included our findings in an ancillary file to the arXiv version of this paper. This includes a full list of diagrams for the TSs appearing at 4--3--2 and a comprehensive list of all the CCs; here, we will content ourselves with presenting a few highlights. Once again we substitute the known values of the \msbar $ \beta $-function coefficients at order 3--2--1 into the system of equations, before eliminating the coefficients of $ T_{IJ} $ and the $ A $-function. Without this step, even Mathematica struggles with what would be a system of 814 non-linear equations.

Four of the CCs involve only the $ \cofq{2}{} $ coefficients, which are known, and read 
	\begin{align}
	\cofq{2}{4} &= \dfrac{1}{2} \cofq{2}{5} - 12, & \cofq{2}{11} &= \dfrac{1}{2} \cofq{2}{12} +2, & \cofq{2}{21} &= -6 \cofq{2}{30} - 8, & \cofq{2}{29} &= 2\cofq{2}{30} + 4.
	\end{align}
We see that they are indeed satisfied by the coefficients of App. \ref{app:coefficients}, providing one more check of the implementation of the routines in \grafer. 
Turning now to the unknown coefficients, it turns out that 4 of the 3-loop Yukawa coefficients are predicted outright: 
	\begin{align}
	\cofy{3}{2} &= -3, & \cofy{3}{4} &= -\dfrac{3}{2}, & \cofy{3}{47} &= \dfrac{3}{4}, & \cofy{3}{49} &= 0.
	\end{align}
If we take the additional step of using the long-known results for the 2-loop quartic $ \beta $-function, we get an additional 10 predicted coefficients, including 4-loop gauge coefficients:
	\begin{align}
	\cofg{4}{55} &= -\dfrac{27}{2}, & \cofg{4}{60} &= \dfrac{1}{2}, & \cofg{4}{67} &= -\dfrac{1}{12}, & \cofg{4}{110} &= -4, & \cofg{4}{158} &=0,\nn
	\cofy{3}{39} &= -12, & \cofy{3}{122} &= 0, & \cofy{3}{127} &= 0, & \cofy{3}{226} &= 4, & \cofy{3}{228} &=2. 
	\end{align}
These provide some easy predictions right out of the box, but by themselves the 14 now-known coefficients are too few to do much more than check eventual direct loop computations. From a more practical point of view there are 133 CCs involving only the $ \cofy{3}{} $ coefficients (after substituting in the known $ \cofq{2}{} $). Conceivably this might be enough to extend the explicit 3-loop Yukawa computations for the SM and 2HDM~\cite{Bednyakov:2014pia,Herren:2017uxn} to the general case. Just as the 2-loop Yukawa $ \beta $-function was computed before the 3-loop gauge $ \beta $-function, it stands to reason that it is easier to obtain the 3-loop Yukawa than the 4-loop gauge; once the 3-loop Yukawa is known, we may substitute the $ \cofy{3}{} $ coefficients into the CCs and obtain 128 CCs for the 4-loop gauge $ \beta $-function coefficients. Given our success at three loops, we therefore expect that these CCs will prove a valuable asset to the general 4-loop computation. Indeed, as discussed in the last section, the corresponding CCs for TSs from Feynman diagrams with non-trivial contributions from $ \gamma_5 $ are enough to resolve the corresponding ambiguity in the computation of the 4-loop gauge $ \beta $-function. 

We now return to the conjecture that $T_{IJ}$ can be symmetrized (see Sec. \ref{sec:sym}). Imposing symmetry on $ T_{IJ} $ up to terms relevant for $A^{(5)}$, we have re-derived the Weyl CCs. Whereas symmetry of $ T_{IJ} $ did not provide any new CCs at order 3--2--1, it turns out that this is not the case at 4--3--2: we find an additional 10 new CCs that appear only after enforcing symmetry. Of particular interest is that these extra conditions are enough to give solutions for 3 of the coefficients in $ \beta^{(2)}_{abcd} $. We find 
	\begin{equation}
	\cofq{2}{21} = 10, \qquad \cofq{2}{29} = -2, \andeq \cofq{2}{30} = -3,
	\label{eq:sympred}
	\end{equation}  
whereas without symmetrization there was one unknown parameter parameterizing all three coefficients. Since only the 2-loop quartic $ \beta $-function is known to this order, these are the only predictions we can check. Indeed we find agreement between the predictions found with the symmetrized $ T_{IJ} $, and the \msbar result (see App. \ref{app:coefficients}), providing a hint that $ T_{IJ} $ can be made symmetric.

\section{Conclusions and outlook} \label{sec:conclusion}
In this paper, we have attempted to provide a comprehensive analysis of the constraints imposed by Weyl Consistency Conditions on the $\beta$-functions of a completely general, renormalizable, four-dimensional QFT. These constraints are a direct consequence of \gfe: their existence was noted as early as \cite{Osborn:1989td,Jack:1990eb}, yet the first investigations of these constraints for their own sake did not occur for over two decades \cite{Jack:2013sha,Antipin:2013sga}. Furthermore, despite steady increases in generality (in particular, higher loops and QFTs in other spacetime dimensions), a full derivation of the constraints on QFTs of phenomenological relevance had still not been done. Quite frankly, we find this astonishing, as the work done in non-phenomenological cases has led to multiple, non-trivial implications\footnote{This list is not exhaustive---as mentioned earlier, it has also been shown that CCs imply the existence of 1PR contributions to the $\beta$-functions in commonly-used non-minimal schemes such as momentum subtraction, despite explicit calculations giving zero for the coefficients of such terms. The resolution of this paradox relies on absorbing such terms into the antisymmetric part of the anomalous dimension \cite{Jack:2016tpp}, but the effect of this on \gfe (and in particular the antisymmetric tensor $S$) is still unknown.}:
\begin{itemize}
\item Due to the CCs relating $\beta$-function coefficients at different loop-orders, \emph{i.e.} the 3--2--1 ordering, one can use lower-order coefficients to predict higher-order coefficients;
\item Using coupling-redefinitions to parametrize perturbative renormalization schemes, the scheme-independence of all CCs can be directly shown, hence they form a rare example of physical information \emph{along} RG flows;
\item The antisymmetric tensor $S$, parameterizing the extension $\beta\rightarrow B$ of \gfe in the presence of relevant operators, can in fact be related to terms in the standard $\beta$-functions using CCs;
\item By parameterizing the contribution of an unknown Feynman integral to the $\beta$-functions, one can use CCs to determine its value, assuming that sufficiently-many other coefficients are already known.
\end{itemize}
The first and last implications are of most obvious interest to phenomenologists, as they provide a potential alternative to directly computing tens of millions of Feynman diagrams, or master integrals of ever-increasing complexity.

A derivation of the CCs for a four-dimensional QFT with a simple gauge group was pursued in \cite{Jack:2014pua}, albeit only up to order 3--2--1, and the relations were verified in \msbar using known $\beta$-function coefficients. Consequently, the only requirement to extend this analysis to a general gauge theory was a suitable notation for multiple gauge groups, in line with the tensor-coupling notation used to represent the Yukawa and quartic interactions. We have developed such a notation (both tensorial and graphical), which treats the general, unified gauge $\beta$-function $\beta_{g} \equiv \beta_{AB}$ as a rank-two tensors in coupling-space, with the corresponding gauge couplings, group Casimirs and Dynkin indices arranged into a (block-)diagonal form. 
With identities from gauge invariance (and, of course, the Jacobi identity for gauge theories), the possible tensors that depend on the gauge coupling in each $\beta$-function are reduced to a basis,
and we have attempted to further simplify the tensors by maximizing the number of Casimirs and Dynkin indices, in line with the preferences of the community. Essentially, our notation is a formalized and generally applicable version of the substitution rules described by Machacek and Vaughn~\cite{Machacek:1983tz}, capable of including multiple $U(1)$ groups via a coupling-matrix. That being said, the notation is highly efficient, and indeed allows us to derive CCs for a completely general four-dimensional QFT in direct analogy with the Yukawa and quartic notation employed in \cite{Jack:2014pua}; furthermore, the 3--2--1 ordering is manifestly a simple topological consequence of attempting to form scalars of a given loop order out of two-, three-, and four-index tensors, and is thus trivially preserved at higher loops.

Having established the notation, we developed our custom code, \grafer, in order to automate the derivation of CCs. Beginning at order 2--1--0, we have shown how the results for a single gauge group extend trivially to the general case with a compact gauge group, and derived two results that have so far been missed in previous analyses: a CC involving 2-loop gauge and 1-loop Yukawa coefficients, where the calculated values happen to be 0; and a simple linear relation between leading-order $T_{IJ}$ coefficients, which is indeed satisfied by the known values \cite{Jack:1990eb}. Next, we generated the results at order 3--2--1, again extending the results to arbitrary gauge groups and resolving the group structures. As a cross-check of \grafer's output, we have applied various checks:
\begin{itemize}
\item The CCs obtained are self consistent;
\item The conditions are satisfied by all known \msbar coefficients for a general theory;
\item The 19 old CCs for a theory with a simple gauge group are linear combinations of our 26 new, more general CCs for a theory with a compact gauge group.
\end{itemize}
Armed with this, we have shown how the 3-loop gauge $\beta$-function for a single gauge group can be matched to our general notation, up to two free parameters, and how this algebraic subsystem is then uniquely solved by the inclusion of our CCs, thereby deriving the complete 3-loop gauge $\beta$-function for an arbitrary, renormalizable, four-dimensional QFT. As referenced in section \ref{sec:results}, this result has in fact been calculated previously via other methods \cite{Molgaard:xxx}; nevertheless, our method provides the first novel application of Weyl CCs as a tool for \emph{deriving} general results from a known special case, demonstrating its viability as a technique at higher loop orders. To our knowledge this also provides the first 3-loop result for the kinetic mixing of the Abelian group factors. We have also demonstrated that the linear relation between $T_{IJ}$ coefficients is extended to include the quartic term, thus all three leading-order coefficients appear in a fixed ratio. Again, while this is somewhat irrelevant in light of the explicit calculations, it is interesting that the leading-order positive-definiteness of Osborn's $A$-function for a completely general theory actually follows from its own self-consistency, plus positive-definiteness of a pure-gauge, pure-Yukawa, or pure-quartic theory.

Finally, we have used \grafer to generate the CCs at order 4--3--2, extending the previous forefront of $\beta$-function constraints, and setting up a framework with which one can eventually predict much of the 4-loop gauge $\beta$-function for a completely general, renormalizable, four-dimensional QFT. Were one to reconstruct the general 3-loop Yukawa $\beta$-function, and augment the total system with a subset of the 4-loop gauge calculations, it should be possible to construct the general 4-loop gauge $\beta$-function. The complete $\beta$-functions for \emph{all three} SM gauge couplings in the unbroken phase, including all matter content, could then be extracted by simply substituting in the SM coupling matrices and generators for the unbroken phase. Given that the current 4-loop SM calculations \cite{Bednyakov:2015ooa,Zoller:2015tha} have been done only for the strong-coupling, incorporating just the top-Yukawa coupling and the Higgs self-coupling, this would be a monumental advance. Additionally, we have drawn attention to a particularly simple subset of CCs that arises at this order and resolves the explicitly-demonstrated ambiguity related to the treatment of $ \gamma_5 $ in the 4-loop strong-coupling $\beta$-function by relating it to unambiguous 3-loop Yukawa coefficients; in our previous paper \cite{Poole:2019txl}, we fixed this value by simply reconstructing the general coefficients for the relevant terms in $\beta_{aij}^{(3)}$, demonstrating that the RG information contained within these CCs may eventually lead to an unambiguous treatment of $\gamma_{5}$ in dimensionally-regularized chiral gauge theories. 

We have also reached the point at which we may again test various non-trivial modifications to \gfe, namely the $S$-tensor parameterizing the shift $\beta\rightarrow B$, and our conjectured possibility of imposing symmetry on $T_{IJ}$. By rearranging the 4--3--2 CCs, we have derived explicit expressions for all but one of the nine possible $S$ coefficients in a general theory; unfortunately, the undetermined coefficients all contribute to the fermionic shift $S^{(3)}_{ij}$, for which we cannot find any explicit calculation in current literature that would allow us to fix the final coefficient. Imposing symmetry of $T_{IJ}$ at order 4--3--2 leads to ten additional CCs, which we have extracted by simple comparison with the CCs when symmetry is not imposed. Our intent was to isolate easily-identifiable criteria for which our conjecture could be falsified. Of the ten conditions, the exact predictions\footnote{Of course, these are actually post-dictions, but as with the derivation of a ratio between leading-order $T_{IJ}$ coefficients, the method is such that these exact values arise without any reference to explicit $\beta_{abcd}^{(2)}$ calculations, and clearly demonstrate the potential for similar results at higher orders.} listed in \eqref{eq:sympred} are indeed the correct \msbar values, providing some corroboration of the conjecture. 

Presenting explicit results in this paper has been challenging, mostly due to the sheer quantity of explicit tensors and CCs as one goes to higher orders, as demonstrated in Table \ref{tab:results}. We have opted to sacrifice the scheme-independence of written results at 3--2--1 and above, in favour of using lower-order \msbar coefficients to present relations valid in \msbar only; the full list of $A$-function tensors, $\beta/B$-function parametrizations, $T_{IJ}$ contributions, and \msbar CCs are given in the ancillary file. Consequently, we feel that we should again emphasize the scheme-independent nature of $\beta$-function constraints derived from Weyl CCs: all one needs to do is refrain from inserting the lower-order \msbar coefficients and include possible 1PR contributions to the $ \beta $-functions. Previous general investigations into $\beta$-function constraints have maintained this scheme-independent approach; the diagrammatic representation for the $\beta$-functions is well-suited to deriving the effects of a scheme change on the general coefficients, allowing one to directly verify scheme-independence of the constraints\footnote{One can also easily reproduce well-known results regarding scheme-independence of the $\beta$-functions, namely that the quartic and Yukawa $\beta$-functions are scheme-independent at one loop, while the gauge $\beta$-function is scheme-independent at one and two loops.}. Once certain coefficients are known to be scheme-independent, one can include their explicit \msbar values without sacrificing overall scheme-independence of the system, while still leading to simplified CCs; this argument also guarantees that the fixed ratio between leading-order $T_{IJ}$ coefficients is a completely general result.

In principle, now that we have an adequate diagrammatic representation of the general gauge $\beta$-function, our procedure is capable of extracting the CCs for an arbitrary renormalizable four-dimensional QFT at any order. If one requests a parametrization of Osborn's $A$-function at some loop order $\tilde{A}^{(n)}$, \grafer derives the complete set of CCs at order ($n-1$)--($n-2$)--($n-3$), with the only limit being available computing power; our hope is that the CCs will therefore play a r\^ole in future $\beta$-function calculations, whenever attempts at such calculations are eventually made\footnote{The 3--2--1 ordering refers only to the \emph{maximum} order of the $\beta$-functions appearing in the CCs. It is worth emphasizing again that the utility of our procedure is not restricted to these maximal loop orders: there exist relations not only between the different $\beta$-functions, but between different orders of each individual $\beta$-function.}. 

While this work might seem to suggest an end to any further theoretical work on the CCs themselves (beyond seeking results for specific higher-loop diagrams, such as those containing $\gamma_{5}$ ambiguities), this paper only covers the case of four-dimensional theories. As mentioned in Section \ref{sec:LRG}, equivalent versions of \gfe have been argued to hold in any even-dimensional spacetime \cite{Grinstein:2013cka}. To obtain the equivalent CCs, we need only change the marginal couplings; for example, a general scalar theory in six dimensions has a marginal interaction $\mathcal{L}_{\text{Int.}} = -\tfrac{1}{3!}g_{ijk}\phi^{i}\phi^{j}\phi^{k}$, hence one can represent the tensor coupling as a three-point vertex with identical lines, and the analogous constructions of $\tilde{A}$, $T_{gg}$ and $\beta_{ijk}$ follow as in four dimensions \cite{Gracey:2015fia}. Because of this, all consequences of the four-dimensional version immediately carry over to other dimensions: one can again predict higher-order $\beta$-function coefficients and evaluate undetermined Feynman integrals. Furthermore, a series of exploratory calculations was performed for general 3-dimensional theories \cite{Jack:2015tka,Jack:2016utw}, where there is no proven analogue of \gfe, yet the associated $\beta$-functions still appear to satisfy an equation of the same form for some as-yet-unknown odd-dimensional function $\tilde{A}$\footnote{Conjectures have been advanced that the three-dimensional function $\tilde{A}$ being constructed is in fact the $F$-function \cite{Jafferis:2011zi,Klebanov:2011gs}, which at RG fixed points reduces to the finite part of $F$, the free-energy of a CFT compactified on a 3-sphere. At leading-order, the $F$-function indeed appears to have the same properties as the $A$-function.}. The associated marginal interaction term for a three-dimensional scalar-fermion theory is $\mathcal{L}_{\text{Int.}} = -\tfrac{1}{4} Y_{abij}\psi^{a}\psi^{b}\phi^{i}\phi^{j} - \tfrac{1}{6!}h_{ijklmn}\phi^{i}\phi^{j}\phi^{k}\phi^{l}\phi^{m}\phi^{n}$ (using both real scalars and fermions), thus all one needs to do is define a graphical representation for the marginal couplings $Y_{abij}$ and $h_{ijklmn}$. Attempts to include gauge interactions have also been made, successfully extending the leading-order construction to three-dimensional theories with a single gauge group, in exact analogy with a four-dimensional theory. In all cases, the procedure is the same, and one can further test our conjecture that symmetry of the analogous $T_{IJ}$ may always be imposed.

Another topic that has not been considered is supersymmetry. Compared to general theories, our understanding of $\tilde{A}$ in supersymmetric theories is considerably more advanced, ranging from an almost immediate proof of the weak $a$-theorem using $a$-maximization \cite{Intriligator:2003jj}\footnote{The full proof required an extension of $a$-maximization \cite{Kutasov:2003ux,Barnes:2004jj}, which holds away from RG fixed points.}, to a proposed non-perturbative expression for the $A$-function \cite{Freedman:1998rd} that satisfies the strong $a$-theorem subject to a constraint on the chiral superfield anomalous dimension \cite{Jack:2013sha,Jack:2014pua}. From our perspective, supersymmetric QFTs are nothing more than special cases of the general four-dimensional QFT, with $\mathcal{N}=1$ supersymmetry determined by specific relations between fields, charges, and couplings, and $\mathcal{N} > 1$ supersymmetry determined by further relations. Consequently, one can test the exact expression at some order of perturbation theory by defining the required relations as in Section 5 of \cite{Jack:2014pua}, suitably extended to multiple gauge groups, and substituting into the solution provided by \grafer. While there are technicalities to this procedure, such as having to evaluate the non-supersymmetric $\beta$-functions using dimensional reduction rather than the more conventional dimensional regularization\footnote{A full discussion of the challenges faced when using dimensional reduction in non-supersymmetric theories, plus its relation to the standard dimensional regularization procedure via a coupling-redefinition, can be found in \cite{Jack:1993ws,Jack:1994bn}.}, the method has been shown to work up to $\tilde{A}^{(4)}$ for theories with a single gauge group \cite{Jack:2014pua}. Conversely, supersymmetric reductions augment the use of Weyl CCs to reconstruct the $\beta$-functions of a general theory, since there exist non-renormalization theorems limiting the form of supersymmetric $\beta$-functions, and even exact results for certain schemes (such as the NSVZ $\beta$-function \cite{Jones:1983vk,Novikov:1983uc,Novikov:1985rd,Shifman:1986zi}), which place similar restrictions on the $\beta$-function coefficients.

We appreciate that our paper is rather long, so if you've skipped to the end looking for a quick summary of the results, think of Grothendieck's nut-cracking analogy for the two styles of mathematics \cite{McLarty:xxx}. Direct $\beta$-function calculations are the ``hammer and chisel" approach---you will get the answer for your particular theory, but it might be a colossal pain to actually set up the calculation, or take too long to obtain the results with the available computing resources. Weyl CCs are more of a ``soak it in water" approach\footnote{Grothendieck also came up with a ``rising sea" analogy for this latter option, which is delightfully apocalyptic, and adds a much-needed artistic touch to our mundane linear-algebraic labors: given an island surrounded by water, a slowly rising sea appears to have no effect on its structural soundness... until the island is eventually submerged and dismantled by the tides.}---eventually, through sheer overwhelming consistency, the $\beta$-functions will be solved in general, and you can simply extract the results for your special case.

\subsection*{Acknowledgments}
We would like to thank Esben Mølgaard for pointing out to us that the 3-loop semi-simple gauge $ \beta $-function can be extracted from \cite{Pickering:2001aq} by comparison with the known SM results, and for independently cross-checking the calculation. We would also like to thank Florian Herren for helping us understand the nature of $\gamma_{5}$ contributions in general theories, allowing us to clarify our previous work. CP would like to thank Ian Jack for his careful reading of the manuscript and invaluable feedback on our summary of the LRG, and Hugh Osborn for helpful comments. AET would like to thank Fermilab for hosting him during part of this work, and gratefully acknowledges financial support from the Danish Ministry of Higher Education and Science through an EliteForsk Travel Grant. This work is partially supported by the Danish National Research Foundation grant DNRF90. All diagrams have been drawn using TikZ-Feynman \cite{Ellis:2016jkw}.

\appendix
\renewcommand{\thesection}{\Alph{section}}
\section{Appendices} 
These appendices contain the full parametrization of the \msbar $ \beta $-function for the couplings of a completely general theory, with Lagrangian density \eqref{eq:general_Lagrangian}, up to order 3--2--2. We include the full TS basis for $ T^{(3)}_{IJ} $ and $ \tilde{A}^{(4)} $. Finally we provide a comparison of our results for kinetic mixing with those of \cite{Luo:2002iq}, and of our 3--2--1 CCs with those derived in \cite{Jack:2014pua}.

\subsection{Tensor contractions in the $ \beta $-functions at order 3--2--2} \label{app:beta_tensors}
Here we give our parametrization of the $ \beta $-functions using regular dummy-index notation. We reiterate that fermion indices are always treated in matrix notation and never made explicit. Tildes on more advanced objects are always defined in the usual manner: $ \tilde{X} = \sigma_1 X \sigma_1 $ in the space of fermion indices. Several 1- and 2-loop 2-point function occurring as substructures in the $ \beta $-functions are used throughout. They are: 

\noindent 1-loop gauge \footnote{Be advised that $ S_2(F) $ is twice the trace normalization of the fermions, due to both fermions and their complex conjugated automatically being included in the notation.}: 
	\begin{align}
	\tf{AB} &= \Tr{T^{A} T^{B}}, & 
	\ts{AB} &= (T_\phi^{A} T_\phi^{B})_{aa}, &
	\cg{AB} &= f^{ACD} f^{CDB}.	& 
	\end{align}
\noindent 1-loop fermion:
	\begin{align}
	\cf &= T^{A}T^{B} G^{2}_{AB}, & \yf &= y_a \tilde{y}_a,
	\end{align}
1-loop scalar:
	\begin{align}
	\cs{ab} &= (T_\phi^{A} T_\phi^{B} )_{ab} G^{2}_{AB}, & \ys{ab} &= \Tr{y_a \tilde{y_b}}.
	\end{align}
2-loop gauge:
	\begin{align}
	\tfe{C_F}{AB} &= \Tr{T^{A} T^{B} \cf}, &
	\tfe{Y_F}{AB} &= \Tr{T^{A} T^{B} \yft}, \nn
	\tse{C_S}{AB} &= (T^{A} T^{B})_{ab} \cs{ba}, &
	\tse{Y_S}{AB} &= (T^{A} T^{B})_{ab} \ys{ba} .
	\end{align}
2-loop fermion:
	\begin{align}
	\cfe{G} &= T^{A}T^{B} \left(G^{2} \cg{} G^2 \right)_{AB}, & 
	\cfe{S} &= T^{A}T^{B} \left(G^{2} \ts{} G^2 \right)_{AB},  \nn 
	\cfe{F} &= T^{A}T^{B} \left(G^{2} \tf{} G^2 \right)_{AB}, &
	\yfe{C_F} &= y_a \cf \tilde{y}_a, \nn
	\yfe{C_S} &= y_a \tilde{y}_b \cs{ab}, &
	\yfe{Y_F} &= y_a \yft \tilde{y}_a, \nn
	\yfe{Y_S} &= y_a \tilde{y}_b \ys{ab}, &
	\yyf &= y_a \tilde{y}_b y_a \tilde{y}_b.
	\end{align}
2-loop scalar:
	\begin{align}
	\cse{G}{ab} &= (T_\phi^{A} T_\phi^{B})_{ab} \left(G^{2} \cg{} G^2 \right)_{AB}, &
	\cse{S}{ab} &= (T_\phi^{A} T_\phi^{B})_{ab} \left(G^{2} \ts{} G^2 \right)_{AB}, \nn
	\cse{F}{ab} &= (T_\phi^{A} T_\phi^{B})_{ab} \left(G^{2} \tf{} G^2 \right)_{AB}, &
	\yse{C_F}{ab} &= \Tr{y_a \cf \tilde{y}_b}, \nn
	\yse{Y_F}{ab} &= \Tr{y_a \yft \tilde{y}_b}, &
	\yys{ab} &= \Tr{y_a \tilde{y}_c y_b \tilde{y}_c}.
	\end{align}
Having defined the 2-point structures, we can proceed with the parametrization of the $ \beta $-functions. 

\subsubsection{Gauge $ \beta $-function}
The first three loop orders of the gauge $ \beta $-functions are given by 
	\begin{equation}
	\beta^{(1)}_{AB} = \cofg{1}{1} \cg{AB} \quad + \quad \cofg{1}{2} \tf{AB} \quad + \quad  \cofg{1}{3} \ts{AB},
	\end{equation}
and 
	\begin{align}
	\beta^{(2)}_{AB} = \cofg{2}{1} & \tfe{C_F}{AB}
	& + \cofg{2}{2} & \tse{C_S}{AB} 
	\nn+ \cofg{2}{3} & \left(\cg{} G^{2} \cg{}\right)_{AB}
	& + \cofg{2}{4} & \left(\cg{} G^{2} \tf{} \right)_{AB} 
	\nn+ \cofg{2}{5} & \left(\cg{} G^{2}_{} \ts{} \right)_{AB}
	& + \cofg{2}{6} & \tfe{Y_F}{AB}
	\nn + \cofg{2}{7} & \tse{Y_S}{AB} ,
	\end{align}
and
	\begin{align}
	\beta^{(3)}_{AB} = \cofg{3}{1} & \Tr{T^{A} T^{B}\cf \cf}
	& + \cofg{3}{2} & (T^{A}_{\phi} T^{B}_{\phi})_{ab} \cs{bc} \cs{ca} 
	\nn+ \cofg{3}{3} & \Tr{T^{A} T^{B}\cfe{G}}
	& + \cofg{3}{4} & (T^{A}_{\phi} T^{B}_{\phi})_{ab} \cse{G}{ba} 
	\nn + \cofg{3}{5} & \Tr{T^{A} T^{B}\cfe{F}}
	& + \cofg{3}{6} & (T^{A}_{\phi} T^{B}_{\phi})_{ab} \cse{F}{ba} 
	\nn + \cofg{3}{7} & \Tr{T^{A} T^{B}\cfe{S}}
	& + \cofg{3}{8} & (T^{A}_{\phi} T^{B}_{\phi})_{ab} \cse{S}{ba} 
	\nn+ \cofg{3}{9} & \left(\cg{} G^{2} \, \tfe{C_F}{} \right)_{AB}
	& + \cofg{3}{10} & \left(\cg{} G^{2}\, \tse{C_S}{} \right)_{AB}	
	\nn+ \cofg{3}{11} & \left(\cg{} G^{2}\, \cg{} G^{2}\, \cg{} \right)_{AB}
	& + \cofg{3}{12} & \left(\cg{} G^{2}\, \tf{} G^{2}\, \tf{} \right)_{AB}
	\nn+ \cofg{3}{13} & \left(\cg{} G^{2}\, \ts{} G^{2} \,\ts{} \right)_{AB}
	& + \cofg{3}{14} & \left(\cg{} G^{2}\, \cg{} G^{2}\, \tf{} \right)_{AB}
	\nn+ \cofg{3}{15} & \left(\cg{} G^{2}_{}\, \cg{} G^{2}\, \ts{} \right)_{AB}
	& + \cofg{3}{16} & \left(\cg{} G^{2}\, \tf{} G^{2}\, \ts{} \right)_{AB}
	\nn+ \cofg{3}{17} & (T^{A}_\phi T^{C}_\phi)_{ab} \lambda_{abcd} (T^{B}_\phi T^{D}_\phi)_{cd} G^{2}_{CD}
	& + \cofg{3}{18} & (T^{A}_\phi T^{B}_\phi)_{ab} \lambda_{bcde} \lambda_{cdea}
	\nn+ \cofg{3}{19} & \Tr{T^{A} T^{B}\cf \yft}
	& + \cofg{3}{20} & \Tr{\tilde{T}^{A} \tilde{T}^{B} \yfe{C_F}}
	\nn+ \cofg{3}{21} & (T^{A}_\phi T^{B}_\phi)_{ab} \yse{C_F}{ba}
	& + \cofg{3}{22} & \Tr{\tilde{T}^{A} \tilde{T}^{B} \yfe{C_S}}
	\nn+ \cofg{3}{23} & (T^{A}_{\phi} T^{B}_{\phi})_{ab} \cs{bc} \ys{ca} 
	& + \cofg{3}{24} & \cg{AC} G^{2}_{CD} \tfe{Y_F}{DB}
	\nn+ \cofg{3}{25} & \cg{AC} G^{2}_{CD} \tse{Y_S}{DB}	
	& + \cofg{3}{26} & \Tr{y_a T^{A} \tilde{y}_a y_b T^{B} \tilde{y}_b}
	\nn+ \cofg{3}{27} & (T^{A}_\phi T^{B}_\phi)_{ab} \yys{ba}
	& + \cofg{3}{28} & \Tr{\tilde{T}^{A} \tilde{T}^{B} \yyf}
	\nn+ \cofg{3}{29} & \Tr{\tilde{T}^{A} \tilde{T}^{B} \yfe{Y_S}}
	& + \cofg{3}{30} & (T^{A}_\phi T^{B}_\phi)_{ab} \yse{Y_F}{ba}
	\nn+ \cofg{3}{31} & \Tr{\tilde{T}^{A} \tilde{T}^{B}\yf \yf}
	& + \cofg{3}{32} & \Tr{\tilde{T}^{A} \tilde{T}^{B} \yfe{Y_F}}
	\nn+ \cofg{3}{33} & (T^{A}_{\phi} T^{B}_{\phi})_{ab} \ys{bc} \ys{ca} .
	\end{align}
	
\subsubsection{Yukawa $ \beta $-function}
For the 1-loop Yukawa $ \beta $-function we have 
	\begin{equation}
	\beta^{(1)}_{a} = \cofy{1}{1} y_{b} \cs{ba} + \cofy{1}{2} y_{a} \cf + \cofy{1}{3} y_b \tilde{y}_a y_b + \cofy{1}{4} y_a \yft + \cofy{1}{5} y_b \ys{ba} ,
	\end{equation}
while at 2-loop orders we have
	\begin{align}
	\beta^{(2)}_{a} = \cofy{2}{1} & y_c \cs{cb} \cs{ba}
	& + \cofy{2}{2} & y_b \cf \cs{ba}
	& + \cofy{2}{3} & \cft y_a \cf 
	\nn+ \cofy{2}{4} & y_a \cf \cf
	& + \cofy{2}{5} & y_b \cse{G}{ba}
	& + \cofy{2}{6} & y_b \cse{S}{ba}
	\nn+ \cofy{2}{7} & y_b \cse{F}{ba}
	& + \cofy{2}{8} & y_a \cfe{G}
	& + \cofy{2}{9} & y_a \cfe{S}
	\nn+ \cofy{2}{10} & y_a \cfe{F}
	& + \cofy{2}{11} & y_b \lambda_{bcde} \lambda_{cdea}
	& + \cofy{2}{12} & y_b T^{A} \tilde{y}_b y_a T^{B} G^{2}_{AB}
	\nn+ \cofy{2}{13} & \yf \tilde{T}^{A} y_a T^{B} G^{2}_{AB}
	& + \cofy{2}{14} & y_b \tilde{y}_a y_c \cs{bc}
	& + \cofy{2}{15} & y_b \tilde{y}_c y_b \cs{ca}
	\nn+ \cofy{2}{16} & y_b \cf \tilde{y}_a y_b
	& + \cofy{2}{17} & y_b \tilde{y}_a y_b \cf
	& + \cofy{2}{18} & y_b \yse{C_F}{ba}
	\nn+ \cofy{2}{19} & \yfe{C_S} y_a
	& + \cofy{2}{20} & \yfe{C_F} y_a
	& + \cofy{2}{21} & y_a \yf \cf
	\nn+ \cofy{2}{22} & y_c \ys{cb} \cs{ba}
	& + \cofy{2}{23} & y_b \tilde{y}_c y_d \lambda_{bcda}
	& + \cofy{2}{24} & y_b \tilde{y_c} y_a \tilde{y}_b y_c
	\nn+ \cofy{2}{25} & y_b \tilde{y_a} y_c \tilde{y}_b y_c
	& + \cofy{2}{26} & y_b \tilde{y_c} y_a \tilde{y}_c y_b
	& + \cofy{2}{27} & \yyf y_a
	\nn+ \cofy{2}{28} & y_b \yys{ba}
	& + \cofy{2}{29} & y_b \tilde{y}_a \yf y_b
	& + \cofy{2}{30} & y_a \yfe{Y_F}
	\nn+ \cofy{2}{31} & y_b \yse{Y_F}{ba}
	& + \cofy{2}{32} & y_b \tilde{y}_a y_c \ys{bc}
	& + \cofy{2}{33} & \yfe{Y_S} y_a .
	\end{align}
	
\subsubsection{Quartic $ \beta $-function}	
For the quartic $ \beta $-function we have 
	\begin{align}
	\beta^{(1)}_{abcd} = \cofq{1}{1} & (T_\phi^{A} T_\phi^{C})_{ab} G^{2}_{AB} G^{2}_{CD} (T_\phi^{B} T_\phi^{D})_{cd} 
	& + \cofq{1}{2} & \cs{ae} \lambda_{ebcd}
	& + \cofq{1}{3} & \lambda_{abef} \lambda_{efcd}
	\nn+ \cofq{1}{4} & \ys{ae} \lambda_{ebcd}
	& + \cofq{1}{5} & \Tr{y_a \tilde{y}_b y_c \tilde{y}_d}
	\end{align}
at 1-loop order. For completeness we include the 2-loop quartic $ \beta $-function as this too is known:
	\begin{align}
	\beta^{(2)}_{abcd} = \cofq{2}{1} & G^{2}_{AB} (T_\phi^{A} T_\phi^{C} T_\phi^{E} T^{B}_\phi)_{ab} G^{2}_{CD} G^{2}_{EF} (T_\phi^{D} T_\phi^{F})_{cd} 
	\nn + \cofq{2}{2} & (\cs{} T_\phi^{A} T^{C}_\phi)_{ab} G^{2}_{AB} G^{2}_{CD} (T_\phi^{B} T_\phi^{D})_{cd} 
	\nn + \cofq{2}{3} & (T_\phi^{A} T^{C}_\phi)_{ab} G^{2}_{AB} \left(G^{2} \cg{} G^2 \right)_{CD} (T_\phi^{B} T_\phi^{D})_{cd}
	\nn + \cofq{2}{4} & (T_\phi^{A} T^{C}_\phi)_{ab} G^{2}_{AB} \left(G^{2} \ts{} G^2 \right)_{CD} (T_\phi^{B} T_\phi^{D})_{cd}
	\nn + \cofq{2}{5} & (T_\phi^{A} T^{C}_\phi)_{ab} G^{2}_{AB} \left(G^{2} \tf{} G^2 \right)_{CD} (T_\phi^{B} T_\phi^{D})_{cd}
	\nn + \cofq{2}{6} & (T_\phi^{A} T^{C}_\phi)_{ae} G^{2}_{AB} G^{2}_{CD} (T_\phi^{B} T_\phi^{D})_{bf} \lambda_{efcd}
	\nn + \cofq{2}{7} & (T_\phi^{A} T^{C}_\phi)_{ab} G^{2}_{AB} G^{2}_{CD} (T_\phi^{B} T_\phi^{D})_{ef} \lambda_{efcd}
	\nn + \cofq{2}{8} & \cs{ae} \cs{bf} \lambda_{efcd} 
	\qquad\quad  +\cofq{2}{9} \cs{ae} \cs{ef} \lambda_{ebcd}
	\nn +\cofq{2}{10} & \cse{G}{ae} \lambda_{ebcd} 
	\qquad \quad +\cofq{2}{11} \cse{S}{ae} \lambda_{ebcd}
	\nn	+\cofq{2}{12} & \cse{F}{ae} \lambda_{ebcd} 
	\qquad \quad + \cofq{2}{13} (T_{\phi}^{A})_{ae} (T_{\phi}^{B})_{bf} G^{2}_{AB} \lambda_{efgh} \lambda_{ghcd}
	\nn + \cofq{2}{14} & \lambda_{abef} \cs{fg} \lambda_{egcd} 
	\qquad \quad + \cofq{2}{15} \cs{ae} \lambda_{ebfg} \lambda_{fgcd} 
	\nn + \cofq{2}{16} & \lambda_{aefg} \lambda_{efgh} \lambda_{hbcd}
	\qquad \quad + \cofq{2}{17} \lambda_{abef} \lambda_{eghc} \lambda_{fghd} 
	\qquad \quad + \cofq{2}{18} \lambda_{abef} \lambda_{efgh} \lambda_{ghcd}
	\nn + \cofq{2}{19} & (T_\phi^{A} T^{C}_\phi)_{ab} G^{2}_{AB} G^{2}_{CD} \Tr{T^{D}\, T^{B}\, \tilde{y}_c\, y_d}
	\nn + \cofq{2}{20} & (\ys{} T_\phi^{A} T^{C}_\phi)_{ab} G^{2}_{AB} G^{2}_{CD} (T_\phi^{B} T_\phi^{D})_{cd} 
	\nn +\cofq{2}{21} & \yse{C_F}{ae} \lambda_{ebcd} 
	\qquad \quad + \cofq{2}{22} \cs{ae} \ys{ef} \lambda_{fbcd}
	\nn + \cofq{2}{23} & \lambda_{abef} \ys{fg} \lambda_{egcd} 
	\qquad \quad + \cofq{2}{24} \Tr{y_a\, T^{A}\, \tilde{y}_b\, y_c\, T^{B}\, \tilde{y}_d} G^{2}_{AB}
	\nn +\cofq{2}{25} & \cs{ae} \Tr{y_e\, \tilde{y}_b\, y_c\, \tilde{y}_d}
	\qquad \quad +\cofq{2}{26}  \Tr{y_a\, \tilde{y}_b\, y_c\, \tilde{y}_d\, \cft}
	\nn +\cofq{2}{27} & \Tr{y_a\, \tilde{y}_e\, y_b\, \tilde{y}_f} \lambda_{efcd}
	\qquad \quad +\cofq{2}{28} \Tr{y_e\, \tilde{y}_b\, y_e\, \tilde{y}_f} \lambda_{efcd} 
	\nn +\cofq{2}{29} & \yys{ae} \,\lambda_{ebcd}
	\qquad \quad +\cofq{2}{30} \yse{Y_F}{ae} \,\lambda_{ebcd} 
	\nn +\cofq{2}{31} & \Tr{y_a\, \tilde{y}_b\, y_c\, \tilde{y}_e\, y_d\, \tilde{y}_e} 
	\qquad \quad + \cofq{2}{32} \Tr{y_a\, \tilde{y}_b\, y_e\, \tilde{y}_c\, y_d\, \tilde{y}_e}
	\nn +\cofq{2}{33} & \Tr{y_a \,\tilde{y}_b\, y_c\, \tilde{y}_d\, \yf }. 
	\end{align}
At this point the reader may begin to appreciate the use of the graphical notation employed throughout the paper.

\subsection{Coefficients of the $ \beta $-functions at order 3--2--2} \label{app:coefficients}
The basis we have employed for the $ \beta $-functions in this work deviates from previous works in some places. Part of this is simply due to \msbar-specific computations not including terms with vanishing coefficients. It requires some work to match our basis onto \cite{Machacek:1983tz}, \cite{Luo:2002ti} and \cite{Pickering:2001aq}, the latter of which has (to our knowledge) never before been extended to a compact gauge group in published literature. We will therefore list all the known coefficients in our basis here.
\\1-loop gauge coefficients:
	\begin{align}
	&
	\cofg{1}{1}=-\frac{22}{3}
	,&&
	\cofg{1}{2}=\frac{2}{3}
	,&&
	\cofg{1}{3}=\frac{1}{3}.
	\end{align}
2-loop gauge coefficients:
	\begin{align}
	&
	\cofg{2}{1}=2
	,&&
	\cofg{2}{2}=4
	,&&
	\cofg{2}{3}=-\frac{68}{3}
	,&&
	\cofg{2}{4}=\frac{10}{3}
	,&&
	\cofg{2}{5}=\frac{2}{3}
	,&&
	\cofg{2}{6}=-1
	,&&
	\cofg{2}{7}=0.
	\end{align}
3-loop gauge coefficients:
	\begin{align}
	&
	\cofg{3}{1}=-1
	,&&
	\cofg{3}{2}=\frac{29}{2}
	,&&
	\cofg{3}{3}=\frac{133}{18}
	,&&
	\cofg{3}{4}=\frac{679}{36}
	,&&
	\cofg{3}{5}=-\frac{11}{18}
	,&&
	\cofg{3}{6}=-\frac{25}{18}
	\nonumber \\ &
	\cofg{3}{7}=-\frac{23}{36}
	,&&
	\cofg{3}{8}=-\frac{49}{36}
	,&&
	\cofg{3}{9}=4
	,&&
	\cofg{3}{10}=\frac{25}{2}
	,&&
	\cofg{3}{11}=-\frac{2857}{27}
	,&&
	\cofg{3}{12}=-\frac{79}{108}
	\nonumber \\ &
	\cofg{3}{13}=\frac{1}{54}
	,&&
	\cofg{3}{14}=\frac{1415}{54}
	,&&
	\cofg{3}{15}=\frac{545}{108}
	,&&
	\cofg{3}{16}=-\frac{29}{54}
	,&&
	\cofg{3}{17}=1
	,&&
	\cofg{3}{18}=-\frac{1}{12}
	\nonumber \\ &
	\cofg{3}{19}=-\frac{5}{4}
	,&&
	\cofg{3}{20}=-\frac{1}{4}
	,&&
	\cofg{3}{21}=-1
	,&&
	\cofg{3}{22}=-7
	,&&
	\cofg{3}{23}=-\frac{7}{2}
	,&&
	\cofg{3}{24}=-6
	\nonumber \\ &
	\cofg{3}{25}=\frac{9}{4}
	,&&
	\cofg{3}{26}=1
	,&&
	\cofg{3}{27}=-1
	,&&
	\cofg{3}{28}=\frac{3}{2}
	,&&
	\cofg{3}{29}=\frac{7}{8}
	,&&
	\cofg{3}{30}=\frac{1}{2}
	\nonumber \\ &
	\cofg{3}{31}=\frac{1}{8}
	,&&
	\cofg{3}{32}=\frac{3}{8}
	,&&
	\cofg{3}{33}=-\frac{1}{8}.
	\end{align}
1-loop Yukawa coefficients:
	\begin{align}
	&
	\cofy{1}{1}=0
	,&&
	\cofy{1}{2}=-6
	,&&
	\cofy{1}{3}=2
	,&&
	\cofy{1}{4}=1
	,&&
	\cofy{1}{5}=\frac{1}{2}.
	\end{align}
2-loop Yukawa coefficients:
	\begin{align}
	&
	\cofy{2}{1}=-\frac{21}{2}
	,&&
	\cofy{2}{2}=12
	,&&
	\cofy{2}{3}=0
	,&&
	\cofy{2}{4}=-3
	,&&
	\cofy{2}{5}=\frac{49}{4}
	,&&
	\cofy{2}{6}=-\frac{1}{4}
	\nonumber \\ &
	\cofy{2}{7}=-\frac{1}{2}
	,&&
	\cofy{2}{8}=-\frac{97}{3}
	,&&
	\cofy{2}{9}=\frac{11}{6}
	,&&
	\cofy{2}{10}=\frac{5}{3}
	,&&
	\cofy{2}{11}=\frac{1}{12}
	,&&
	\cofy{2}{12}=12
	\nonumber \\ &
	\cofy{2}{13}=0
	,&&
	\cofy{2}{14}=6
	,&&
	\cofy{2}{15}=-12
	,&&
	\cofy{2}{16}=10
	,&&
	\cofy{2}{17}=6
	,&&
	\cofy{2}{18}=\frac{5}{2}
	\nonumber \\ &
	\cofy{2}{19}=9
	,&&
	\cofy{2}{20}=-\frac{1}{2}
	,&&
	\cofy{2}{21}=-\frac{7}{2}
	,&&
	\cofy{2}{22}=0
	,&&
	\cofy{2}{23}=-2
	,&&
	\cofy{2}{24}=2
	\nonumber \\ &
	\cofy{2}{25}=0
	,&&
	\cofy{2}{26}=-2
	,&&
	\cofy{2}{27}=0
	,&&
	\cofy{2}{28}=-\frac{1}{2}
	,&&
	\cofy{2}{29}=-2
	,&&
	\cofy{2}{30}=-\frac{1}{4}
	\nonumber \\ &
	\cofy{2}{31}=-\frac{3}{4}
	,&&
	\cofy{2}{32}=-1
	,&&
	\cofy{2}{33}=-\frac{3}{4}.
	\end{align}
1-loop quartic coefficients:
	\begin{align}
	&
	\cofq{1}{1}=36
	,&&
	\cofq{1}{2}=-12
	,&&
	\cofq{1}{3}=3
	,&&
	\cofq{1}{4}=2
	,&&
	\cofq{1}{5}=-12.
	\end{align}
2-loop quartic coefficients:
	\begin{align}
	&
	\cofq{2}{1}=324
	,&&
	\cofq{2}{2}=-684
	,&&
	\cofq{2}{3}=646
	,&&
	\cofq{2}{4}=-28
	,&&
	\cofq{2}{5}=-32
	,&&
	\cofq{2}{6}=12
	\nonumber \\ &
	\cofq{2}{7}=60
	,&&
	\cofq{2}{8}=0
	,&&
	\cofq{2}{9}=6
	,&&
	\cofq{2}{10}=-\frac{143}{3}
	,&&
	\cofq{2}{11}=\frac{11}{3}
	,&&
	\cofq{2}{12}=\frac{10}{3}
	\nonumber \\ &
	\cofq{2}{13}=-18
	,&&
	\cofq{2}{14}=24
	,&&
	\cofq{2}{15}=-18
	,&&
	\cofq{2}{16}=\frac{1}{3}
	,&&
	\cofq{2}{17}=-6
	,&&
	\cofq{2}{18}=0
	\nonumber \\ &
	\cofq{2}{19}=-144
	,&&
	\cofq{2}{20}=60
	,&&
	\cofq{2}{21}=10
	,&&
	\cofq{2}{22}=0
	,&&
	\cofq{2}{23}=-3
	,&&
	\cofq{2}{24}=0
	\nonumber \\ &
	\cofq{2}{25}=24
	,&&
	\cofq{2}{26}=-48
	,&&
	\cofq{2}{27}=12
	,&&
	\cofq{2}{28}=0
	,&&
	\cofq{2}{29}=-2
	,&&
	\cofq{2}{30}=-3
	\nonumber \\ &
	\cofq{2}{31}=48
	,&&
	\cofq{2}{32}=24
	,&&
	\cofq{2}{33}=24.
	\end{align}

\subsection{$A$-function and $T_{IJ}$ tensors at order 3--2--1} \label{app:A4}
\begin{figure}
	\centering
	\includegraphics[width=\textwidth]{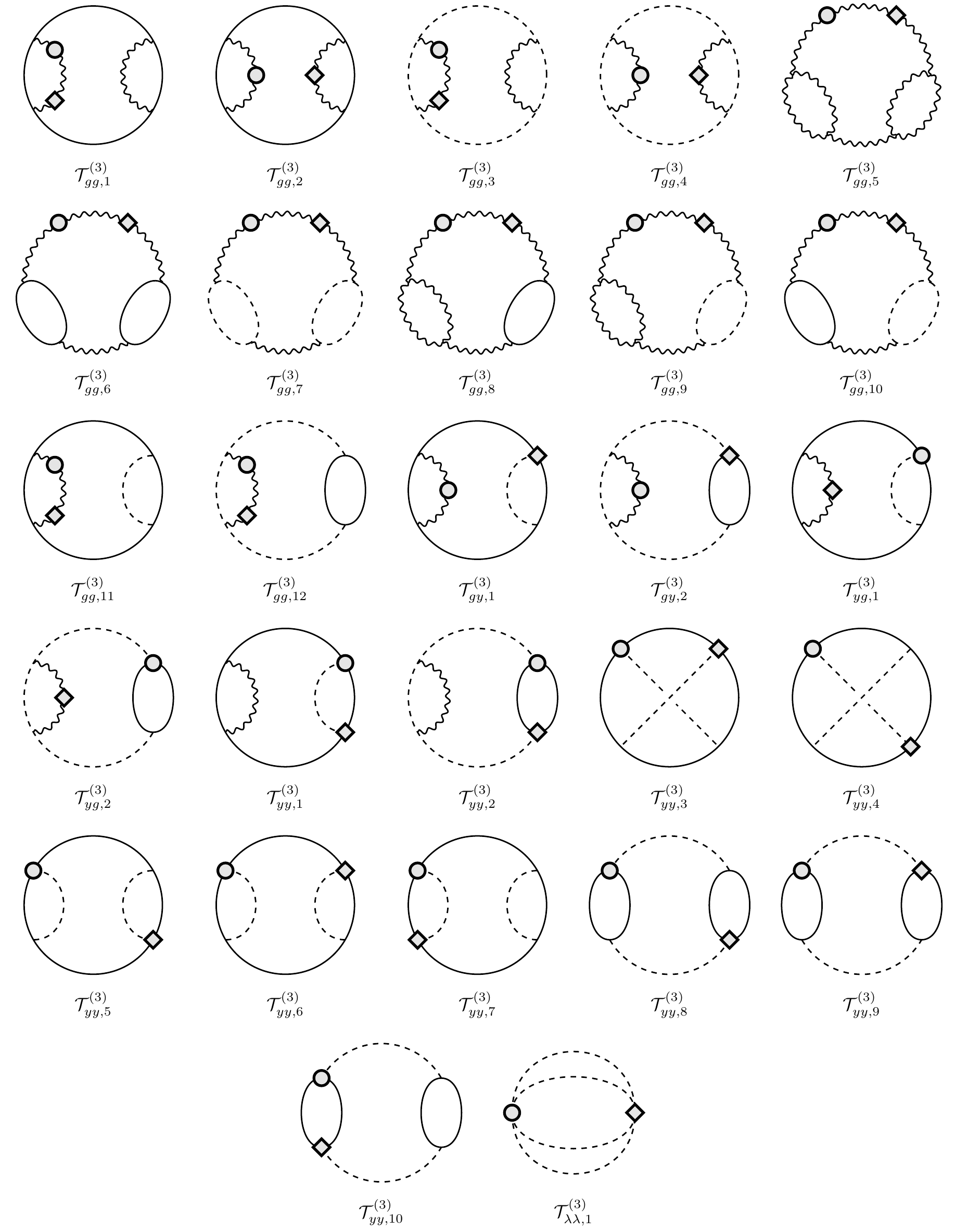}
	\caption{Graph representation of the 27 3-loop contributions to the metric $ T_{IJ} $. A circle indicates an open index $ I $ and a diamond an index $ J $. Note that the scalar lines in $\cofT{3}{yy,3}$ and $\cofT{3}{yy,4}$ do not intersect.}
	\label{fig:T_3-loop}
\end{figure}

\begin{figure}
	\centering
	\includegraphics[width=\textwidth]{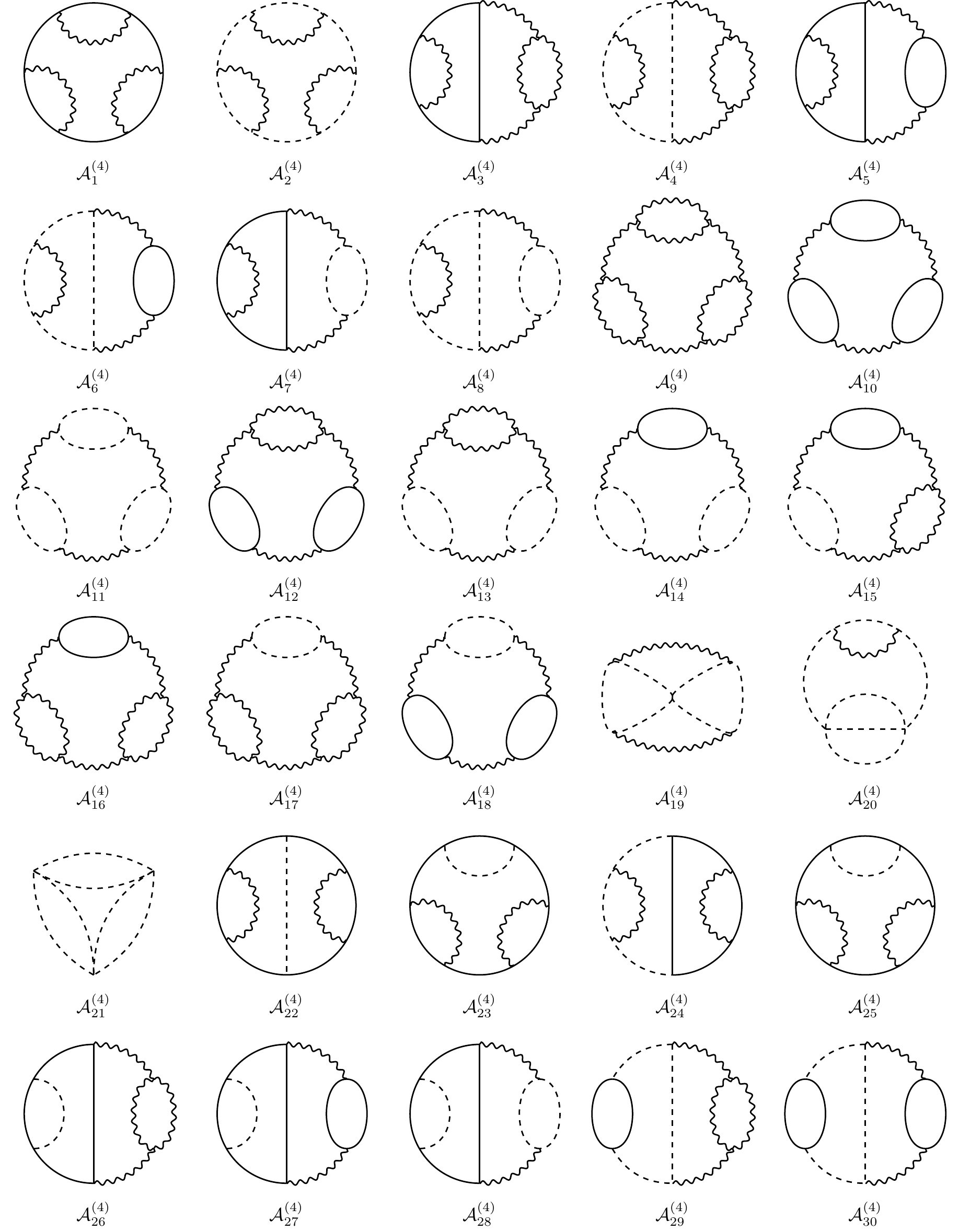}
	\caption{Graph representation of the first 30 contributions to the 4-loop $A$-function.}
	\label{fig:a_4-loop1}
\end{figure}

\begin{figure}
	\centering
	\includegraphics[width=\textwidth]{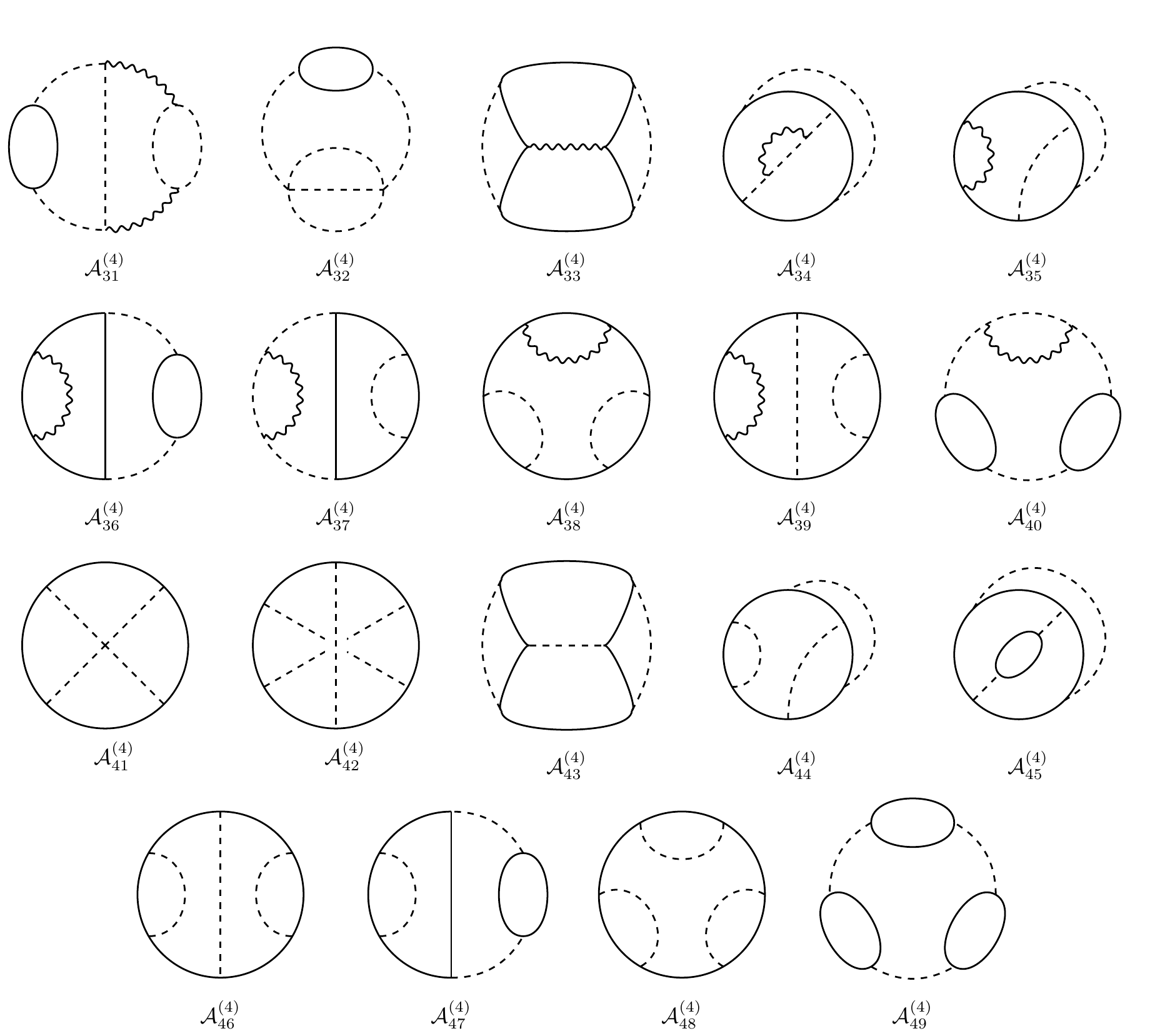}
	\caption{Graph representation of the last 19 contributions to the 4-loop $A$-function.}
	\label{fig:a_4-loop2}
\end{figure}

The structure of $T_{IJ}^{(3)}$ can be inferred from the 3-loop $A$-function, by considering all different ways of replacing couplings with open indices. The 27 resulting TSs are shown in Fig. \ref{fig:T_3-loop}. The 49 TSs making up the 4-loop $A$-function are shown in Figs. \ref{fig:a_4-loop1} and \ref{fig:a_4-loop2}. 

\subsection{Comparison of kinetic mixing results with previous work} \label{app:kin_mixing}
We may consider two different $ \beta $-functions related to the gauge coupling matrix $ G^2 $. With the generic formula for $ \beta $-functions \eqref{eq:generic_beta_function}, one finds 
	\begin{equation}
	\partial_t G^{-2} = (\rho_J g^{J} \partial_J + 1) \left. \delta G^{-2} \right|_{1/\epsilon},
	\end{equation}
using $ t=\ln \mu $ for the RG time, and the inverse (normal $ \beta $-function)
	\begin{equation} \label{eq:betag_from_betag_inv}
	\partial_t G^2 = (\rho_J g^{J} \partial_J - 1) \left. \delta G^{2} \right|_{1/\epsilon}  = - G^2 (\partial_t G^{-2}) G^2, 
	\end{equation}
from Eq. \eqref{eq:G2_ct}. 

With our $ \beta $-functions under control we can compare to \citet{Luo:2002iq}, where they parametrize the kinetic term of their Lagrangian by
	\begin{equation}
	\mathcal{L} \supset -\dfrac{1}{4} F_{\mu\nu} \xi F^{\mu\nu}, \qquad \text{where} \quad \xi_{m m} = 1.
	\end{equation} 
That is, they keep their gauge couplings in the interaction terms, and let $ \xi $ include the effects of the kinetic mixing in off-diagonal terms. Gathering all their gauge couplings in a diagonal matrix $ g = \diag(g_1, g_2, \ldots) $ we may match their notation to the one developed in the present paper\footnote{We can keep the non-Abelian entires in $ G^2 $, but all explicit indices in this section refer to $ \U(1) $ factors.}:
	\begin{equation}
	G^{-2} = g^{-1} \xi g^{-1} \implies \xi = g G^{-2} g.
	\end{equation}
Luo and Xiao define their $ \beta $-functions as 
	\begin{equation}
	\gamma_{uv} \equiv \partial_t \xi_{uv} \andeq \beta_u \equiv \gamma_u g = \partial_t g_u,
	\end{equation}
so that $ \gamma_{uu} = 0 $. We then find that
	\begin{equation}
	\begin{split}
	\gamma_{uv} &= \left[\beta_g G^{-2} g + g \partial_t G^{-2} g + g G^{-2} \beta_g \right]_{uv} = (\gamma_u + \gamma_v) \xi_{uv} + g_u \partial_t G^{-2}_{uv} g_v \\
	&\equiv (\gamma_u + \gamma_v) \xi_{uv} + \gamma^{\xi}_{uv}
	\end{split}
	\end{equation}
for $ u \neq v $. Meanwhile the regular $ \beta $-functions are determined from
	\begin{equation}
	\begin{gathered}
	\partial_t G^{-2}_{uu} = \left[ g^{-1} \gamma g^{-1} - g^{-2} \beta_g \xi g^{-1} - g^{-1} \xi \beta_g g^{-2} \right]_{uu} = -2 \beta_u g_u^{-3}\\
	\implies \beta_u = - \dfrac{1}{2} g_u^{3} \partial_t G^{-2}_{uu}.
	\end{gathered}
	\end{equation}

Finally for direct comparison  we typically arrange the $\beta$-function by loop order. It follows from Eqs. \eqref{eq:betag_from_betag_inv} and \eqref{eq:beta_parametrization} that 
	\begin{equation}
	\partial_t G^{-2}_{AB} = - \dfrac{1}{2} \sum_{\mathrm{perm}} \sum_{\ell} \dfrac{ \beta^{(\ell)}_{AB} }{(4 \pi)^{2\ell}}.
	\end{equation}
Thus, 
	\begin{equation}
	\beta_u^{(n)} = \dfrac{1}{2} g_u^3 \beta^{(n)}_{uu} \andeq \gamma^{\xi\, (n)}_{uv} = - \tfrac{1}{2} g_u \left( \beta^{(n)}_{uv} + \beta^{(n)}_{vu} \right) g_v.
	\end{equation}
With the known 2-loop results from appendices \ref{app:beta_tensors} and \ref{app:coefficients}, we reproduce the results of \citet{Luo:2002iq}; using a similar extraction procedure, our general 3-loop result encompasses the analogous 3-loop kinetic mixing results.

\subsection{Comparison of 3--2--1 consistency conditions with previous work}	\label{app:321}
While the calculation method used in this paper is conceptually identical to \cite{Jack:2014pua}, there are various differences in the intermediate steps that render direct comparison difficult; furthermore, although implicit in their results, the authors did not explicitly include CCs involving the coefficients of $\beta_{g}^{(3)}$, instead opting to phrase the information as a consistency check on the \msbar calculation of \cite{Pickering:2001aq}. Not only does \cite{Jack:2014pua} use $\beta_{g} = \tfrac{\text{d} g}{\text{d} \ln\mu}$ rather than our $\beta_{AB} = \tfrac{\text{d} G^{2}_{AB}}{\text{d}\ln\mu}$ (which simply leads to our coefficients being larger by a factor 2), but they use a different basis for the gauge-dependent tensors in $\beta^{(2)}_{aij}$, which does not directly correspond to the single-gauge case of our own basis. Thus, to facilitate comparison, we have translated the 19 conditions obtained (implicitly) in \cite{Jack:2014pua} into our notation and conventions, arranged such that each condition is equal to zero:
\begin{subequations}
	\begin{align}
	\cofq{1}{4} \cofy{2}{23} - 4\cofq{1}{5} \cofy{2}{11} &=0\tag{$\text{Y}_{1}$}\\ 
	\tfrac{1}{2}\cofy{2}{25} - \cofy{2}{26} - 2\cofy{2}{27} - 4\cofy{2}{32} + 8\cofy{2}{33} &=0\tag{$\text{Y}_{2}$}\\ 
	\tfrac{1}{2}\cofy{2}{25} - \cofy{2}{26} + 2\cofy{2}{27} + 2\cofy{2}{29} - 8\cofy{2}{30} &=0\tag{$\text{Y}_{3}$}\\ 
	\cofy{2}{28} + \tfrac{1}{2}\cofy{2}{29} - 2\cofy{2}{31} &=0\tag{$\text{Y}_{4}$}\\ 
	\tfrac{1}{2}\cofy{2}{13} + \cofy{2}{18} - \cofy{2}{20} - 6\cofy{2}{30} + 6\cofy{2}{31} &=0\tag{$\text{Y}_{5}$}\\ 
	\cofy{2}{13} - \cofy{2}{14} + \cofy{2}{15} + 2\cofy{2}{19} &=0\tag{$\text{Y}_{6}$}\\ 
	\tfrac{1}{2}\cofy{2}{16} - \tfrac{1}{2}\cofy{2}{17} - 2\cofy{2}{18} - 6\cofy{2}{28} &=0\tag{$\text{Y}_{7}$}\\ 
	6\cofq{1}{5}\cofg{3}{17} - \cofq{1}{1}\cofy{2}{23} &=0\tag{$\text{Y}_{8}$}\\ 
	\cofq{1}{2}\cofg{3}{17} - 4\cofq{1}{1}\cofg{3}{18} &=0\tag{$\text{Y}_{9}$}\\ 
	\tfrac{1}{2}\cofy{2}{12} - 6\cofg{3}{26} &=0\tag{$\text{Y}_{10}$}\\ 
	\cofy{2}{1} - 3\cofg{3}{23} &=0\tag{$\text{Y}_{11}$}\\ 
	\cofy{2}{15} + \cofy{2}{13} + 12\cofg{3}{30} - 6\cofg{3}{27} &=0\tag{$\text{Y}_{12}$}\\ 
	3\cofg{3}{19} + \cofy{2}{3} - 3\cofg{3}{20} - \cofy{2}{4} &=0\tag{$\text{Y}_{13}$}\\ 
	\cofy{2}{15} + \cofy{2}{14} - 4\cofy{2}{22} - 12\cofg{3}{27} + 48\cofg{3}{33} &=0\tag{$\text{Y}_{14}$}\\ 
	6\cofy{2}{15} - 3\cofy{2}{14} + 2\cofy{2}{2} - 6\cofg{3}{21} - 6\cofg{3}{22} - 18\cofg{3}{27} &=0\tag{$\text{Y}_{15}$}\\ 
	\tfrac{1}{2}\cofy{2}{20} - \tfrac{1}{2}\cofy{2}{21} + 6\cofg{3}{31} - 6\cofg{3}{32} &=0\tag{$\text{Y}_{16}$}\\ 
	6\cofy{2}{27} - 6\cofy{2}{29} + 4\cofy{2}{20} - \cofy{2}{16} - 24\cofg{3}{32} + 6\cofg{3}{28} &=0\tag{$\text{Y}_{17}$}\\ 
	18\cofy{2}{33} - \tfrac{3}{2}\cofy{2}{13} + 3\cofy{2}{20} - \cofy{2}{4} + 18\cofg{3}{29} + 3\cofg{3}{19} &=0\tag{$\text{Y}_{18}$}\\ 
	9\cofy{2}{30} + \tfrac{3}{2}\cofy{2}{20} + \tfrac{1}{2}\cofy{2}{4} - 18\cofy{2}{31} - 3\cofy{2}{18} - 9\cofg{3}{32} - \tfrac{3}{2}\cofg{3}{19} &=0 \tag{$\text{Y}_{19}$}
	\end{align}
\end{subequations}
Having done the translation, we can verify that these conditions are indeed linear combinations of our more general conditions. Defining conditions $X_{i}$ as the corresponding equations \eqref{eq:321} with all terms moved to the LHS, we find that
	\begin{align*}
	\text{Y}_{1} &=  \tfrac{2\mathfrak{q}_{5}}{\mathfrak{q}_{1}}\text{X}_{25}-\tfrac{\mathfrak{q}_{4}}{\mathfrak{q}_{1}} \text{X}_{26} & \text{Y}_{11} &= -3 \text{X}_{3}\\
	\text{Y}_{2} &= -\tfrac{1}{3} \text{X}_{16} - \tfrac{2}{3}\text{X}_{17} + 4 \text{X}_{22} + \tfrac{4}{3} \text{X}_{23} & \text{Y}_{12} &= -\text{X}_{10} + \text{X}_{12}\\
	\text{Y}_{3} &= -\tfrac{1}{3} \text{X}_{16} + \tfrac{2}{3}\text{X}_{17} + \tfrac{2}{3} \text{X}_{19} - \tfrac{4}{3} \text{X}_{20} & \text{Y}_{13} &= \text{X}_{4}\\
	\text{Y}_{4} &= -\tfrac{1}{12} (\text{X}_{18}+\text{X}_{21}) & \text{Y}_{14} &= \text{X}_{11} + \text{X}_{12} + 2 \text{X}_{15}\\
	\text{Y}_{5} &= -\tfrac{1}{2} \text{X}_{10} + \tfrac{1}{2}\text{X}_{19} - \text{X}_{20} + \tfrac{1}{4} \text{X}_{21} & \text{Y}_{15} &= -3 \text{X}_{11} + 6 \text{X}_{12}\\
	\text{Y}_{6} &= -\text{X}_{10} - \text{X}_{11} + \text{X}_{12} - 2\text{X}_{13} & \text{Y}_{16} &= \tfrac{1}{2} \text{X}_{14}\\
	\text{Y}_{7} &= \tfrac{1}{2} \text{X}_{18} & \text{Y}_{17} &= 2 \text{X}_{17} - 2 \text{X}_{19}\\
	\text{Y}_{8} &= \text{X}_{26} & \text{Y}_{18} &= \text{X}_{4} + \tfrac{3}{2} \text{X}_{10} + 3\text{X}_{23}\\
	\text{Y}_{9} &= \text{X}_{24} & \text{Y}_{19} &= -\tfrac{1}{2} \text{X}_{4} - \tfrac{3}{2}\text{X}_{19} + \tfrac{3}{2} \text{X}_{20} - \tfrac{3}{4} \text{X}_{21}\\
	\text{Y}_{10} &= -6 \text{X}_{9} & &
	\end{align*}
From this, we can isolate the new information contained in our calculation. Clearly, equations \eqref{eq:X1}, \eqref{eq:X2}, \eqref{eq:X5}, \eqref{eq:X6}, \eqref{eq:X7} and \eqref{eq:X8} are completely new, as they do not appear in the linear combinations. The final piece of information fixes an under-determination involving equations $\text{X}_{i},\; i \in \left\{10,11,12,13,15,17,19,20,23 \right\}$: attempting to solve $\text{X}_{i}$ in terms of $\text{Y}_{i}$ leaves one free parameter, hence knowing any one of these nine conditions determines the rest.

%\newpage
\bibliography{Bibliography} 

\begin{thebibliography}{72}%
\makeatletter
\providecommand \@ifxundefined [1]{%
 \@ifx{#1\undefined}
}%
\providecommand \@ifnum [1]{%
 \ifnum #1\expandafter \@firstoftwo
 \else \expandafter \@secondoftwo
 \fi
}%
\providecommand \@ifx [1]{%
 \ifx #1\expandafter \@firstoftwo
 \else \expandafter \@secondoftwo
 \fi
}%
\providecommand \natexlab [1]{#1}%
\providecommand \enquote  [1]{``#1''}%
\providecommand \bibnamefont  [1]{#1}%
\providecommand \bibfnamefont [1]{#1}%
\providecommand \citenamefont [1]{#1}%
\providecommand \href@noop [0]{\@secondoftwo}%
\providecommand \href [0]{\begingroup \@sanitize@url \@href}%
\providecommand \@href[1]{\@@startlink{#1}\@@href}%
\providecommand \@@href[1]{\endgroup#1\@@endlink}%
\providecommand \@sanitize@url [0]{\catcode `\\12\catcode `\$12\catcode
  `\&12\catcode `\#12\catcode `\^12\catcode `\_12\catcode `\%12\relax}%
\providecommand \@@startlink[1]{}%
\providecommand \@@endlink[0]{}%
\providecommand \url  [0]{\begingroup\@sanitize@url \@url }%
\providecommand \@url [1]{\endgroup\@href {#1}{\urlprefix }}%
\providecommand \urlprefix  [0]{URL }%
\providecommand \Eprint [0]{\href }%
\providecommand \doibase [0]{http://dx.doi.org/}%
\providecommand \selectlanguage [0]{\@gobble}%
\providecommand \bibinfo  [0]{\@secondoftwo}%
\providecommand \bibfield  [0]{\@secondoftwo}%
\providecommand \translation [1]{[#1]}%
\providecommand \BibitemOpen [0]{}%
\providecommand \bibitemStop [0]{}%
\providecommand \bibitemNoStop [0]{.\EOS\space}%
\providecommand \EOS [0]{\spacefactor3000\relax}%
\providecommand \BibitemShut  [1]{\csname bibitem#1\endcsname}%
\let\auto@bib@innerbib\@empty
%</preamble>
\bibitem [{\citenamefont {Wilson}\ and\ \citenamefont
  {Kogut}(1974)}]{Wilson:1973jj}%
  \BibitemOpen
  \bibfield  {author} {\bibinfo {author} {\bibfnamefont {K.~G.}\ \bibnamefont
  {Wilson}}\ and\ \bibinfo {author} {\bibfnamefont {J.~B.}\ \bibnamefont
  {Kogut}},\ }\href {\doibase 10.1016/0370-1573(74)90023-4} {\bibfield
  {journal} {\bibinfo  {journal} {Phys. Rept.}\ }\textbf {\bibinfo {volume}
  {12}},\ \bibinfo {pages} {75} (\bibinfo {year} {1974})}\BibitemShut {NoStop}%
%%CITATION = PRPLC,12,75;%%
\bibitem [{\citenamefont {Machacek}\ and\ \citenamefont
  {Vaughn}(1983)}]{Machacek:1983tz}%
  \BibitemOpen
  \bibfield  {author} {\bibinfo {author} {\bibfnamefont {M.~E.}\ \bibnamefont
  {Machacek}}\ and\ \bibinfo {author} {\bibfnamefont {M.~T.}\ \bibnamefont
  {Vaughn}},\ }\href {\doibase 10.1016/0550-3213(83)90610-7} {\bibfield
  {journal} {\bibinfo  {journal} {Nucl. Phys.}\ }\textbf {\bibinfo {volume}
  {B222}},\ \bibinfo {pages} {83} (\bibinfo {year} {1983})}\BibitemShut
  {NoStop}%
%%CITATION = NUPHA,B222,83;%%
\bibitem [{\citenamefont {Machacek}\ and\ \citenamefont
  {Vaughn}(1984)}]{Machacek:1983fi}%
  \BibitemOpen
  \bibfield  {author} {\bibinfo {author} {\bibfnamefont {M.~E.}\ \bibnamefont
  {Machacek}}\ and\ \bibinfo {author} {\bibfnamefont {M.~T.}\ \bibnamefont
  {Vaughn}},\ }\href {\doibase 10.1016/0550-3213(84)90533-9} {\bibfield
  {journal} {\bibinfo  {journal} {Nucl. Phys.}\ }\textbf {\bibinfo {volume}
  {B236}},\ \bibinfo {pages} {221} (\bibinfo {year} {1984})}\BibitemShut
  {NoStop}%
%%CITATION = NUPHA,B236,221;%%
\bibitem [{\citenamefont {Machacek}\ and\ \citenamefont
  {Vaughn}(1985)}]{Machacek:1984zw}%
  \BibitemOpen
  \bibfield  {author} {\bibinfo {author} {\bibfnamefont {M.~E.}\ \bibnamefont
  {Machacek}}\ and\ \bibinfo {author} {\bibfnamefont {M.~T.}\ \bibnamefont
  {Vaughn}},\ }\href {\doibase 10.1016/0550-3213(85)90040-9} {\bibfield
  {journal} {\bibinfo  {journal} {Nucl. Phys.}\ }\textbf {\bibinfo {volume}
  {B249}},\ \bibinfo {pages} {70} (\bibinfo {year} {1985})}\BibitemShut
  {NoStop}%
%%CITATION = NUPHA,B249,70;%%
\bibitem [{\citenamefont {Jack}\ and\ \citenamefont
  {Osborn}(2014)}]{Jack:2013sha}%
  \BibitemOpen
  \bibfield  {author} {\bibinfo {author} {\bibfnamefont {I.}~\bibnamefont
  {Jack}}\ and\ \bibinfo {author} {\bibfnamefont {H.}~\bibnamefont {Osborn}},\
  }\href {\doibase 10.1016/j.nuclphysb.2014.03.018} {\bibfield  {journal}
  {\bibinfo  {journal} {Nucl. Phys.}\ }\textbf {\bibinfo {volume} {B883}},\
  \bibinfo {pages} {425} (\bibinfo {year} {2014})},\ \Eprint
  {http://arxiv.org/abs/1312.0428} {arXiv:1312.0428 [hep-th]} \BibitemShut
  {NoStop}%
%%CITATION = ARXIV:1312.0428;%%
\bibitem [{\citenamefont {Jack}\ and\ \citenamefont
  {Poole}(2015)}]{Jack:2014pua}%
  \BibitemOpen
  \bibfield  {author} {\bibinfo {author} {\bibfnamefont {I.}~\bibnamefont
  {Jack}}\ and\ \bibinfo {author} {\bibfnamefont {C.}~\bibnamefont {Poole}},\
  }\href {\doibase 10.1007/JHEP01(2015)138} {\bibfield  {journal} {\bibinfo
  {journal} {JHEP}\ }\textbf {\bibinfo {volume} {01}},\ \bibinfo {pages} {138}
  (\bibinfo {year} {2015})},\ \Eprint {http://arxiv.org/abs/1411.1301}
  {arXiv:1411.1301 [hep-th]} \BibitemShut {NoStop}%
%%CITATION = ARXIV:1411.1301;%%
\bibitem [{\citenamefont {Antipin}\ \emph {et~al.}(2013)\citenamefont
  {Antipin}, \citenamefont {Gillioz}, \citenamefont {Krog}, \citenamefont
  {Mølgaard},\ and\ \citenamefont {Sannino}}]{Antipin:2013sga}%
  \BibitemOpen
  \bibfield  {author} {\bibinfo {author} {\bibfnamefont {O.}~\bibnamefont
  {Antipin}}, \bibinfo {author} {\bibfnamefont {M.}~\bibnamefont {Gillioz}},
  \bibinfo {author} {\bibfnamefont {J.}~\bibnamefont {Krog}}, \bibinfo {author}
  {\bibfnamefont {E.}~\bibnamefont {Mølgaard}}, \ and\ \bibinfo {author}
  {\bibfnamefont {F.}~\bibnamefont {Sannino}},\ }\href {\doibase
  10.1007/JHEP08(2013)034} {\bibfield  {journal} {\bibinfo  {journal} {JHEP}\
  }\textbf {\bibinfo {volume} {08}},\ \bibinfo {pages} {034} (\bibinfo {year}
  {2013})},\ \Eprint {http://arxiv.org/abs/1306.3234} {arXiv:1306.3234
  [hep-ph]} \BibitemShut {NoStop}%
%%CITATION = ARXIV:1306.3234;%%
\bibitem [{\citenamefont {Bednyakov}\ and\ \citenamefont
  {Pikelner}(2016)}]{Bednyakov:2015ooa}%
  \BibitemOpen
  \bibfield  {author} {\bibinfo {author} {\bibfnamefont {A.~V.}\ \bibnamefont
  {Bednyakov}}\ and\ \bibinfo {author} {\bibfnamefont {A.~F.}\ \bibnamefont
  {Pikelner}},\ }\href {\doibase 10.1016/j.physletb.2016.09.007} {\bibfield
  {journal} {\bibinfo  {journal} {Phys. Lett.}\ }\textbf {\bibinfo {volume}
  {B762}},\ \bibinfo {pages} {151} (\bibinfo {year} {2016})},\ \Eprint
  {http://arxiv.org/abs/1508.02680} {arXiv:1508.02680 [hep-ph]} \BibitemShut
  {NoStop}%
%%CITATION = ARXIV:1508.02680;%%
\bibitem [{\citenamefont {Zoller}(2016)}]{Zoller:2015tha}%
  \BibitemOpen
  \bibfield  {author} {\bibinfo {author} {\bibfnamefont {M.~F.}\ \bibnamefont
  {Zoller}},\ }\href {\doibase 10.1007/JHEP02(2016)095} {\bibfield  {journal}
  {\bibinfo  {journal} {JHEP}\ }\textbf {\bibinfo {volume} {02}},\ \bibinfo
  {pages} {095} (\bibinfo {year} {2016})},\ \Eprint
  {http://arxiv.org/abs/1508.03624} {arXiv:1508.03624 [hep-ph]} \BibitemShut
  {NoStop}%
%%CITATION = ARXIV:1508.03624;%%
\bibitem [{\citenamefont {Poole}\ and\ \citenamefont
  {Thomsen}(2019)}]{Poole:2019txl}%
  \BibitemOpen
  \bibfield  {author} {\bibinfo {author} {\bibfnamefont {C.}~\bibnamefont
  {Poole}}\ and\ \bibinfo {author} {\bibfnamefont {A.~E.}\ \bibnamefont
  {Thomsen}},\ }\href@noop {} {\  (\bibinfo {year} {2019})},\ \Eprint
  {http://arxiv.org/abs/1901.02749} {arXiv:1901.02749 [hep-th]} \BibitemShut
  {NoStop}%
%%CITATION = ARXIV:1901.02749;%%
\bibitem [{\citenamefont {Wallace}\ and\ \citenamefont
  {Zia}(1974)}]{Wallace:1974dx}%
  \BibitemOpen
  \bibfield  {author} {\bibinfo {author} {\bibfnamefont {D.~J.}\ \bibnamefont
  {Wallace}}\ and\ \bibinfo {author} {\bibfnamefont {R.~K.~P.}\ \bibnamefont
  {Zia}},\ }\href {\doibase 10.1016/0375-9601(74)90449-6} {\bibfield  {journal}
  {\bibinfo  {journal} {Phys. Lett.}\ }\textbf {\bibinfo {volume} {A48}},\
  \bibinfo {pages} {325} (\bibinfo {year} {1974})}\BibitemShut {NoStop}%
%%CITATION = PHLTA,A48,325;%%
\bibitem [{\citenamefont {Wallace}\ and\ \citenamefont
  {Zia}(1975)}]{Wallace:1974dy}%
  \BibitemOpen
  \bibfield  {author} {\bibinfo {author} {\bibfnamefont {D.~J.}\ \bibnamefont
  {Wallace}}\ and\ \bibinfo {author} {\bibfnamefont {R.~K.~P.}\ \bibnamefont
  {Zia}},\ }\href {\doibase 10.1016/0003-4916(75)90267-5} {\bibfield  {journal}
  {\bibinfo  {journal} {Annals Phys.}\ }\textbf {\bibinfo {volume} {92}},\
  \bibinfo {pages} {142} (\bibinfo {year} {1975})}\BibitemShut {NoStop}%
%%CITATION = APNYA,92,142;%%
\bibitem [{\citenamefont {Osborn}(1989)}]{Osborn:1989td}%
  \BibitemOpen
  \bibfield  {author} {\bibinfo {author} {\bibfnamefont {H.}~\bibnamefont
  {Osborn}},\ }\href {\doibase 10.1016/0370-2693(89)90729-6} {\bibfield
  {journal} {\bibinfo  {journal} {Phys. Lett.}\ }\textbf {\bibinfo {volume}
  {B222}},\ \bibinfo {pages} {97} (\bibinfo {year} {1989})}\BibitemShut
  {NoStop}%
%%CITATION = PHLTA,B222,97;%%
\bibitem [{\citenamefont {Jack}\ and\ \citenamefont
  {Osborn}(1990)}]{Jack:1990eb}%
  \BibitemOpen
  \bibfield  {author} {\bibinfo {author} {\bibfnamefont {I.}~\bibnamefont
  {Jack}}\ and\ \bibinfo {author} {\bibfnamefont {H.}~\bibnamefont {Osborn}},\
  }\href {\doibase 10.1016/0550-3213(90)90584-Z} {\bibfield  {journal}
  {\bibinfo  {journal} {Nucl. Phys.}\ }\textbf {\bibinfo {volume} {B343}},\
  \bibinfo {pages} {647} (\bibinfo {year} {1990})}\BibitemShut {NoStop}%
%%CITATION = NUPHA,B343,647;%%
\bibitem [{\citenamefont {Osborn}(1991)}]{Osborn:1991gm}%
  \BibitemOpen
  \bibfield  {author} {\bibinfo {author} {\bibfnamefont {H.}~\bibnamefont
  {Osborn}},\ }\href {\doibase 10.1016/0550-3213(91)80030-P} {\bibfield
  {journal} {\bibinfo  {journal} {Nucl. Phys.}\ }\textbf {\bibinfo {volume}
  {B363}},\ \bibinfo {pages} {486} (\bibinfo {year} {1991})}\BibitemShut
  {NoStop}%
%%CITATION = NUPHA,B363,486;%%
\bibitem [{\citenamefont {Zamolodchikov}(1986)}]{Zamolodchikov:1986gt}%
  \BibitemOpen
  \bibfield  {author} {\bibinfo {author} {\bibfnamefont {A.~B.}\ \bibnamefont
  {Zamolodchikov}},\ }\href@noop {} {\bibfield  {journal} {\bibinfo  {journal}
  {JETP Lett.}\ }\textbf {\bibinfo {volume} {43}},\ \bibinfo {pages} {730}
  (\bibinfo {year} {1986})},\ \bibinfo {note} {[Pisma Zh. Eksp. Teor.
  Fiz.43,565(1986)]}\BibitemShut {NoStop}%
%%CITATION = JTPLA,43,730;%%
\bibitem [{\citenamefont {Cardy}(1988)}]{Cardy:1988cwa}%
  \BibitemOpen
  \bibfield  {author} {\bibinfo {author} {\bibfnamefont {J.~L.}\ \bibnamefont
  {Cardy}},\ }\href {\doibase 10.1016/0370-2693(88)90054-8} {\bibfield
  {journal} {\bibinfo  {journal} {Phys. Lett.}\ }\textbf {\bibinfo {volume}
  {B215}},\ \bibinfo {pages} {749} (\bibinfo {year} {1988})}\BibitemShut
  {NoStop}%
%%CITATION = PHLTA,B215,749;%%
\bibitem [{\citenamefont {Komargodski}\ and\ \citenamefont
  {Schwimmer}(2011)}]{Komargodski:2011vj}%
  \BibitemOpen
  \bibfield  {author} {\bibinfo {author} {\bibfnamefont {Z.}~\bibnamefont
  {Komargodski}}\ and\ \bibinfo {author} {\bibfnamefont {A.}~\bibnamefont
  {Schwimmer}},\ }\href {\doibase 10.1007/JHEP12(2011)099} {\bibfield
  {journal} {\bibinfo  {journal} {JHEP}\ }\textbf {\bibinfo {volume} {12}},\
  \bibinfo {pages} {099} (\bibinfo {year} {2011})},\ \Eprint
  {http://arxiv.org/abs/1107.3987} {arXiv:1107.3987 [hep-th]} \BibitemShut
  {NoStop}%
%%CITATION = ARXIV:1107.3987;%%
\bibitem [{\citenamefont {Shore}(2017)}]{Shore:2016xor}%
  \BibitemOpen
  \bibfield  {author} {\bibinfo {author} {\bibfnamefont {G.~M.}\ \bibnamefont
  {Shore}},\ }\href {\doibase 10.1007/978-3-319-54000-9} {\emph {\bibinfo
  {title} {{The c and a-theorems and the Local Renormalisation Group}}}},\
  SpringerBriefs in Physics\ (\bibinfo  {publisher} {Springer},\ \bibinfo
  {address} {Cham},\ \bibinfo {year} {2017})\ \Eprint
  {http://arxiv.org/abs/1601.06662} {arXiv:1601.06662 [hep-th]} \BibitemShut
  {NoStop}%
%%CITATION = ARXIV:1601.06662;%%
\bibitem [{\citenamefont {Fortin}\ \emph {et~al.}(2013)\citenamefont {Fortin},
  \citenamefont {Grinstein},\ and\ \citenamefont {Stergiou}}]{Fortin:2012hn}%
  \BibitemOpen
  \bibfield  {author} {\bibinfo {author} {\bibfnamefont {J.-F.}\ \bibnamefont
  {Fortin}}, \bibinfo {author} {\bibfnamefont {B.}~\bibnamefont {Grinstein}}, \
  and\ \bibinfo {author} {\bibfnamefont {A.}~\bibnamefont {Stergiou}},\ }\href
  {\doibase 10.1007/JHEP01(2013)184} {\bibfield  {journal} {\bibinfo  {journal}
  {JHEP}\ }\textbf {\bibinfo {volume} {01}},\ \bibinfo {pages} {184} (\bibinfo
  {year} {2013})},\ \Eprint {http://arxiv.org/abs/1208.3674} {arXiv:1208.3674
  [hep-th]} \BibitemShut {NoStop}%
%%CITATION = ARXIV:1208.3674;%%
\bibitem [{\citenamefont {Gracey}\ \emph {et~al.}(2016)\citenamefont {Gracey},
  \citenamefont {Jack},\ and\ \citenamefont {Poole}}]{Gracey:2015fia}%
  \BibitemOpen
  \bibfield  {author} {\bibinfo {author} {\bibfnamefont {J.~A.}\ \bibnamefont
  {Gracey}}, \bibinfo {author} {\bibfnamefont {I.}~\bibnamefont {Jack}}, \ and\
  \bibinfo {author} {\bibfnamefont {C.}~\bibnamefont {Poole}},\ }\href
  {\doibase 10.1007/JHEP01(2016)174} {\bibfield  {journal} {\bibinfo  {journal}
  {JHEP}\ }\textbf {\bibinfo {volume} {01}},\ \bibinfo {pages} {174} (\bibinfo
  {year} {2016})},\ \Eprint {http://arxiv.org/abs/1507.02174} {arXiv:1507.02174
  [hep-th]} \BibitemShut {NoStop}%
%%CITATION = ARXIV:1507.02174;%%
\bibitem [{\citenamefont {Jack}\ and\ \citenamefont
  {Poole}(2017)}]{Jack:2016utw}%
  \BibitemOpen
  \bibfield  {author} {\bibinfo {author} {\bibfnamefont {I.}~\bibnamefont
  {Jack}}\ and\ \bibinfo {author} {\bibfnamefont {C.}~\bibnamefont {Poole}},\
  }\href {\doibase 10.1103/PhysRevD.95.025010} {\bibfield  {journal} {\bibinfo
  {journal} {Phys. Rev.}\ }\textbf {\bibinfo {volume} {D95}},\ \bibinfo {pages}
  {025010} (\bibinfo {year} {2017})},\ \Eprint
  {http://arxiv.org/abs/1607.00236} {arXiv:1607.00236 [hep-th]} \BibitemShut
  {NoStop}%
%%CITATION = ARXIV:1607.00236;%%
\bibitem [{\citenamefont {Gracey}\ \emph {et~al.}(2017)\citenamefont {Gracey},
  \citenamefont {Jack}, \citenamefont {Poole},\ and\ \citenamefont
  {Schröder}}]{Gracey:2016tuh}%
  \BibitemOpen
  \bibfield  {author} {\bibinfo {author} {\bibfnamefont {J.~A.}\ \bibnamefont
  {Gracey}}, \bibinfo {author} {\bibfnamefont {I.}~\bibnamefont {Jack}},
  \bibinfo {author} {\bibfnamefont {C.}~\bibnamefont {Poole}}, \ and\ \bibinfo
  {author} {\bibfnamefont {Y.}~\bibnamefont {Schröder}},\ }\href {\doibase
  10.1103/PhysRevD.95.025005} {\bibfield  {journal} {\bibinfo  {journal} {Phys.
  Rev.}\ }\textbf {\bibinfo {volume} {D95}},\ \bibinfo {pages} {025005}
  (\bibinfo {year} {2017})},\ \Eprint {http://arxiv.org/abs/1609.06458}
  {arXiv:1609.06458 [hep-th]} \BibitemShut {NoStop}%
%%CITATION = ARXIV:1609.06458;%%
\bibitem [{\citenamefont {Gross}\ and\ \citenamefont
  {Wilczek}(1973)}]{Gross:1973id}%
  \BibitemOpen
  \bibfield  {author} {\bibinfo {author} {\bibfnamefont {D.~J.}\ \bibnamefont
  {Gross}}\ and\ \bibinfo {author} {\bibfnamefont {F.}~\bibnamefont
  {Wilczek}},\ }\href {\doibase 10.1103/PhysRevLett.30.1343} {\bibfield
  {journal} {\bibinfo  {journal} {Phys. Rev. Lett.}\ }\textbf {\bibinfo
  {volume} {30}},\ \bibinfo {pages} {1343} (\bibinfo {year} {1973})},\ \bibinfo
  {note} {[,271(1973)]}\BibitemShut {NoStop}%
%%CITATION = PRLTA,30,1343;%%
\bibitem [{\citenamefont {Politzer}(1973)}]{Politzer:1973fx}%
  \BibitemOpen
  \bibfield  {author} {\bibinfo {author} {\bibfnamefont {H.~D.}\ \bibnamefont
  {Politzer}},\ }\href {\doibase 10.1103/PhysRevLett.30.1346} {\bibfield
  {journal} {\bibinfo  {journal} {Phys. Rev. Lett.}\ }\textbf {\bibinfo
  {volume} {30}},\ \bibinfo {pages} {1346} (\bibinfo {year} {1973})},\ \bibinfo
  {note} {[,274(1973)]}\BibitemShut {NoStop}%
%%CITATION = PRLTA,30,1346;%%
\bibitem [{\citenamefont {Litim}\ and\ \citenamefont
  {Sannino}(2014)}]{Litim:2014uca}%
  \BibitemOpen
  \bibfield  {author} {\bibinfo {author} {\bibfnamefont {D.~F.}\ \bibnamefont
  {Litim}}\ and\ \bibinfo {author} {\bibfnamefont {F.}~\bibnamefont
  {Sannino}},\ }\href {\doibase 10.1007/JHEP12(2014)178} {\bibfield  {journal}
  {\bibinfo  {journal} {JHEP}\ }\textbf {\bibinfo {volume} {12}},\ \bibinfo
  {pages} {178} (\bibinfo {year} {2014})},\ \Eprint
  {http://arxiv.org/abs/1406.2337} {arXiv:1406.2337 [hep-th]} \BibitemShut
  {NoStop}%
%%CITATION = ARXIV:1406.2337;%%
\bibitem [{\citenamefont {Herzog}\ \emph {et~al.}(2017)\citenamefont {Herzog},
  \citenamefont {Ruijl}, \citenamefont {Ueda}, \citenamefont {Vermaseren},\
  and\ \citenamefont {Vogt}}]{Herzog:2017ohr}%
  \BibitemOpen
  \bibfield  {author} {\bibinfo {author} {\bibfnamefont {F.}~\bibnamefont
  {Herzog}}, \bibinfo {author} {\bibfnamefont {B.}~\bibnamefont {Ruijl}},
  \bibinfo {author} {\bibfnamefont {T.}~\bibnamefont {Ueda}}, \bibinfo {author}
  {\bibfnamefont {J.~A.~M.}\ \bibnamefont {Vermaseren}}, \ and\ \bibinfo
  {author} {\bibfnamefont {A.}~\bibnamefont {Vogt}},\ }\href {\doibase
  10.1007/JHEP02(2017)090} {\bibfield  {journal} {\bibinfo  {journal} {JHEP}\
  }\textbf {\bibinfo {volume} {02}},\ \bibinfo {pages} {090} (\bibinfo {year}
  {2017})},\ \Eprint {http://arxiv.org/abs/1701.01404} {arXiv:1701.01404
  [hep-ph]} \BibitemShut {NoStop}%
%%CITATION = ARXIV:1701.01404;%%
\bibitem [{\citenamefont {Mihaila}\ \emph {et~al.}(2012)\citenamefont
  {Mihaila}, \citenamefont {Salomon},\ and\ \citenamefont
  {Steinhauser}}]{Mihaila:2012pz}%
  \BibitemOpen
  \bibfield  {author} {\bibinfo {author} {\bibfnamefont {L.~N.}\ \bibnamefont
  {Mihaila}}, \bibinfo {author} {\bibfnamefont {J.}~\bibnamefont {Salomon}}, \
  and\ \bibinfo {author} {\bibfnamefont {M.}~\bibnamefont {Steinhauser}},\
  }\href {\doibase 10.1103/PhysRevD.86.096008} {\bibfield  {journal} {\bibinfo
  {journal} {Phys. Rev.}\ }\textbf {\bibinfo {volume} {D86}},\ \bibinfo {pages}
  {096008} (\bibinfo {year} {2012})},\ \Eprint {http://arxiv.org/abs/1208.3357}
  {arXiv:1208.3357 [hep-ph]} \BibitemShut {NoStop}%
%%CITATION = ARXIV:1208.3357;%%
\bibitem [{\citenamefont {Bednyakov}\ \emph
  {et~al.}(2014{\natexlab{a}})\citenamefont {Bednyakov}, \citenamefont
  {Pikelner},\ and\ \citenamefont {Velizhanin}}]{Bednyakov:2013cpa}%
  \BibitemOpen
  \bibfield  {author} {\bibinfo {author} {\bibfnamefont {A.~V.}\ \bibnamefont
  {Bednyakov}}, \bibinfo {author} {\bibfnamefont {A.~F.}\ \bibnamefont
  {Pikelner}}, \ and\ \bibinfo {author} {\bibfnamefont {V.~N.}\ \bibnamefont
  {Velizhanin}},\ }\href {\doibase 10.1016/j.nuclphysb.2013.12.012} {\bibfield
  {journal} {\bibinfo  {journal} {Nucl. Phys.}\ }\textbf {\bibinfo {volume}
  {B879}},\ \bibinfo {pages} {256} (\bibinfo {year} {2014}{\natexlab{a}})},\
  \Eprint {http://arxiv.org/abs/1310.3806} {arXiv:1310.3806 [hep-ph]}
  \BibitemShut {NoStop}%
%%CITATION = ARXIV:1310.3806;%%
\bibitem [{\citenamefont {Bednyakov}\ \emph
  {et~al.}(2014{\natexlab{b}})\citenamefont {Bednyakov}, \citenamefont
  {Pikelner},\ and\ \citenamefont {Velizhanin}}]{Bednyakov:2014pia}%
  \BibitemOpen
  \bibfield  {author} {\bibinfo {author} {\bibfnamefont {A.~V.}\ \bibnamefont
  {Bednyakov}}, \bibinfo {author} {\bibfnamefont {A.~F.}\ \bibnamefont
  {Pikelner}}, \ and\ \bibinfo {author} {\bibfnamefont {V.~N.}\ \bibnamefont
  {Velizhanin}},\ }\href {\doibase 10.1016/j.physletb.2014.08.049} {\bibfield
  {journal} {\bibinfo  {journal} {Phys. Lett.}\ }\textbf {\bibinfo {volume}
  {B737}},\ \bibinfo {pages} {129} (\bibinfo {year} {2014}{\natexlab{b}})},\
  \Eprint {http://arxiv.org/abs/1406.7171} {arXiv:1406.7171 [hep-ph]}
  \BibitemShut {NoStop}%
%%CITATION = ARXIV:1406.7171;%%
\bibitem [{\citenamefont {Luo}\ and\ \citenamefont {Xiao}(2003)}]{Luo:2002iq}%
  \BibitemOpen
  \bibfield  {author} {\bibinfo {author} {\bibfnamefont {M.-x.}\ \bibnamefont
  {Luo}}\ and\ \bibinfo {author} {\bibfnamefont {Y.}~\bibnamefont {Xiao}},\
  }\href {\doibase 10.1016/S0370-2693(03)00076-5} {\bibfield  {journal}
  {\bibinfo  {journal} {Phys. Lett.}\ }\textbf {\bibinfo {volume} {B555}},\
  \bibinfo {pages} {279} (\bibinfo {year} {2003})},\ \Eprint
  {http://arxiv.org/abs/hep-ph/0212152} {arXiv:hep-ph/0212152 [hep-ph]}
  \BibitemShut {NoStop}%
%%CITATION = HEP-PH/0212152;%%
\bibitem [{\citenamefont {Pickering}\ \emph {et~al.}(2001)\citenamefont
  {Pickering}, \citenamefont {Gracey},\ and\ \citenamefont
  {Jones}}]{Pickering:2001aq}%
  \BibitemOpen
  \bibfield  {author} {\bibinfo {author} {\bibfnamefont {A.~G.~M.}\
  \bibnamefont {Pickering}}, \bibinfo {author} {\bibfnamefont {J.~A.}\
  \bibnamefont {Gracey}}, \ and\ \bibinfo {author} {\bibfnamefont {D.~R.~T.}\
  \bibnamefont {Jones}},\ }\href {\doibase 10.1016/S0370-2693(02)01779-3,
  10.1016/S0370-2693(01)00624-4} {\bibfield  {journal} {\bibinfo  {journal}
  {Phys. Lett.}\ }\textbf {\bibinfo {volume} {B510}},\ \bibinfo {pages} {347}
  (\bibinfo {year} {2001})},\ \bibinfo {note} {[Erratum: Phys.
  Lett.B535,377(2002)]},\ \Eprint {http://arxiv.org/abs/hep-ph/0104247}
  {arXiv:hep-ph/0104247 [hep-ph]} \BibitemShut {NoStop}%
%%CITATION = HEP-PH/0104247;%%
\bibitem [{\citenamefont {Nakayama}(2015)}]{Nakayama:2013is}%
  \BibitemOpen
  \bibfield  {author} {\bibinfo {author} {\bibfnamefont {Y.}~\bibnamefont
  {Nakayama}},\ }\href {\doibase 10.1016/j.physrep.2014.12.003} {\bibfield
  {journal} {\bibinfo  {journal} {Phys. Rept.}\ }\textbf {\bibinfo {volume}
  {569}},\ \bibinfo {pages} {1} (\bibinfo {year} {2015})},\ \Eprint
  {http://arxiv.org/abs/1302.0884} {arXiv:1302.0884 [hep-th]} \BibitemShut
  {NoStop}%
%%CITATION = ARXIV:1302.0884;%%
\bibitem [{\citenamefont {Grinstein}\ \emph {et~al.}(2013)\citenamefont
  {Grinstein}, \citenamefont {Stergiou},\ and\ \citenamefont
  {Stone}}]{Grinstein:2013cka}%
  \BibitemOpen
  \bibfield  {author} {\bibinfo {author} {\bibfnamefont {B.}~\bibnamefont
  {Grinstein}}, \bibinfo {author} {\bibfnamefont {A.}~\bibnamefont {Stergiou}},
  \ and\ \bibinfo {author} {\bibfnamefont {D.}~\bibnamefont {Stone}},\ }\href
  {\doibase 10.1007/JHEP11(2013)195} {\bibfield  {journal} {\bibinfo  {journal}
  {JHEP}\ }\textbf {\bibinfo {volume} {11}},\ \bibinfo {pages} {195} (\bibinfo
  {year} {2013})},\ \Eprint {http://arxiv.org/abs/1308.1096} {arXiv:1308.1096
  [hep-th]} \BibitemShut {NoStop}%
%%CITATION = ARXIV:1308.1096;%%
\bibitem [{\citenamefont {Stergiou}\ \emph {et~al.}(2016)\citenamefont
  {Stergiou}, \citenamefont {Stone},\ and\ \citenamefont
  {Vitale}}]{Stergiou:2016uqq}%
  \BibitemOpen
  \bibfield  {author} {\bibinfo {author} {\bibfnamefont {A.}~\bibnamefont
  {Stergiou}}, \bibinfo {author} {\bibfnamefont {D.}~\bibnamefont {Stone}}, \
  and\ \bibinfo {author} {\bibfnamefont {L.~G.}\ \bibnamefont {Vitale}},\
  }\href {\doibase 10.1007/JHEP08(2016)010} {\bibfield  {journal} {\bibinfo
  {journal} {JHEP}\ }\textbf {\bibinfo {volume} {08}},\ \bibinfo {pages} {010}
  (\bibinfo {year} {2016})},\ \Eprint {http://arxiv.org/abs/1604.01782}
  {arXiv:1604.01782 [hep-th]} \BibitemShut {NoStop}%
%%CITATION = ARXIV:1604.01782;%%
\bibitem [{\citenamefont {Bonora}\ \emph {et~al.}(1983)\citenamefont {Bonora},
  \citenamefont {Cotta-Ramusino},\ and\ \citenamefont {Reina}}]{Bonora:1983ff}%
  \BibitemOpen
  \bibfield  {author} {\bibinfo {author} {\bibfnamefont {L.}~\bibnamefont
  {Bonora}}, \bibinfo {author} {\bibfnamefont {P.}~\bibnamefont
  {Cotta-Ramusino}}, \ and\ \bibinfo {author} {\bibfnamefont {C.}~\bibnamefont
  {Reina}},\ }\href {\doibase 10.1016/0370-2693(83)90169-7} {\bibfield
  {journal} {\bibinfo  {journal} {Phys. Lett.}\ }\textbf {\bibinfo {volume}
  {126B}},\ \bibinfo {pages} {305} (\bibinfo {year} {1983})}\BibitemShut
  {NoStop}%
%%CITATION = PHLTA,126B,305;%%
\bibitem [{\citenamefont {Nakayama}(2012)}]{Nakayama:2012gu}%
  \BibitemOpen
  \bibfield  {author} {\bibinfo {author} {\bibfnamefont {Y.}~\bibnamefont
  {Nakayama}},\ }\href {\doibase 10.1016/j.nuclphysb.2012.02.006} {\bibfield
  {journal} {\bibinfo  {journal} {Nucl. Phys.}\ }\textbf {\bibinfo {volume}
  {B859}},\ \bibinfo {pages} {288} (\bibinfo {year} {2012})},\ \Eprint
  {http://arxiv.org/abs/1201.3428} {arXiv:1201.3428 [hep-th]} \BibitemShut
  {NoStop}%
%%CITATION = ARXIV:1201.3428;%%
\bibitem [{\citenamefont {Bonora}\ \emph {et~al.}(2017)\citenamefont {Bonora},
  \citenamefont {Cvitan}, \citenamefont {Dominis~Prester}, \citenamefont
  {Duarte~Pereira}, \citenamefont {Giaccari},\ and\ \citenamefont
  {Štemberga}}]{Bonora:2017gzz}%
  \BibitemOpen
  \bibfield  {author} {\bibinfo {author} {\bibfnamefont {L.}~\bibnamefont
  {Bonora}}, \bibinfo {author} {\bibfnamefont {M.}~\bibnamefont {Cvitan}},
  \bibinfo {author} {\bibfnamefont {P.}~\bibnamefont {Dominis~Prester}},
  \bibinfo {author} {\bibfnamefont {A.}~\bibnamefont {Duarte~Pereira}},
  \bibinfo {author} {\bibfnamefont {S.}~\bibnamefont {Giaccari}}, \ and\
  \bibinfo {author} {\bibfnamefont {T.}~\bibnamefont {Štemberga}},\ }\href
  {\doibase 10.1140/epjc/s10052-017-5071-7} {\bibfield  {journal} {\bibinfo
  {journal} {Eur. Phys. J.}\ }\textbf {\bibinfo {volume} {C77}},\ \bibinfo
  {pages} {511} (\bibinfo {year} {2017})},\ \Eprint
  {http://arxiv.org/abs/1703.10473} {arXiv:1703.10473 [hep-th]} \BibitemShut
  {NoStop}%
%%CITATION = ARXIV:1703.10473;%%
\bibitem [{\citenamefont {Bastianelli}\ and\ \citenamefont
  {Broccoli}(2019)}]{Bastianelli:2018osv}%
  \BibitemOpen
  \bibfield  {author} {\bibinfo {author} {\bibfnamefont {F.}~\bibnamefont
  {Bastianelli}}\ and\ \bibinfo {author} {\bibfnamefont {M.}~\bibnamefont
  {Broccoli}},\ }\href {\doibase 10.1140/epjc/s10052-019-6799-z} {\bibfield
  {journal} {\bibinfo  {journal} {Eur. Phys. J.}\ }\textbf {\bibinfo {volume}
  {C79}},\ \bibinfo {pages} {292} (\bibinfo {year} {2019})},\ \Eprint
  {http://arxiv.org/abs/1808.03489} {arXiv:1808.03489 [hep-th]} \BibitemShut
  {NoStop}%
%%CITATION = ARXIV:1808.03489;%%
\bibitem [{\citenamefont {Keren-Zur}(2014)}]{Keren-Zur:2014sva}%
  \BibitemOpen
  \bibfield  {author} {\bibinfo {author} {\bibfnamefont {B.}~\bibnamefont
  {Keren-Zur}},\ }\href {\doibase 10.1007/JHEP09(2014)011} {\bibfield
  {journal} {\bibinfo  {journal} {JHEP}\ }\textbf {\bibinfo {volume} {09}},\
  \bibinfo {pages} {011} (\bibinfo {year} {2014})},\ \Eprint
  {http://arxiv.org/abs/1406.0869} {arXiv:1406.0869 [hep-th]} \BibitemShut
  {NoStop}%
%%CITATION = ARXIV:1406.0869;%%
\bibitem [{\citenamefont {Mølgaard}(2014)}]{Molgaard:2014hpa}%
  \BibitemOpen
  \bibfield  {author} {\bibinfo {author} {\bibfnamefont {E.}~\bibnamefont
  {Mølgaard}},\ }\href {\doibase 10.1140/epjp/i2014-14159-2} {\bibfield
  {journal} {\bibinfo  {journal} {Eur. Phys. J. Plus}\ }\textbf {\bibinfo
  {volume} {129}},\ \bibinfo {pages} {159} (\bibinfo {year} {2014})},\ \Eprint
  {http://arxiv.org/abs/1404.5550} {arXiv:1404.5550 [hep-th]} \BibitemShut
  {NoStop}%
%%CITATION = ARXIV:1404.5550;%%
\bibitem [{\citenamefont {Dreiner}\ \emph {et~al.}(2010)\citenamefont
  {Dreiner}, \citenamefont {Haber},\ and\ \citenamefont
  {Martin}}]{Dreiner:2008tw}%
  \BibitemOpen
  \bibfield  {author} {\bibinfo {author} {\bibfnamefont {H.~K.}\ \bibnamefont
  {Dreiner}}, \bibinfo {author} {\bibfnamefont {H.~E.}\ \bibnamefont {Haber}},
  \ and\ \bibinfo {author} {\bibfnamefont {S.~P.}\ \bibnamefont {Martin}},\
  }\href {\doibase 10.1016/j.physrep.2010.05.002} {\bibfield  {journal}
  {\bibinfo  {journal} {Phys. Rept.}\ }\textbf {\bibinfo {volume} {494}},\
  \bibinfo {pages} {1} (\bibinfo {year} {2010})},\ \Eprint
  {http://arxiv.org/abs/0812.1594} {arXiv:0812.1594 [hep-ph]} \BibitemShut
  {NoStop}%
%%CITATION = ARXIV:0812.1594;%%
\bibitem [{\citenamefont {Holdom}(1986)}]{Holdom:1985ag}%
  \BibitemOpen
  \bibfield  {author} {\bibinfo {author} {\bibfnamefont {B.}~\bibnamefont
  {Holdom}},\ }\href {\doibase 10.1016/0370-2693(86)91377-8} {\bibfield
  {journal} {\bibinfo  {journal} {Phys. Lett.}\ }\textbf {\bibinfo {volume}
  {166B}},\ \bibinfo {pages} {196} (\bibinfo {year} {1986})}\BibitemShut
  {NoStop}%
%%CITATION = PHLTA,166B,196;%%
\bibitem [{\citenamefont {Fonseca}\ \emph {et~al.}(2013)\citenamefont
  {Fonseca}, \citenamefont {Malinský},\ and\ \citenamefont
  {Staub}}]{Fonseca:2013bua}%
  \BibitemOpen
  \bibfield  {author} {\bibinfo {author} {\bibfnamefont {R.~M.}\ \bibnamefont
  {Fonseca}}, \bibinfo {author} {\bibfnamefont {M.}~\bibnamefont {Malinský}},
  \ and\ \bibinfo {author} {\bibfnamefont {F.}~\bibnamefont {Staub}},\ }\href
  {\doibase 10.1016/j.physletb.2013.09.042} {\bibfield  {journal} {\bibinfo
  {journal} {Phys. Lett.}\ }\textbf {\bibinfo {volume} {B726}},\ \bibinfo
  {pages} {882} (\bibinfo {year} {2013})},\ \Eprint
  {http://arxiv.org/abs/1308.1674} {arXiv:1308.1674 [hep-ph]} \BibitemShut
  {NoStop}%
%%CITATION = ARXIV:1308.1674;%%
\bibitem [{\citenamefont {Mølgaard}()}]{Molgaard:xxx}%
  \BibitemOpen
  \bibfield  {author} {\bibinfo {author} {\bibfnamefont {E.}~\bibnamefont
  {Mølgaard}},\ }\href@noop {} {}\bibinfo {howpublished} {{private
  communication}}\BibitemShut {NoStop}%
\bibitem [{\citenamefont {Cheng}\ \emph {et~al.}(1974)\citenamefont {Cheng},
  \citenamefont {Eichten},\ and\ \citenamefont {Li}}]{Cheng:1973nv}%
  \BibitemOpen
  \bibfield  {author} {\bibinfo {author} {\bibfnamefont {T.~P.}\ \bibnamefont
  {Cheng}}, \bibinfo {author} {\bibfnamefont {E.}~\bibnamefont {Eichten}}, \
  and\ \bibinfo {author} {\bibfnamefont {L.-F.}\ \bibnamefont {Li}},\ }\href
  {\doibase 10.1103/PhysRevD.9.2259} {\bibfield  {journal} {\bibinfo  {journal}
  {Phys. Rev.}\ }\textbf {\bibinfo {volume} {D9}},\ \bibinfo {pages} {2259}
  (\bibinfo {year} {1974})}\BibitemShut {NoStop}%
%%CITATION = PHRVA,D9,2259;%%
\bibitem [{\citenamefont {Jack}\ and\ \citenamefont
  {Osborn}(1985)}]{Jack:1984vj}%
  \BibitemOpen
  \bibfield  {author} {\bibinfo {author} {\bibfnamefont {I.}~\bibnamefont
  {Jack}}\ and\ \bibinfo {author} {\bibfnamefont {H.}~\bibnamefont {Osborn}},\
  }\href {\doibase 10.1016/0550-3213(85)90088-4} {\bibfield  {journal}
  {\bibinfo  {journal} {Nucl. Phys.}\ }\textbf {\bibinfo {volume} {B249}},\
  \bibinfo {pages} {472} (\bibinfo {year} {1985})}\BibitemShut {NoStop}%
%%CITATION = NUPHA,B249,472;%%
\bibitem [{\citenamefont {DeWitt}(1967)}]{DeWitt:1967ub}%
  \BibitemOpen
  \bibfield  {author} {\bibinfo {author} {\bibfnamefont {B.~S.}\ \bibnamefont
  {DeWitt}},\ }\href {\doibase 10.1103/PhysRev.162.1195} {\bibfield  {journal}
  {\bibinfo  {journal} {Phys. Rev.}\ }\textbf {\bibinfo {volume} {162}},\
  \bibinfo {pages} {1195} (\bibinfo {year} {1967})}\BibitemShut {NoStop}%
%%CITATION = PHRVA,162,1195;%%
\bibitem [{\citenamefont {Abbott}(1982)}]{Abbott:1981ke}%
  \BibitemOpen
  \bibfield  {author} {\bibinfo {author} {\bibfnamefont {L.~F.}\ \bibnamefont
  {Abbott}},\ }\href@noop {} {\bibfield  {journal} {\bibinfo  {journal} {Acta
  Phys. Polon.}\ }\textbf {\bibinfo {volume} {B13}},\ \bibinfo {pages} {33}
  (\bibinfo {year} {1982})}\BibitemShut {NoStop}%
%%CITATION = APPOA,B13,33;%%
\bibitem [{\citenamefont {Jack}\ and\ \citenamefont
  {Osborn}(2016)}]{Jack:2016tpp}%
  \BibitemOpen
  \bibfield  {author} {\bibinfo {author} {\bibfnamefont {I.}~\bibnamefont
  {Jack}}\ and\ \bibinfo {author} {\bibfnamefont {H.}~\bibnamefont {Osborn}},\
  }\href@noop {} {\  (\bibinfo {year} {2016})},\ \Eprint
  {http://arxiv.org/abs/1606.02571} {arXiv:1606.02571 [hep-th]} \BibitemShut
  {NoStop}%
%%CITATION = ARXIV:1606.02571;%%
\bibitem [{\citenamefont {Luo}\ \emph {et~al.}(2003)\citenamefont {Luo},
  \citenamefont {Wang},\ and\ \citenamefont {Xiao}}]{Luo:2002ti}%
  \BibitemOpen
  \bibfield  {author} {\bibinfo {author} {\bibfnamefont {M.-x.}\ \bibnamefont
  {Luo}}, \bibinfo {author} {\bibfnamefont {H.-w.}\ \bibnamefont {Wang}}, \
  and\ \bibinfo {author} {\bibfnamefont {Y.}~\bibnamefont {Xiao}},\ }\href
  {\doibase 10.1103/PhysRevD.67.065019} {\bibfield  {journal} {\bibinfo
  {journal} {Phys. Rev.}\ }\textbf {\bibinfo {volume} {D67}},\ \bibinfo {pages}
  {065019} (\bibinfo {year} {2003})},\ \Eprint
  {http://arxiv.org/abs/hep-ph/0211440} {arXiv:hep-ph/0211440 [hep-ph]}
  \BibitemShut {NoStop}%
%%CITATION = HEP-PH/0211440;%%
\bibitem [{\citenamefont {Herren}\ \emph {et~al.}(2018)\citenamefont {Herren},
  \citenamefont {Mihaila},\ and\ \citenamefont {Steinhauser}}]{Herren:2017uxn}%
  \BibitemOpen
  \bibfield  {author} {\bibinfo {author} {\bibfnamefont {F.}~\bibnamefont
  {Herren}}, \bibinfo {author} {\bibfnamefont {L.}~\bibnamefont {Mihaila}}, \
  and\ \bibinfo {author} {\bibfnamefont {M.}~\bibnamefont {Steinhauser}},\
  }\href {\doibase 10.1103/PhysRevD.97.015016} {\bibfield  {journal} {\bibinfo
  {journal} {Phys. Rev.}\ }\textbf {\bibinfo {volume} {D97}},\ \bibinfo {pages}
  {015016} (\bibinfo {year} {2018})},\ \Eprint
  {http://arxiv.org/abs/1712.06614} {arXiv:1712.06614 [hep-ph]} \BibitemShut
  {NoStop}%
%%CITATION = ARXIV:1712.06614;%%
\bibitem [{\citenamefont {van Ritbergen}\ \emph {et~al.}(1997)\citenamefont
  {van Ritbergen}, \citenamefont {Vermaseren},\ and\ \citenamefont
  {Larin}}]{vanRitbergen:1997va}%
  \BibitemOpen
  \bibfield  {author} {\bibinfo {author} {\bibfnamefont {T.}~\bibnamefont {van
  Ritbergen}}, \bibinfo {author} {\bibfnamefont {J.~A.~M.}\ \bibnamefont
  {Vermaseren}}, \ and\ \bibinfo {author} {\bibfnamefont {S.~A.}\ \bibnamefont
  {Larin}},\ }\href {\doibase 10.1016/S0370-2693(97)00370-5} {\bibfield
  {journal} {\bibinfo  {journal} {Phys. Lett.}\ }\textbf {\bibinfo {volume}
  {B400}},\ \bibinfo {pages} {379} (\bibinfo {year} {1997})},\ \Eprint
  {http://arxiv.org/abs/hep-ph/9701390} {arXiv:hep-ph/9701390 [hep-ph]}
  \BibitemShut {NoStop}%
%%CITATION = HEP-PH/9701390;%%
\bibitem [{\citenamefont {'t~Hooft}\ and\ \citenamefont
  {Veltman}(1972)}]{tHooft:1972tcz}%
  \BibitemOpen
  \bibfield  {author} {\bibinfo {author} {\bibfnamefont {G.}~\bibnamefont
  {'t~Hooft}}\ and\ \bibinfo {author} {\bibfnamefont {M.~J.~G.}\ \bibnamefont
  {Veltman}},\ }\href {\doibase 10.1016/0550-3213(72)90279-9} {\bibfield
  {journal} {\bibinfo  {journal} {Nucl. Phys.}\ }\textbf {\bibinfo {volume}
  {B44}},\ \bibinfo {pages} {189} (\bibinfo {year} {1972})}\BibitemShut
  {NoStop}%
%%CITATION = NUPHA,B44,189;%%
\bibitem [{\citenamefont {Jegerlehner}(2001)}]{Jegerlehner:2000dz}%
  \BibitemOpen
  \bibfield  {author} {\bibinfo {author} {\bibfnamefont {F.}~\bibnamefont
  {Jegerlehner}},\ }\href {\doibase 10.1007/s100520100573} {\bibfield
  {journal} {\bibinfo  {journal} {Eur. Phys. J.}\ }\textbf {\bibinfo {volume}
  {C18}},\ \bibinfo {pages} {673} (\bibinfo {year} {2001})},\ \Eprint
  {http://arxiv.org/abs/hep-th/0005255} {arXiv:hep-th/0005255 [hep-th]}
  \BibitemShut {NoStop}%
%%CITATION = HEP-TH/0005255;%%
\bibitem [{\citenamefont {Chetyrkin}\ and\ \citenamefont
  {Zoller}(2012)}]{Chetyrkin:2012rz}%
  \BibitemOpen
  \bibfield  {author} {\bibinfo {author} {\bibfnamefont {K.~G.}\ \bibnamefont
  {Chetyrkin}}\ and\ \bibinfo {author} {\bibfnamefont {M.~F.}\ \bibnamefont
  {Zoller}},\ }\href {\doibase 10.1007/JHEP06(2012)033} {\bibfield  {journal}
  {\bibinfo  {journal} {JHEP}\ }\textbf {\bibinfo {volume} {06}},\ \bibinfo
  {pages} {033} (\bibinfo {year} {2012})},\ \Eprint
  {http://arxiv.org/abs/1205.2892} {arXiv:1205.2892 [hep-ph]} \BibitemShut
  {NoStop}%
%%CITATION = ARXIV:1205.2892;%%
\bibitem [{\citenamefont {Bednyakov}\ \emph {et~al.}(2013)\citenamefont
  {Bednyakov}, \citenamefont {Pikelner},\ and\ \citenamefont
  {Velizhanin}}]{Bednyakov:2012en}%
  \BibitemOpen
  \bibfield  {author} {\bibinfo {author} {\bibfnamefont {A.~V.}\ \bibnamefont
  {Bednyakov}}, \bibinfo {author} {\bibfnamefont {A.~F.}\ \bibnamefont
  {Pikelner}}, \ and\ \bibinfo {author} {\bibfnamefont {V.~N.}\ \bibnamefont
  {Velizhanin}},\ }\href {\doibase 10.1016/j.physletb.2013.04.038} {\bibfield
  {journal} {\bibinfo  {journal} {Phys. Lett.}\ }\textbf {\bibinfo {volume}
  {B722}},\ \bibinfo {pages} {336} (\bibinfo {year} {2013})},\ \Eprint
  {http://arxiv.org/abs/1212.6829} {arXiv:1212.6829 [hep-ph]} \BibitemShut
  {NoStop}%
%%CITATION = ARXIV:1212.6829;%%
\bibitem [{\citenamefont {Jack}\ \emph {et~al.}(2015)\citenamefont {Jack},
  \citenamefont {Jones},\ and\ \citenamefont {Poole}}]{Jack:2015tka}%
  \BibitemOpen
  \bibfield  {author} {\bibinfo {author} {\bibfnamefont {I.}~\bibnamefont
  {Jack}}, \bibinfo {author} {\bibfnamefont {D.~R.~T.}\ \bibnamefont {Jones}},
  \ and\ \bibinfo {author} {\bibfnamefont {C.}~\bibnamefont {Poole}},\ }\href
  {\doibase 10.1007/JHEP09(2015)061} {\bibfield  {journal} {\bibinfo  {journal}
  {JHEP}\ }\textbf {\bibinfo {volume} {09}},\ \bibinfo {pages} {061} (\bibinfo
  {year} {2015})},\ \Eprint {http://arxiv.org/abs/1505.05400} {arXiv:1505.05400
  [hep-th]} \BibitemShut {NoStop}%
%%CITATION = ARXIV:1505.05400;%%
\bibitem [{\citenamefont {Jafferis}\ \emph {et~al.}(2011)\citenamefont
  {Jafferis}, \citenamefont {Klebanov}, \citenamefont {Pufu},\ and\
  \citenamefont {Safdi}}]{Jafferis:2011zi}%
  \BibitemOpen
  \bibfield  {author} {\bibinfo {author} {\bibfnamefont {D.~L.}\ \bibnamefont
  {Jafferis}}, \bibinfo {author} {\bibfnamefont {I.~R.}\ \bibnamefont
  {Klebanov}}, \bibinfo {author} {\bibfnamefont {S.~S.}\ \bibnamefont {Pufu}},
  \ and\ \bibinfo {author} {\bibfnamefont {B.~R.}\ \bibnamefont {Safdi}},\
  }\href {\doibase 10.1007/JHEP06(2011)102} {\bibfield  {journal} {\bibinfo
  {journal} {JHEP}\ }\textbf {\bibinfo {volume} {06}},\ \bibinfo {pages} {102}
  (\bibinfo {year} {2011})},\ \Eprint {http://arxiv.org/abs/1103.1181}
  {arXiv:1103.1181 [hep-th]} \BibitemShut {NoStop}%
%%CITATION = ARXIV:1103.1181;%%
\bibitem [{\citenamefont {Klebanov}\ \emph {et~al.}(2011)\citenamefont
  {Klebanov}, \citenamefont {Pufu},\ and\ \citenamefont
  {Safdi}}]{Klebanov:2011gs}%
  \BibitemOpen
  \bibfield  {author} {\bibinfo {author} {\bibfnamefont {I.~R.}\ \bibnamefont
  {Klebanov}}, \bibinfo {author} {\bibfnamefont {S.~S.}\ \bibnamefont {Pufu}},
  \ and\ \bibinfo {author} {\bibfnamefont {B.~R.}\ \bibnamefont {Safdi}},\
  }\href {\doibase 10.1007/JHEP10(2011)038} {\bibfield  {journal} {\bibinfo
  {journal} {JHEP}\ }\textbf {\bibinfo {volume} {10}},\ \bibinfo {pages} {038}
  (\bibinfo {year} {2011})},\ \Eprint {http://arxiv.org/abs/1105.4598}
  {arXiv:1105.4598 [hep-th]} \BibitemShut {NoStop}%
%%CITATION = ARXIV:1105.4598;%%
\bibitem [{\citenamefont {Intriligator}\ and\ \citenamefont
  {Wecht}(2003)}]{Intriligator:2003jj}%
  \BibitemOpen
  \bibfield  {author} {\bibinfo {author} {\bibfnamefont {K.~A.}\ \bibnamefont
  {Intriligator}}\ and\ \bibinfo {author} {\bibfnamefont {B.}~\bibnamefont
  {Wecht}},\ }\href {\doibase 10.1016/S0550-3213(03)00459-0} {\bibfield
  {journal} {\bibinfo  {journal} {Nucl. Phys.}\ }\textbf {\bibinfo {volume}
  {B667}},\ \bibinfo {pages} {183} (\bibinfo {year} {2003})},\ \Eprint
  {http://arxiv.org/abs/hep-th/0304128} {arXiv:hep-th/0304128 [hep-th]}
  \BibitemShut {NoStop}%
%%CITATION = HEP-TH/0304128;%%
\bibitem [{\citenamefont {Kutasov}(2003)}]{Kutasov:2003ux}%
  \BibitemOpen
  \bibfield  {author} {\bibinfo {author} {\bibfnamefont {D.}~\bibnamefont
  {Kutasov}},\ }\href@noop {} {\enquote {\bibinfo {title} {{New results on the
  'a theorem' in four-dimensional supersymmetric field theory}},}\ } (\bibinfo
  {year} {2003}),\ \Eprint {http://arxiv.org/abs/hep-th/0312098}
  {arXiv:hep-th/0312098 [hep-th]} \BibitemShut {NoStop}%
%%CITATION = HEP-TH/0312098;%%
\bibitem [{\citenamefont {Barnes}\ \emph {et~al.}(2004)\citenamefont {Barnes},
  \citenamefont {Intriligator}, \citenamefont {Wecht},\ and\ \citenamefont
  {Wright}}]{Barnes:2004jj}%
  \BibitemOpen
  \bibfield  {author} {\bibinfo {author} {\bibfnamefont {E.}~\bibnamefont
  {Barnes}}, \bibinfo {author} {\bibfnamefont {K.~A.}\ \bibnamefont
  {Intriligator}}, \bibinfo {author} {\bibfnamefont {B.}~\bibnamefont {Wecht}},
  \ and\ \bibinfo {author} {\bibfnamefont {J.}~\bibnamefont {Wright}},\ }\href
  {\doibase 10.1016/j.nuclphysb.2004.09.016} {\bibfield  {journal} {\bibinfo
  {journal} {Nucl. Phys.}\ }\textbf {\bibinfo {volume} {B702}},\ \bibinfo
  {pages} {131} (\bibinfo {year} {2004})},\ \Eprint
  {http://arxiv.org/abs/hep-th/0408156} {arXiv:hep-th/0408156 [hep-th]}
  \BibitemShut {NoStop}%
%%CITATION = HEP-TH/0408156;%%
\bibitem [{\citenamefont {Freedman}\ and\ \citenamefont
  {Osborn}(1998)}]{Freedman:1998rd}%
  \BibitemOpen
  \bibfield  {author} {\bibinfo {author} {\bibfnamefont {D.~Z.}\ \bibnamefont
  {Freedman}}\ and\ \bibinfo {author} {\bibfnamefont {H.}~\bibnamefont
  {Osborn}},\ }\href {\doibase 10.1016/S0370-2693(98)00649-2} {\bibfield
  {journal} {\bibinfo  {journal} {Phys. Lett.}\ }\textbf {\bibinfo {volume}
  {B432}},\ \bibinfo {pages} {353} (\bibinfo {year} {1998})},\ \Eprint
  {http://arxiv.org/abs/hep-th/9804101} {arXiv:hep-th/9804101 [hep-th]}
  \BibitemShut {NoStop}%
%%CITATION = HEP-TH/9804101;%%
\bibitem [{\citenamefont {Jack}\ \emph
  {et~al.}(1994{\natexlab{a}})\citenamefont {Jack}, \citenamefont {Jones},\
  and\ \citenamefont {Roberts}}]{Jack:1993ws}%
  \BibitemOpen
  \bibfield  {author} {\bibinfo {author} {\bibfnamefont {I.}~\bibnamefont
  {Jack}}, \bibinfo {author} {\bibfnamefont {D.~R.~T.}\ \bibnamefont {Jones}},
  \ and\ \bibinfo {author} {\bibfnamefont {K.~L.}\ \bibnamefont {Roberts}},\
  }\href {\doibase 10.1007/BF01559535} {\bibfield  {journal} {\bibinfo
  {journal} {Z. Phys.}\ }\textbf {\bibinfo {volume} {C62}},\ \bibinfo {pages}
  {161} (\bibinfo {year} {1994}{\natexlab{a}})},\ \Eprint
  {http://arxiv.org/abs/hep-ph/9310301} {arXiv:hep-ph/9310301 [hep-ph]}
  \BibitemShut {NoStop}%
%%CITATION = HEP-PH/9310301;%%
\bibitem [{\citenamefont {Jack}\ \emph
  {et~al.}(1994{\natexlab{b}})\citenamefont {Jack}, \citenamefont {Jones},\
  and\ \citenamefont {Roberts}}]{Jack:1994bn}%
  \BibitemOpen
  \bibfield  {author} {\bibinfo {author} {\bibfnamefont {I.}~\bibnamefont
  {Jack}}, \bibinfo {author} {\bibfnamefont {D.~R.~T.}\ \bibnamefont {Jones}},
  \ and\ \bibinfo {author} {\bibfnamefont {K.~L.}\ \bibnamefont {Roberts}},\
  }\href {\doibase 10.1007/BF01577555} {\bibfield  {journal} {\bibinfo
  {journal} {Z. Phys.}\ }\textbf {\bibinfo {volume} {C63}},\ \bibinfo {pages}
  {151} (\bibinfo {year} {1994}{\natexlab{b}})},\ \Eprint
  {http://arxiv.org/abs/hep-ph/9401349} {arXiv:hep-ph/9401349 [hep-ph]}
  \BibitemShut {NoStop}%
%%CITATION = HEP-PH/9401349;%%
\bibitem [{\citenamefont {Jones}\ and\ \citenamefont
  {Mezincescu}(1984)}]{Jones:1983vk}%
  \BibitemOpen
  \bibfield  {author} {\bibinfo {author} {\bibfnamefont {D.~R.~T.}\
  \bibnamefont {Jones}}\ and\ \bibinfo {author} {\bibfnamefont
  {L.}~\bibnamefont {Mezincescu}},\ }\href {\doibase
  10.1016/0370-2693(84)91154-7} {\bibfield  {journal} {\bibinfo  {journal}
  {Phys. Lett.}\ }\textbf {\bibinfo {volume} {136B}},\ \bibinfo {pages} {242}
  (\bibinfo {year} {1984})}\BibitemShut {NoStop}%
%%CITATION = PHLTA,136B,242;%%
\bibitem [{\citenamefont {Novikov}\ \emph {et~al.}(1983)\citenamefont
  {Novikov}, \citenamefont {Shifman}, \citenamefont {Vainshtein},\ and\
  \citenamefont {Zakharov}}]{Novikov:1983uc}%
  \BibitemOpen
  \bibfield  {author} {\bibinfo {author} {\bibfnamefont {V.~A.}\ \bibnamefont
  {Novikov}}, \bibinfo {author} {\bibfnamefont {M.~A.}\ \bibnamefont
  {Shifman}}, \bibinfo {author} {\bibfnamefont {A.~I.}\ \bibnamefont
  {Vainshtein}}, \ and\ \bibinfo {author} {\bibfnamefont {V.~I.}\ \bibnamefont
  {Zakharov}},\ }\href {\doibase 10.1016/0550-3213(83)90338-3} {\bibfield
  {journal} {\bibinfo  {journal} {Nucl. Phys.}\ }\textbf {\bibinfo {volume}
  {B229}},\ \bibinfo {pages} {381} (\bibinfo {year} {1983})}\BibitemShut
  {NoStop}%
%%CITATION = NUPHA,B229,381;%%
\bibitem [{\citenamefont {Novikov}\ \emph {et~al.}(1986)\citenamefont
  {Novikov}, \citenamefont {Shifman}, \citenamefont {Vainshtein},\ and\
  \citenamefont {Zakharov}}]{Novikov:1985rd}%
  \BibitemOpen
  \bibfield  {author} {\bibinfo {author} {\bibfnamefont {V.~A.}\ \bibnamefont
  {Novikov}}, \bibinfo {author} {\bibfnamefont {M.~A.}\ \bibnamefont
  {Shifman}}, \bibinfo {author} {\bibfnamefont {A.~I.}\ \bibnamefont
  {Vainshtein}}, \ and\ \bibinfo {author} {\bibfnamefont {V.~I.}\ \bibnamefont
  {Zakharov}},\ }\href {\doibase 10.1016/0370-2693(86)90810-5} {\bibfield
  {journal} {\bibinfo  {journal} {Phys. Lett.}\ }\textbf {\bibinfo {volume}
  {166B}},\ \bibinfo {pages} {329} (\bibinfo {year} {1986})},\ \bibinfo {note}
  {[Yad. Fiz.43,459(1986)]}\BibitemShut {NoStop}%
%%CITATION = PHLTA,166B,329;%%
\bibitem [{\citenamefont {Shifman}\ and\ \citenamefont
  {Vainshtein}(1986)}]{Shifman:1986zi}%
  \BibitemOpen
  \bibfield  {author} {\bibinfo {author} {\bibfnamefont {M.~A.}\ \bibnamefont
  {Shifman}}\ and\ \bibinfo {author} {\bibfnamefont {A.~I.}\ \bibnamefont
  {Vainshtein}},\ }\href {\doibase 10.1016/0550-3213(86)90451-7} {\bibfield
  {journal} {\bibinfo  {journal} {Nucl. Phys.}\ }\textbf {\bibinfo {volume}
  {B277}},\ \bibinfo {pages} {456} (\bibinfo {year} {1986})},\ \bibinfo {note}
  {[Zh. Eksp. Teor. Fiz.91,723(1986)]}\BibitemShut {NoStop}%
%%CITATION = NUPHA,B277,456;%%
\bibitem [{\citenamefont {{C.~McLarty}}(2003)}]{McLarty:xxx}%
  \BibitemOpen
  \bibfield  {author} {\bibinfo {author} {\bibnamefont {{C.~McLarty}}},\ }\href
  {http://www.landsburg.com/grothendieck/mclarty1.pdf} {\enquote {\bibinfo
  {title} {The rising sea: Grothendieck on simplicity and generality},}\ }
  (\bibinfo {year} {2003}),\ \bibinfo {note} {[Online; accessed
  7-June-2019]}\BibitemShut {NoStop}%
\bibitem [{\citenamefont {Ellis}(2017)}]{Ellis:2016jkw}%
  \BibitemOpen
  \bibfield  {author} {\bibinfo {author} {\bibfnamefont {J.}~\bibnamefont
  {Ellis}},\ }\href {\doibase 10.1016/j.cpc.2016.08.019} {\bibfield  {journal}
  {\bibinfo  {journal} {Comput. Phys. Commun.}\ }\textbf {\bibinfo {volume}
  {210}},\ \bibinfo {pages} {103} (\bibinfo {year} {2017})},\ \Eprint
  {http://arxiv.org/abs/1601.05437} {arXiv:1601.05437 [hep-ph]} \BibitemShut
  {NoStop}%
%%CITATION = ARXIV:1601.05437;%%
\end{thebibliography}%
	
\end{document}